\documentclass[prd,onecolumn,amsmath,amssymb,superscriptaddress,nofootinbib,preprintnumbers]{revtex4-2}
\usepackage[T1]{fontenc}
\usepackage{hyperref}
\usepackage{customcommands}
\usepackage{amsmath}
\usepackage{amsthm}
\usepackage{amsfonts}
\usepackage{xcolor}
\usepackage{slashed}
\usepackage{mathtools}

\def\eq#1 { \begin{equation} #1 \end{equation} }
\def\eqn#1{ \begin{eqnarray} #1 \end{eqnarray} }
\def\nn { \nonumber }
\def\nl { \nonumber \\}

\def\cK{\mathcal{K}}

\def\bb{\vec{b}}
\def\bk{\vec{k}}

\def\bK{\vec{K}}

\def\bl{\vec{\ell}}

\def\bx{\vec{x}}

\def\bv{\vec{v}}
\def\bw{\vec{w}}
\def\bb{\vec{b}}
\def\bc{\vec{c}}
\def\bl{\vec{\ell}}

\begin{document}

\preprint{Imperial/TP/2023/TW/01 \qquad 
APCTP Pre2023 - 006}

\title{Analog gravity and the continuum effective theory of the graphene tight binding lattice model}

\author{Matthew M.~Roberts}
\email{matthew.roberts@apctp.org}

	\affiliation{Asia Pacific Center for Theoretical Physics, Pohang, 37673, Korea}
	\affiliation{Department of Physics, Pohang University of Science and Technology, Pohang 37673, Korea}

\author{Toby Wiseman}
\email{t.wiseman@imperial.ac.uk}

\affiliation{Theoretical Physics Group, Blackett Laboratory, Imperial College, London SW7 2AZ, United Kingdom}


\begin{abstract}

We consider the nearest-neighbour lattice tight-binding model of graphene with slowly spatially varying hopping functions. We develop a systematic low energy approximation as a derivative expansion in a Dirac spinor field that is perturbative in the strength of the hopping function deformation.
The well known leading description in both the derivative and perturbative expansions is the Dirac equation in flat 2+1-dimensional spacetime with magnetic (strain-)gauge field. 
Prior work has considered subleading corrections written as non-trivial frame and spin connection terms. We have argued previously that such corrections cannot be considered consistently without taking all the terms at the same order of approximation, which due to the unconventional power counting originating from the large gauge field, involve also higher covariant derivative terms.
Here we confirm this, explicitly computing subleading terms in this approximation.
To the order we explore, the theory is elegantly determined by the non-trivial gauge field and frame, both given by the slowly varying hopping functions, the torsion free spin connection of the frame, together with coefficients for the higher derivative terms derived from lattice invariants. We stress the importance of the local frame and gauge symmetries that are inherited from matching to the lattice model.
For the first time we compute the metric that the Dirac field sees -- the `electrometric' -- to quadratic order in the hopping function deformation. 
This allows us to describe the subleading corrections to the dispersion relation for inhomogeneous deformations that originate from corrections to the frame.
Focussing on purely in-plane inhomogeneous strain, 
we use a simple model to relate the hopping functions to the strain field, finding the electrometric becomes curved at this quadratic order.
Thus this lattice model yields an effective `analog gravity' description as a curved space Dirac theory, with large magnetic field, and Lorentz violating higher covariant derivative terms. 
We check our calculations by a simple numerical diagonalization of the lattice model for a periodic arm-chair deformation, 
confirming that frame corrections contribute at the same order of approximation as higher covariant derivative terms. 
Finally, based on our lattice model derivation, we conjecture a form for the effective theory 
for monolayer graphene, written in terms of the strain tensor, and consistent up to quadratic order in the electrometric deformation.

\end{abstract} 

\maketitle

\tableofcontents


\section{Introduction}

Monolayer graphene is well known for having a band structure with two massless Dirac cones, protected by inversion and time reversal, and when undoped the chemical potential sits precisely at the Dirac points, giving a low energy spectrum of massless Dirac fermions \cite{novoselov2005two,novoselov2005two,Zhang:2005zz} that governs transport.
It forms a flexible membrane~\cite{lee2008measurement,naumis2017electronic}, and thus it is natural to think that if this monolayer membrane is deformed to be curved there will be a corresponding curved spacetime Dirac equation description of transport. If it is true that elastically deformed graphene has a low energy description in terms of relativistic fields in curved spacetime, it would then be a prime candidate of an `analog gravity' model and provide an important link between physics in curved spacetimes and transport in graphene.\footnote{We are careful to emphasize that by `analog gravity' we mean a relativistic curved spacetime description, rather than a theory with a dynamical spacetime governed by an Einstein-like equation.} 
This question can be addressed by considering the nearest-neighbour tight-binding model, where one assumes the conductance and valence electron wavefunctions are localized to the carbon atoms. To model mild lattice distortions the tight-binding model can be generalized to have nearest-neighbour tunneling amplitudes that vary in space. This leads to a continuum limit which at leading order is described by massless Dirac fields in flat space coupled to an effective magnetic field which is proportional to the lattice strain, and is thus often referred to as the ``strain gauge field'' -- not to be confused with the actual Maxwell field of electromagnetism which we will not play a role here. This was originally noted before the discovery of graphene in the study of carbon nanotubes \cite{sasaki2005local,PhysRevB.65.235412} and was quickly generalized to monolayer graphene, which is simply an unrolled nanotube (for a review see \cite{vozmediano:physrev}). It was shortly afterwards seen that some subleading corrections to this description can be written as frame and spin connection terms \cite{dejuan2012spacedepfermi,Zubkov:2013sja,oliva2015generalizing,yang2015dirac,VOLOVIK2015255,si2016strain,khaidukov2016landau,oliva2017low,wagner2019quantum}, which many people took to imply precisely such  an effective description in terms of a curved space Dirac equation coupled to the strain gauge field. Indeed a large body of literature has studied graphene using some flavor of continuum Dirac equation in curved space as a tool \cite{de2007charge,guinea2008gauge,vozmediano2008gauge,de2013gauge,arias2015gauge,stegmann2016current,castro2017pseudomagnetic,Golkar:2014paa,Golkar:2014wwa}.

Unfortunately, this tempting interpretation suffers from two flaws. First, as we argued in \cite{Roberts:2021vmt}, without fine-tuning of the lattice model there are generically higher covariant derivative terms which contribute at the same order as those of the frame and spin connection. This is due to the fact that the full covariant derivative includes the strain gauge field, which is anomalously large by a factor of the inverse lattice spacing, ruining the expected power counting of higher derivative terms in relativistic theories. This large gauge field was noted in \cite{Zubkov:2013sja,VOLOVIK2014352}, but was interpreted as meaning that while one should include the nontrivial frame, the spin connection (and torsion) should be ignored, leading to a so-called ``Weitzenb\"ock'' geometry.\footnote{For another perspective on whether curved space Dirac is a good description, see
\cite{iorio2015revisiting}, and also \cite{iorio2022comment,roberts2022reply}.
} Our interpretation in~\cite{Roberts:2021vmt} and here is very different - we find a perfectly conventional torsion-free Riemannian geometry, it is simply that the curvature of this geometry is subleading to the large magnetic field. This means that there is no consistent truncation in a gradient expansion which is that of a relativistic Dirac equation in curved spacetime coupled to a magnetic field. To be consistent, the effective low energy theory must include higher covariant derivative terms as well as the term that yields the Dirac equation and gauge field, and these additional terms physically contribute at the same order of approximation as effects from spatial curvature. The spatial metric that governs this theory we term the electronic metric, or `electrometric'.

Second, there is the issue of precisely what type of lattice distortions we wish to study. In undistorted graphene, the $sp^2$ and $p_z$ orbitals are orthogonal and thus the tunneling amplitude between them vanishes. When considering out-of-plane deformations, this is no longer the case, and it is not clear that the Dirac cone structure remains~\cite{PhysRevB.77.205421,kim2008graphene,guinea2008midgap,wehling2008midgap,PhysRevB.84.245444,lin2015feature,lopez2016magnetic,talla2022structural}, and therefore not clear that the nearest-neighbour tight-binding model describing only the $p_z$ orbitals provides a good  approximation.
One can avoid this issue by restricting to the case of pure in-plane deformations, which are still very physically interesting. However in this case to leading order in the elastic distortion the electrometric geometry seen in the curved space Dirac equation of~\cite{dejuan2012spacedepfermi,oliva2015generalizing} is also flat, and just corresponds to a coordinate transformation.

In this paper we address both these issues. Concerning the first, we show explicitly how to construct the low energy effective theory of Dirac points in the nearest-neighbour tight binding model. The leading behaviour is governed by the flat space Dirac equation with strain gauge field. Then subleading corrections involving spatial curvature of the electrometric can be consistently included if higher covariant derivative terms are also added.  The higher the order of perturbation to the flat electrometric that we wish to work to, the more derivatives are required. This enables us to explicitly derive the effective theory to quadratic order in the perturbation to the electrometric, going beyond leading order for inhomogeneous deformations the first time.  
Homogeneous but anisotropic deformations had previous been considered to this order~\cite{oliva2017low} but without spatial variation these do not lead to magnetic strain fields and can only give a rigid coordinate transformation of flat space for the electrometric.
We find our effective theory involves the expected nontrivial ``strain'' gauge field, frame and torsion-free spin connection, and requires including terms with up to \emph{three} covariant derivatives, which we give explicitly. These higher covariant derivative terms are not Lorentz invariant (as the leading covariant Dirac equation term is), instead inheriting index structure coming from lattice invariants, and contribute to the physics at the same order of approximation the curvature of the electrometric.
Working to this subleading order allows us to study the electrometric corrections to the dispersion relation for inhomogeneous deformations of the hopping functions. While corrections to the dispersion relation arise at leading linear order in the electrometric perturbation, translation symmetry of the undeformed system implies that they are not sensitive to inhomogeneity, and this linear correction only responds to the homogeneous part of the deformation.

While our derivation of this effective theory can be phrased purely in terms of spatially varying tunneling amplitude functions of the lattice model, in order to make contact with graphene we discuss deriving such a lattice model from embedding the graphene lattice as a deformed membrane in 3d space. We use a simple bond model to relate the tunneling functions to the length deformation of the embedded lattice links. Restricting ourselves to the case of pure in-plane deformations, the induced geometry of the embedding is trivially flat, simply that of a 2d plane. While to leading order in this strain deformation the electrometric remains flat, with the new tool of our effective theory which is valid to quadratic order, we find that indeed the electrometric geometry generically becomes curved at this order. However, we again emphasize that the effect of this curvature is subleading to the large effective `strain' magnetic field induced by the lattice deformation, and furthermore, it contributes at the same order as the Lorentz violating higher covariant derivative terms.

Our effective theory is coordinate invariant -- the various tensors used to construct it are naturally phrased in lattice coordinates, but then once one has them, one can employ any coordinate system. 
Following the work of~\cite{oliva2013understanding,oliva2015generalizing,balents:twisted_bilayer} we give an explicit discussion of how, given an embedding of the lattice, we may transform to the natural lab coordinates. 
To make an explicit comparison of our effective theory to the tight binding model, we solve this continuum description for a certain class of ``armchair'' distortions that arise from a periodic in-plane strain along a lattice direction. 
Given that we have the theory to quadratic order in the electrometric perturbation, we are able to compute the dispersion relation consistently to include the subleading effects due to electrometric curvature.
We compare our effective theory to explicit numerical diagonalization of the nearest neighbour tight binding model.
Firstly we show that the leading behaviour is indeed governed by the flat space Dirac equation with gauge field. Then we confirm that the subleading behaviour, involving the curvature of the electrometric, is correctly captured by our effective theory at quadratic order in the deformation. In particular we confirm that the higher covariant derivative terms are essential to match the subleading behaviour, and can cannot be truncated away if one wishes to see the effects of curvature of the electrometric.

The paper is structured as follows. Firstly, in Section~\ref{sec:tightbinding} we review the nearest-neighbour tight binding model with spatially varying tunneling amplitudes, and discuss its low energy continuum limit. Since the derivation of the low energy effective theory is somewhat involved, we give a summary of its structure in Section~\ref{sec:summary} showing how it is constructed in lattice coordinates from the varying tunneling amplitudes. We then discuss how these tunneling amplitudes may arise from embedding a lattice in 3d space using a simple bond model in section~\ref{sec:embedding}, and further show how to transform the various elements of the effective theory to the natural lab frame coordinates. In Section~\ref{sec:deriveeffective} we then discuss the structure of the effective theory, and give its explicit construction up to the order which consistently includes quadratic corrections to the electrometric. We then test this effective description in Section~\ref{sec:armchair} by comparing its predictions to a direct numerical diagonalization of the lattice model for a periodic armchair deformation. Finally, given the structure of our effective theory for the lattice model, in Section~\ref{sec:EffectiveTheory} we conjecture the structure of the effective theory for true strained monolayer graphene, again up to quadratic perturbations to the electrometric, before concluding.

\section{Review of the spatially deformed graphene tight-binding model}
\label{sec:tightbinding}

The atoms of the graphene lattice may be described by their positions in either 2-d lattice coordinates $x^i = (x,y)$ or 3-d lab frame coordinates $(X, Y, Z)$. 
For undistorted graphene we will take the plane of atoms to be located at $Z = 0$, and writing $X^I = (X, Y)$, choose our lattice coordinates to be $x^i = \delta^i_I {X}^I$.
The lattice sites subdivide into A and B triangular sublattices, and we label the lattice coordinate position of these as $\vec{x}_A$ and $\vec{x}_B$ respectively.

\begin{figure}[h!]
\begin{center}
\includegraphics[scale=.2]{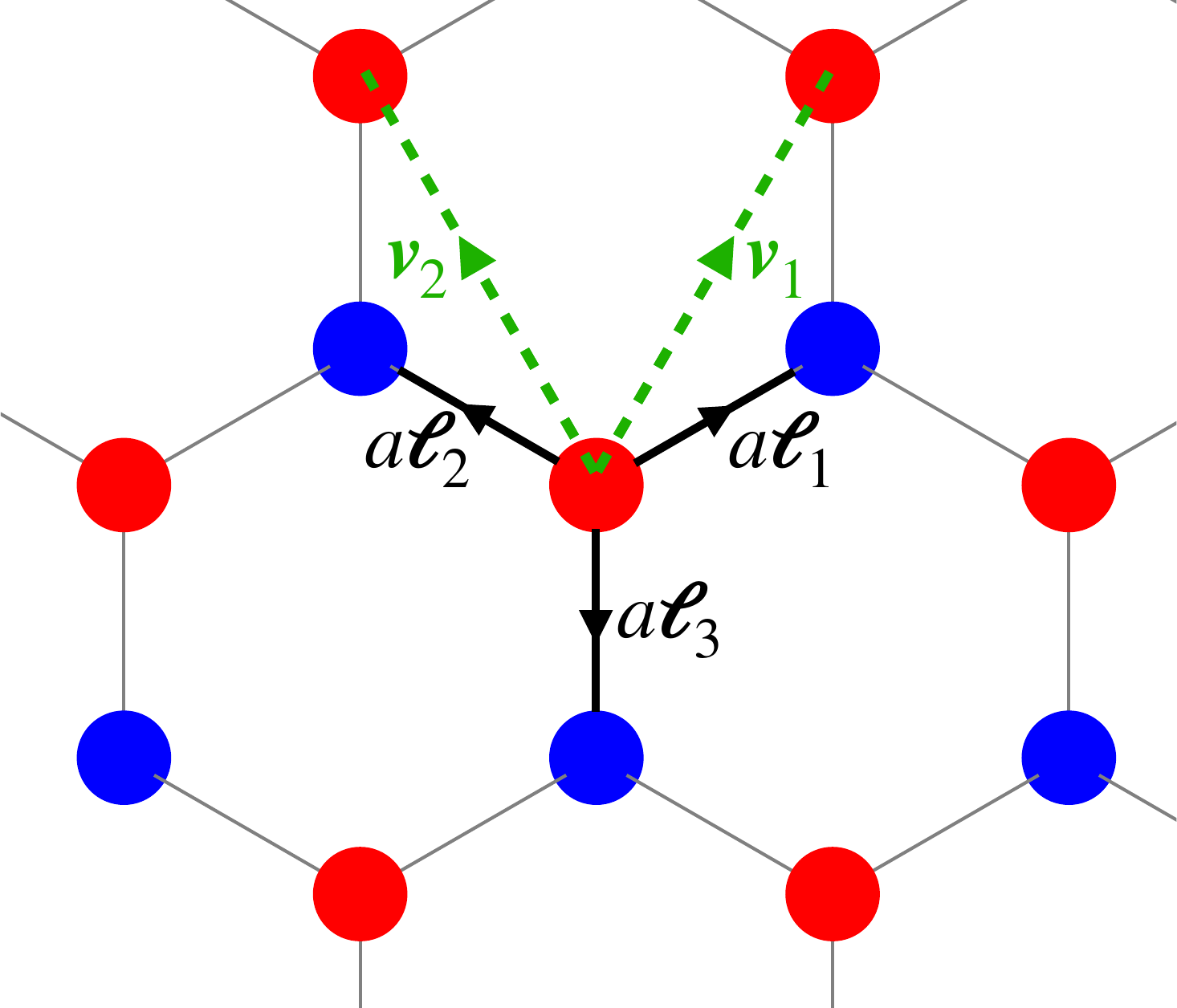} \qquad\qquad\qquad
\includegraphics[scale=.2]{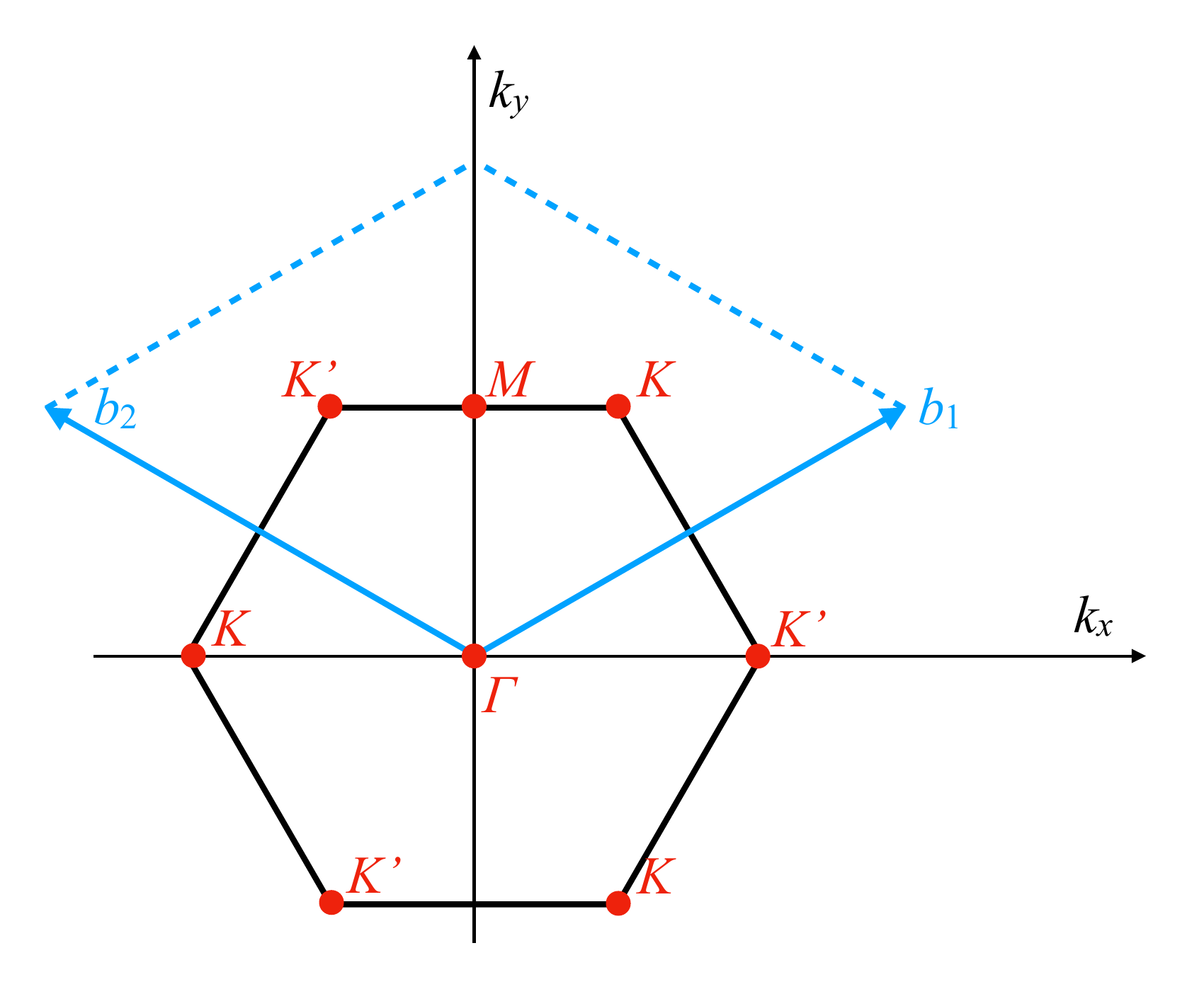}
\caption{Left: the honeycomb lattice, with red A sites and blue B sites, related by translations by $a \vec{\ell}_{1,2,3}$. The lattice symmetry is generated by translations $\vec{v}_{1,2}.$ Right: The standard hexagonal fundamental domain of the Brillouin zone and massless Dirac points at $K$ and $K'$ for the undistorted lattice model.
\label{fig:honeycomb_sym_bz}}
\end{center}
\end{figure}

The lattice sites lie a distance $a$ from their nearest neighbours. The translation vectors between sites may be given in terms of unit vectors,
\eqn{
\bl_1 =  \left( \frac{\sqrt{3}}{2},\frac{1}{2} \right) \;, \quad \bl_2 =  \left(- \frac{\sqrt{3}}{2},\frac{1}{2} \right)  \;, \quad\bl_3  = -\bl_1 - \bl_2 = \left( 0, -1 \right)
}
so that defining $\vec{v}_{1,2} = a \left( \bl_{1,2} - \bl_3 \right)$ then the lattice sites are at lattice coordinates,
\eqn{
\vec{x}_{A,B} =  m \, \vec{v}_1 + n \, \vec{v}_2  \mp \frac{a}{2} \bl_3 
}
generated by $( m,n)  \in \mathbb{Z}^2$, with the sign above giving the A and B triangular sublattices, see figure \ref{fig:honeycomb_sym_bz}. 

Some important relations we will use later are,
\eqn{
\label{eq:Ktensor}
\delta^{ij} = \frac{2}{3}  \sum_n \ell^i_n \ell^j_n \; , \quad  \quad K^{ijk} = \frac{4}{3} \sum_n \ell^i_n \ell^j_n \ell^k_n \; , \quad \frac{8}{3} \sum_n \ell^i_n \ell^j_n \ell^k_n \ell^l_n =  \delta^{ij} \delta^{kl} + \delta^{ik} \delta^{jl}  + \delta^{il} \delta^{jk}
}
where $K^{ijk}$ is a natural invariant traceless symmetric tensor for the lattice, with $K^{112} = 1$ and $K^{222} = -1$.
In the nearest-neighbour tight binding approximation, the $\pi$ electrons are described by the Hamiltonian,
\eqn{
\label{eq:tightbindingundeformed}
H_{undeformed} &=& T \sum_{n,\bx_A} \left( a^\dagger_{\bx_A} b_{\bx_A +a \bl_n} + \mathrm{h.c.}\right)
}
where $T$ gives the tunneling amplitude between $p_z$
orbitals on adjacent lattice sites, and $a_{\bx_A}^\dagger,~b_{\bx_B}^\dagger$ are fermionic creation operators on the respective sublattices $A$ and $B$. The dual lattice generators $\vec{b}_{1,2}$, are defined by $\vec{b}_i \cdot \vec{v}_j = 2 \pi \delta_{ij}$ and one finds the spectrum of this model has two inequivalent Dirac points, labelled $K$ and $K'$, which are illustrated together with the hexagonal fundamental domain of the Brillouin zone in figure~\ref{fig:honeycomb_sym_bz}.

Now a natural generalization of this is to allow the tunneling amplitudes associated to each link of the lattice to vary. Denoting the tunneling amplitude between the A-site at $\bx_A$ and the B-site at $\bx_A + a \ell_n$ as $T_{n, A}$, which we assume to again be real, then yields the Hamiltonian,
\eqn{
\label{eq:tightbindingdeformed}
H_{deformed} &=&  \sum_{n,\bx_A} T_{n,A} \left( a^\dagger_{\bx_A} b_{\bx_A +a \bl_n} + \mathrm{h.c.}\right) \; .
}
A special case is taking $T_{n,A} = T_n$, so they don't depend on the lattice site location. This is an anisotropic but homogeneous deformation.
In principle the $T_{n,A}$ may vary arbitrarily between sites. However in this work we are interested in the situation that:
\begin{itemize}
\item Firstly, the $T_{n,A}$ are only deformed perturbatively from the homogeneous amplitude $T$. We introduce a deformation parameter $\epsilon$ and take the amplitudes to be,
\eqn{
\label{eq:Texp}
T_{n, A} = T \left( 1 + \epsilon \delta_{1}t_{n, A}  + \epsilon^2 \delta_{2}t_{n, A}  + \ldots \right)
}
where we take the $\delta_{k}t_{n, A} \sim O(1)$ so that $\epsilon$ controls the size of the deformation.
\item Secondly we are interested in the situation that the perturbations $\delta_{k}t_{n,A}$ vary slowly on lattice scales so that we may consider a continuum limit where we parametrically separate the lattice length scale, $a$, from the scale of the slow spatial variation. In order to describe the approach to this continuum limit we further expand the $\delta_{k}t_{n, A}$ in $a$,
\eqn{
\delta_{k}t_{n, A} = \delta_{k,0}t_{n, A} + a \delta_{k,1}t_{n, A} + a^2 \delta_{k,2}t_{n, A}  + \ldots
}
and then write the coefficients $\delta_{k,m}t_{n, A}$ in terms of smooth functions, $\delta_{k,m}t_n(\bx)$, which we term the `hopping functions', as,
\eqn{
\label{eq:tfn}
\delta_{k,m}t_{n,A} = \delta_{k,m}t_n(\bx_A + \frac{1}{2} a \bl_n)
}
so that the value $\delta_{k,m}t_{n,A}$ associated to a link between sites $\bx_A$ and $\bx_A + a \ell_n$ is given by the function $\delta_{k,m}t_n(\bx)$ of the lattice coordinates $\bx$, evaluated at the mid point of this link. We also require that these functions $\delta_{k,m}t_n(\bx) \sim O(1)$.
\end{itemize}
In order that the $\delta_{k,m}t_{n,A}$ slowly vary on lattice scales, we then require that the functions $\delta_{k,m}t_n(\bx)$ slowly vary relative to the lattice scale $a$, encoded in the condition that everywhere
$\left| \partial_i \delta_{k,m}t_n \right| \ll 1/a$.
We then describe the continuum limit by specifying the $\delta_{k,m}t_n(\bx)$ as functions of the lattice coordinates, which are fixed with no $\epsilon$ or $a$ dependence, and then we consider the continuum limit by taking $a \to 0$. 

One might think that we should only be concerned with the leading behaviour in the continuum limit, $\delta_{k,m}t_{n,A} = \delta_{k,0}t_{n, A} + O(a)$, and subleading terms will encode irrelevant microscopic detail. However, this is not the case. Remembering that the tight-binding model is intended to be a microscopic description of graphene, here our aim is precisely to go beyond the leading low energy description, the flat space Dirac equation coupled to the strain magnetic field, and elucidate its subleading behaviour which will include the effects of a curved electrometric. Since these are subleading corrections of microscopic origin, we are forced to be careful to include these subleading details in taking the continuum limit.

We may define the characteristic minimum length scale associated to the variations, $L$, by,
\eqn{
1/L = \mathrm{max}\left\{  \left| \partial_i \delta_{k,m}t_n(\bx) \right| \right\} \; .
}
The condition of slowly varying deformations of the lattice then implies $a \ll L$. It is then convenient to choose units so that $L = 1$, and thus $a \ll 1$. From this point on we will employ these units unless otherwise stated. Thus noting that $\delta_{k,m}t_n(\bx) , \partial_i \delta_{k,m}t_n(\bx) \sim O(1)$ in these units for all $\bx$, then this implies all derivatives are also of order one, so
\eqn{
\label{eq:slowt}
\partial_{i_1} \partial_{i_2} \ldots \partial_{i_p} \delta_{k,m}t_n \sim O(1) 
}
for any $p \ge 0$.

Thus we have two expansion parameters in our model, firstly  the amplitude of the hopping strength deformation determined by $\epsilon$, and secondly in the length scale of the variation relative to $a$. It is important to understand the order of limits we consider. 
We will later find that in constructing the effective theory when we fix the length scale of the variation, taking these units $L = 1$, then the natural perturbative couplings are in fact $\epsilon/a$ and $a$, 
meaning that we should hold $\epsilon/a$ fixed as we scale towards the continuum limit $a \to 0$. In terms of orders of limits for perturbation expansions in $\epsilon$ and $a$, this implies that the order of limits we take is to first expand quantities in $\epsilon$, and after this expand in $a$.

\section{Summary of the effective theory for the lattice model}
\label{sec:summary}

Since the derivation of the low energy effective theory for the above lattice model is somewhat technical, we will summarize here its structure, and give results to the order that allows the electrometric to be described at quadratic order in the hopping function perturbation, so to $O(\epsilon^2)$. The full derivation of the results summarized here is given in  Section~\ref{sec:deriveeffective}, but we believe it is beneficial to have an overview of these results before delving into the technicalities.

Following the discussion above we write the lattice tight-binding Hamiltonian with perturbatively deformed hopping functions that are slowly varying as,
\eqn{
\label{eq:Hamiltonian}
H = T  \sum_{n,\bx_A} t_n( \bx_A + \frac{a}{2} \bl_n) \left( a^\dagger_{\bx_A} b_{\bx_A +a \bl_n} + \mathrm{h.c.}\right) 
}
and write the deformation of the hopping functions perturbatively in $\epsilon$ and $a$ as,
\eqn{
\label{eq:texp}
t_n( \bx ) = 1 + \epsilon \delta_1 t_n( \bx)    + \epsilon^2  \delta_2 t_n( \bx)  + \ldots 
} 
having factored out the equilibrium hopping strength $T$ above, where the $O(\epsilon^k)$ coefficient function is derived from our smooth functions $\delta_{k,m} t_n( \bx)$ described above as,
\eqn{
 \delta_k t_n( \bx) & =& \delta_{k,0} t_n( \bx) + a \delta_{k,1} t_n( \bx) + a^2 \delta_{k,2} t_n( \bx) + \ldots  \; .
}
We reiterate again that the $\delta_{k,m} t_n( \bx )$ which describe the continuum limit have no explicit $\epsilon$ or $a$ dependence, and are simply fixed functions of the lattice coordinates $\vec{x}$ as we scale towards the continuum, taking $a \to 0$, and deform the system with $\epsilon$.  
The low energy behaviour of the Dirac points of this lattice model are captured by the continuum effective theory living in 2+1-dimensions, whose truncation to three covariant derivatives takes the explicit form, 
 \eqn{
\label{eq:summarytheory}
0 =  a e^\mu_{~A} \gamma^A D_\mu \Psi \pm i a^2 \, \eta_{AB} \gamma^A e^B_{~\sigma} D_\mu \left( C^{\sigma\mu\nu} D_\nu \Psi \right) 
+ a^3 \,\eta_{AB} \gamma^A e^B_{~\sigma} D^{\sigma\mu\nu\rho}  D_\mu D_\nu D_\rho \Psi 
  + O(\epsilon^4, \epsilon^3 a, \epsilon^2 a^2, \epsilon a^3, a^4)  \nn \\
  }
where $\Psi$ is a 2-component Dirac spinor field, and the sign represents the choice of Dirac point that the theory is to describe. 
This is a free field theory, since the tight-binding lattice model has only hopping terms, and has no electron-electron interactions.
The spinor is normalized such that its particle number density agrees with the microscopic electron number density, as we discuss later in more detail.   Here $e^\mu_{~A}$ is the frame, the inverse of the coframe $e^A_{~\mu}$,  and associated to the spacetime metric $g_{\mu\nu}$ as $g_{\mu\nu} = e^A_{~\mu} e^B_{~\nu}  \eta_{AB}$ with $\eta_{AB} = \mathrm{diag}( -1, +1, +1)$ as usual.
The covariant derivative, $D_\mu$, encodes the strain gauge field and spin connection of the frame. For example, acting on the spinor,
\eqn{
\label{eq:covderiv}
D_\mu \Psi = \partial_\mu \Psi \mp i A_\mu \Psi - \frac{i}{2} \Omega_{\mu AB} S^{AB} \Psi 
}
with $A_\mu$ the gauge field, and the last term comprises the spin-connection $\Omega_{\mu AB}$ and Lorentz generators $S^{AB}$ and makes the theory geometric.\footnote{
Let us briefly comment on coupling the effective theory to an external electromagnetic field. For purely transverse magnetic fields and purely in-plane electric fields, we simply make the replacement $A_{strain} \rightarrow A_{strain}+A_{EM}$ for the $K$ point field, and for $K'$, $A_{strain} \rightarrow A_{strain}-A_{EM}$. It would be interesting to understand precisely how tilted fields would couple to the effective theory.
}
The signs in \eqref{eq:summarytheory} and \eqref{eq:covderiv} should be taken consistently, either choosing the upper or lower signs, and again reflect the choice of Dirac point being described -- thus the two Dirac fields, corresponding to the two distinct Dirac points, have opposite charge but couple to the same geometry.
The spin connection is simply the canonical torsion free one associated to the frame. While one doesn't expect to see torsion without dislocations~\cite{cortijo2007effects}, it is striking that it \emph{really is} the torsion free connection that enters here. We have no rigorous mathematical understanding why, beyond the heuristic of there being no dislocations.
The tensors $C^{\sigma\mu\nu}$ and $D^{\sigma\mu\nu\rho}$ derive from lattice invariants, and are remnants of the lattice structure. 
The truncation above including these higher covariant derivative terms allows us, for the first time, to consistently describe the metric to quadratic order in $O(\epsilon^2)$ which is one of our main goals here. Working to higher order in the metric deformation requires an increasing number of such higher covariant derivative terms. 
In particular while the dispersion relation of the Dirac points are corrected at $O(\epsilon)$, on general grounds they are only sensitive to the homogeneous (but anisotropic) part of the hopping function deformation (as for example studied in~\cite{oliva2013understanding}). They become sensitive to inhomogeneity in the deformation only at $O(\epsilon^2)$, and so our effective theory allows us access to these effects.

The theory is fully coordinate, frame and gauge covariant. However given the origin of the theory, it is natural to take time to be the usual lab time of the tight binding model. Then all the quantities entering above, apart from the dynamical field $\Psi$ itself, are independent of time, so static.\footnote{
One could in theory consider time-dependent elastic deformations of graphene, like in \cite{morales_timedep_graphene2023}, but this is outside of the scope of our analysis. 
} The metric takes the (ultrastatic) form,
\eqn{
ds^2_{effective} = - c_{eff}^2 dt^2 + g_{ij}(\vec{\sigma}) d\sigma^i d\sigma^j
}
for some 2-d spatial coordinates $\sigma^i$, where $c_{eff} = \frac{3 a T}{2 \hbar}$ gives the effective sound speed for the undeformed Dirac point. Writing the 2-d metric $g_{electro} = g_{ij}(\vec{\sigma}) d\sigma^i d\sigma^j$, we term the 2-d geometry given by the Riemannian manifold $( \mathbb{R}^2, g_{electro})$ the `electronic geometry', and $g_{electro}$ the `electronic metric' or more compactly the `electrometric'. We will use the notation $\Sigma_{electro} = ( \mathbb{R}^2, g_{electro})$.
The gauge field is also purely magnetic and static, 
\eqn{
A = A_i(\vec{\sigma}) d\sigma^i
}
and further, with this choice of frame, the tensors $C^{\sigma\mu\nu}$ and $D^{\sigma\mu\nu\rho}$ are orthogonal to the time direction, so they have only spatial components which are static. An interesting consequence of this is that the effective theory above remains second order in time derivatives, even though it has higher numbers of spatial derivatives. In particular, in this frame, the canonical momenta for the spinor is unchanged from that of the leading Dirac equation and the Hamiltonian is given as, 
\eqn{\label{eq:hamilt_genl}
\mathcal{H} = \int d^2x \sqrt{g} \left[  
a \bar\Psi \gamma^A e_A^i D_i \Psi \pm i a^2 \gamma_A e_i^A C^{ijk} D_j \bar\Psi D_k \Psi
+ a^3  \gamma_A e^A_i D^{ijk\ell}  D_j \bar\Psi D_k D_\ell \psi
+\ldots
\right] \; .
}
In order to define these various quantities above we take the spatial coordinates $\vec{\sigma}$ to be the lattice coordinates $\vec{x}$.
Then the electrometric takes the remarkably elegant form,
\eqn{
\label{eq:summarymetric}
g_{ij}(\bx) =  \frac{3}{\Delta^2} \sum_n ( \delta_{ij} - \frac{4}{3} \ell_n^i \ell_n^j ) t_n^2(\bx) + O(\epsilon^3, \epsilon^2 a, \epsilon a^2 ) \; , \quad 
\Delta^2 = ( \sum_n  t_n^2 )^2 - 2 ( \sum_m t_m^4 )
}
where this expression correctly gives the behaviour at orders $O(\epsilon)$, $O(\epsilon a)$ and $O(\epsilon^2)$ (in fact the $O(\epsilon a)$ contribution vanishes) which is consistent with the order that the theory is written to above.
Since we have a local frame invariance, any spatial frame components can be taken consistent with this metric.
An important point that will be discussed in detail later is that the subleading correction in $a$ at order $O(\epsilon)$, so the contribution going as $\sim \epsilon a$, must be included in order to derive the metric at order $O(\epsilon^2)$. Full detail of the frame components including subleading terms in $a$ will be given later. 
Defining, 
\eqn{
\Delta t_n = t_n - 1
}
then the explicit expression for the magnetic part of the gauge field to the order in $\epsilon$ and $a$ that the above truncation applies, is,
\eqn{
\label{eq:summarygauge}
A_i(\bx) &=& \frac{1}{a \Delta^2} \epsilon_{ij} \sum_m \Big[ \ell^j_m \Delta t_m \left( 2  + \sum_n \left( 3 \delta_{mn}  \Delta t_n \right) + \sum_{n,p} \left( \left( \frac{1}{3} + 2 \delta_{mn} - 3 \delta_{np} \right)  \Delta t_n  \Delta t_p  \right) \right) \nl
&& \qquad + a^2 \left( \frac{1}{4} \ell^j_m \ell^k_m \ell^l_m - \frac{3}{8} K^{jkl} + \frac{1}{6}  \delta^{jk}  \ell^l_m \right) \partial_k \partial_l \Delta t_m \Big] + O(\frac{\epsilon^4}{a},\epsilon^3 , \epsilon^2 a, \epsilon a^2)
}
up to a gauge transformation.
This expression encompasses the behaviour at orders $O(\epsilon/a)$, $O(\epsilon)$, $O(\epsilon a)$, $O(\epsilon^2/a)$ and $O(\epsilon^2)$, as well as $O(\epsilon^3/a)$, and as for the metric the first subleading corrections in $a$, here at orders $O(\epsilon)$ and $O(\epsilon^2)$ vanish.  Again these subleading corrections in $a$ at orders $O(\epsilon)$ and $O(\epsilon^2)$ are required in order to consistently solve for the metric.
Finally the tensors $C^{\sigma\mu\nu}$ and $D^{\sigma\mu\nu\rho}$ are given by the expressions, 
\eqn{
\label{eq:invariants}
\sqrt{ | g |} \, C^{ijk}(\bx) = - \frac{1}{3} \epsilon_{kl} \sum_n \ell^i_n \ell^j_n \ell^l_n = - \frac{1}{4} \epsilon_{kl} K^{ijl} \; , \quad \sqrt{ | g |} \, D^{ijkl}(\bx) = \frac{1}{9} \sum_n \ell^i_n \ell^j_n \ell^k_n \ell^l_n = \frac{1}{24} \left( \delta^{ik} \delta^{jl} + \delta^{il} \delta^{jk} + \delta^{ij} \delta^{kl} \right) \nl
}
with all other components (ie. those with a time index) vanishing, and here $| g | = \det( g_{ij} )$ and  $\epsilon_{ij}$ is the antisymmetric spatial Levi-Civita symbol with $1 = \epsilon_{12} = - \epsilon_{21}$. We emphasize that these expressions for the components $g_{ij}$, $A_i$, $C^{ijk}$ and $D^{ijkl}$ are not tensor equations, and hold only when we take lattice coordinates.

A key result that will be discussed in the next section is that using a simple model to map an in-plane distortion of the lattice to deformed hopping functions results in curvature of this electrometric at quadratic order $O(\epsilon^2)$ in the deformation. Thus even though the lattice is only deformed in-plane, the effective metric governing this Dirac theory generally becomes curved.

Usually in such an effective theory power counting goes with covariant derivatives, and so one may truncate to terms with some number of derivatives, and terms with more derivatives are subleading to this, and it is consistent to ignore them. This would be seen due to the increasing powers of $a$ in the coefficients of the higher derivative terms, and thus naively this makes increasingly higher derivative terms increasingly irrelevant in the low energy continuum limit where we take $a \to 0$ (in our units where the deformation scale is $O(1)$).
However the key novel feature of this effective theory is that since the gauge field goes as $A_i \sim O( \epsilon/a )$, there is mixing between covariant derivative orders in this theory due to the inverse factor of $a$. 
We refer to this inverse scaling with $a$ as giving a `large magnetic field' -- more precisely it is large relative to $\epsilon$, but we should tune $\epsilon$ such that its amplitude actually remains small if we are to stay in a regime where we may apply perturbation theory.
Very schematically the leading one derivative term goes as,
\eqn{
\label{eq:onederiv}
a e^\mu_{~A} \gamma^A D_\mu \Psi & \sim & a \partial \Psi  + \textcolor{red}{ \epsilon \tilde{A} \Psi } + \textcolor{blue}{  \epsilon a \partial \Psi }  + \textcolor{blue}{  \epsilon a \Psi } 
}
where we have suppressed all indices and written $A \sim \frac{\epsilon}{a} \tilde{A}$ so that $\tilde{A} \sim O(1)$. This contains the undeformed Dirac term, the first term on the righthand side, and a leading correction (in red) from the gauge field. These constitute the leading effective theory due to inhomogeneous hopping functions. While this red term naively dominates the Dirac term in terms of the expansion due to the gauge field have a factor of $1/a$, it is suppressed by a factor of $\epsilon$.  
As mentioned above, the natural coupling to hold fixed is $\epsilon/a$ as $a \to 0$, rather than simply $\epsilon$, as it is $\epsilon/a$ that controls the relative size of the gauge field contribution compared to the undeformed Dirac term. The blue terms are subleading to the red gauge field contribution, due to the factor of $a$, and come from the non-trivial frame. Now consider the same schematic expansion for the two derivative term; 
\eqn{
\label{eq:twoderiv}
i a^2 \, \eta_{AB} \gamma^A e^B_{~\sigma} D_\mu \left( 
 C^{\sigma\mu\nu} D_\nu \Psi \right) & \sim & 
 a^2 \partial^2 \Psi  + \textcolor{blue}{  \epsilon a \tilde{A}  \partial \Psi } + \textcolor{blue}{  \epsilon a (\partial \tilde{A} )  \Psi }  +  \textcolor{purple}{  \epsilon^2 \tilde{A}^2 \Psi  } +   \epsilon a^2 \partial^2 \Psi  +   \epsilon a^2 \partial \Psi +  \epsilon a^2 \Psi 
}
The key point is that the components of this where one of the covariant derivative contributes a gauge field (those in blue) are of the same order as the blue contribution from the one derivative term above in equation~\eqref{eq:onederiv}. Note that if both covariant derivatives contribute a gauge field (the purple term) the contribution is dominant in $a$, but suppressed now due to two powers of $\epsilon$. The blue contributions coming from both the one and two covariant derivative terms in equations~\eqref{eq:onederiv} and~\eqref{eq:twoderiv} then constitute the next correction to the effective theory at order $O(\epsilon)$ after the leading red term from the gauge field. Hence we see the (blue) frame corrections from the one derivative term mix with these contributions from the two derivative terms at the same order -- thus one cannot consider these frame corrections without also including the two derivative term too.

Due to this mixing we will see later that if we wish to consistently derive the contribution from the gauge field and metric at some order $\sim \epsilon^p a^q$, we are required to include up to $(1+p+q)$ covariant derivative terms, and we need all contributions to the metric and gauge field going as $\sim \epsilon^m a^n$ for $m \le p$ and $m + n \le p + q$, where $m \ge 1$ and for the metric corrections have $n \ge 0$ and for the gauge field they have $n \ge -1$. Thus the structure of the first few truncations is;
\begin{center}
\begin{tabular}{c|c|c}
Covariant derivatives included & Gauge field contributions & Metric contributions \\
\hline 
Dirac term only & $\frac{\epsilon}{a}$ & Trivial flat metric \\
\hline
Dirac + two derivative & $\frac{\epsilon}{a}$, $\epsilon$, $\frac{\epsilon^2}{a}$ & $\epsilon$ \\
\hline
Dirac, two and three derivatives &  $\frac{\epsilon}{a}$, $\epsilon$, $\epsilon a$, $\frac{\epsilon^2}{a}$, $\epsilon^2$  & $\epsilon$, $\epsilon a$, $\epsilon^2$
\end{tabular}
\end{center}
For the leading truncation to one covariant derivative we see there are no metric corrections -- it is simply the flat space Dirac equation with gauge field. Including the two derivative term allows the first consistent corrections to the metric, those at order $O(\epsilon)$. However, as noted, this is not sufficient to describe the corrections to the dispersion relation from inhomogeneous deformations. For that we require the theory given explicitly above, with up to three derivatives, which allows the metric deformation to be described at $O(\epsilon^2)$.

In the special case that the hopping functions are tuned so that the gauge field vanishes, at least at leading order $O(\epsilon/a)$, then conventional relativistic power counting is restored. In this case the leading effective theory is simply the curved spacetime Dirac equation as shown in~\cite{Roberts:2021vmt}. However, as discussed there, this tuning appears very unnatural -- we may think of it as having to fine tune away a relevant operator. Further, one can consider a simple model of elasticity for the graphene membrane and one finds that energetics do not prefer vanishing strain gauge field when distorted. Interestingly the metric above~\eqref{eq:summarymetric} is the one derived in~\cite{Roberts:2021vmt} to all orders in $\epsilon$ for such fine tuning.  Here we only derive it up to the quadratic order in the metric deformation -- however it is natural to wonder whether it holds to all orders in the presence of the gauge field.

\section{Varying hopping from a deformed lattice}
\label{sec:embedding}

The results above describe the low energy physics of the tight-binding model in terms of its lattice coordinates $x^i$ and slowly varying hopping functions.
We will derive these in detail later in the paper.
In order to relate them to a distortion of graphene, we need an embedding map from the lab coordinates to the graphene lattice, and further a bond model that predicts the hopping functions based on this embedded lattice geometry.

The simplest bond model relates the hopping functions to bond lengths, where we think of these lengths as determining the degree of electron orbital overlap. In our lattice coordinates a bond between sites $\bx_A$ and $\bx_A + a \bl_n$ is the line,
\eqn{
\label{eq:bond}
\bx = \bx_A + \lambda a \bl_n \; , \quad \lambda = [0,1]
}
and we denote its length $L_{n,A}$. As the lattice embedding becomes distorted, these bond lengths will deviate from their unperturbed value $a$. A common approximation is that the hopping functions have an exponential dependence on bond length,
\eqn{
\label{eq:expbondmodel}
\frac{T_{n,A}}{T} = e^{- \beta \left( \frac{L_{n,A}}{a} -  1 \right)}
}
and for graphene this constant $\beta$, the Grüneisen parameter, has been estimated to be $\beta \simeq 3$ \cite{ribeiro2009strained}. However here we will take a more general bond model,
\eqn{
\label{eq:bondmodel}
\frac{T_{n,A}}{T} = F \left( \frac{L_{n,A}}{a} -  1 \right)
}
for some function, $F$, although we emphasize that this still assumes that there is no dependence on the bond angles. To the order we will work to here, we will be sensitive to up to two derivatives of this function $F$ about its zero argument, the undistorted bond length, and we denote these as,
\eqn{
F(0) = 1 \; , \quad F'(0) = - \beta \; , \quad F''(0) = (\tau - 1) \beta \; .
}
In order to recover the exponential bond model we then simply take,
\eqn{
\tau = \beta + 1
}
but we will leave it general for now to illustrate in what follows the sensitivity to the precise nature of the bond model.

In order to compute these hopping functions we need the geometry of the lattice embedding into the 3-d Euclidean space of the laboratory, $\mathbb{R}^3_{lab}$ which we describe using the spatial `lab coordinates' $(X, Y, Z)$. Let us denote the collection of lattice sites $\vec{x}_A$, and the full lattice as $\Gamma$.
We then imagine describing the embedding by providing a map $\Gamma \to \mathbb{R}^3_{lab}$, or explicitly $\vec{x}_A \to ( X, Y,  Z )$.
Restricting to smooth slowly varying embeddings so that we may view the lattice as the 2d space $\mathbb{R}^2_{lat}$ described by the lattice coordinates $\vec{x}$, the embedding is defined by the smooth map,
\eqn{
\mathbb{R}^2_{lat} \quad & \to & \quad \mathbb{R}^3_{lab} \nl
\vec{x} \quad & \to & \quad( X(\vec{x}), Y(\vec{x}),  Z(\vec{x}) )
}
so that when it is evaluated on the lattice sites $\vec{x}_A$ it gives the lattice embedding above, and slowly varying implies that all derivatives $\partial_{i_1} \ldots \partial_{i_m} X \sim O(1)$ and similarly for $Y$ and $Z$. We describe the pristine, or undeformed embedding, as $x = X$, $y = Y$ and $Z = 0$, and in this case the geometry induced (ie. pulled back from $\mathbb{R}^3_{lab}$) is simply the 2-d Euclidean geometry with metric $ds^2_{(pristine)} = \delta_{ij} dx^i dx^j$. 

We now consider embeddings which are a perturbative deformations of this pristine embedding. We define a displacement field $v^i(\vec{x})$ and height function $h(x)$. Note that the displacement field is a vector field on $\mathbb{R}^2_{lat}$.
Then introducing the perturbation parameter $\epsilon$, we define the embedding map explicitly using the displacement vector field and height function as,
\eqn{
\label{eq:straindef}
\mathbb{R}^2_{lat} \quad & \to & \quad \mathbb{R}^3_{lab} \nl
\vec{x} \qquad  &\to& \qquad \left\{ \begin{array}{rcl}
X^I(\vec{x}) &= &\delta^I_i \left( x^i + \epsilon v^i(\vec{x}) \right) \\
 Z(\vec{x}) &=&  \sqrt{\epsilon} h(\vec{x})
 \end{array}
 \right.
}
where, as above,  $X^I = (X, Y)$.
We note that we consider $v^i(\bx)$ and $h(\bx)$ to be independent of the perturbation parameter $\epsilon$ -- thus having specified these we think of varying $\epsilon$ as moving us through a one-parameter family of deformations.
Then in our lattice coordinates the induced metric on $\mathbb{R}^2_{lat}$ given by pulling back the lab Euclidean metric is simply,
\eqn{
g^{(ind)}_{ij} = \delta_{ij} + \epsilon \left( \delta_{ik} \frac{\partial v^k}{\partial x^j} + \delta_{jk} \frac{\partial v^k}{\partial x^i} + \frac{\partial h}{\partial x^i}  \frac{\partial h}{\partial x^j} \right) + \epsilon^2 \delta_{kl} \frac{\partial v^k}{\partial x^i} \frac{\partial v^l}{\partial x^j} \; .
}
We will denote the 2-d geometry induced by this embedding $\Sigma_{ind} = (\mathbb{R}^2_{lat}, g^{(ind)})$.
The usual strain tensor is then defined by comparing the induced, and the pristine metrics, so in lattice coordinates,
\eqn{
\label{eq:inducedmetric}
\sigma_{ij} = \frac{1}{2} \left( g^{(ind)}_{ij} - \delta_{ij} \right) 
}
and at this point these expressions for the strain tensor $\sigma_{ij}$ are exact to all orders in $\epsilon$. 

The physical distance between lattice sites at $\bx_A$ and $\bx_A + a \bl_n$ under the distortion is then computed by integrating the length of the line~\eqref{eq:bond} so, 
\eqn{
\label{eq:bondlen}
L_{n,A} 
&=& a \int_0^1 d\lambda \sqrt{ g^{(ind)}_{ij}(\bx_A + a \lambda \vec{\ell}_n) \ell^i_n \ell^j_n} \; .
}
The bonds of the pristine lattice have length $a$. 
Since the metric is slowly varying,  we may Taylor expand the integrand above in $a$, perform the integrals, and then working to $O(a^2)$ at order $O(\epsilon)$, and to $O(a)$ at order $O(\epsilon^2)$,  the fractional difference in bond length due to the deformation is, 
\eqn{
 \frac{L_{n,A} - a}{a}
 & = &  \ell^i_n \ell^j_n \sigma_{ij}( \bx_{n,A} ) + \frac{a^2}{24} \ell^i_n \ell^j_n ( \ell^k_n \partial_k)^2 \sigma_{ij}( \bx_{n,A} ) - \frac{1}{2} \left( \ell^i_n \ell^j_n \sigma_{ij}( \bx_{n,A} ) \right)^2   + O( \epsilon a^3,  \epsilon^2 a^2 ,\epsilon^3 )  
 }
 where we have defined the location of the mid point of the bond, $\bx_{n,A} = \bx_A + \frac{a}{2} \vec{\ell}_n$,
and we emphasize that here there is no sum over the repeated index $n$, and that $\sigma_{ij} \sim O(\epsilon)$. Thus using that the components slowly vary, together with the relations~\eqref{eq:Texp} and~\eqref{eq:tfn}, we then find the bond model~\eqref{eq:bondmodel} determines the hopping functions as, \footnote{In \cite{wagner2019quantum} perturbation theory was carried out to second order when considering the effective geometry in the continuum limit, but it failed to keep track of the higher derivative terms which we demonstrate are of the same order.}
\eqn{
\label{eq:tfnstrain}
t_n(\bx) &=& 1 - \beta \left( \ell^i_n \ell^j_n \sigma_{ij}( \bx ) 
+ \frac{a^2}{24} \ell^i_n \ell^j_n ( \ell^k_n \partial_k)^2 \sigma_{ij}( \bx )  \right) 
+ \frac{\beta \tau}{2} \left( \ell^i_n \ell^j_n \sigma_{ij}( \bx ) \right)^2 
+ O( \epsilon a^3,  \epsilon^2 a^2 ,\epsilon^3 ) 
 \; .
}
Note that while we have kept track of corrections subleading in $a$, these necessarily involve precise details of the deformation on lattice scales, and from an effective field theory point of view should be thought of as corrections from irrelevant operators. Conversely this implies that when one is matching \emph{subleading} corrections, as we are interested in doing here, then they are necessary.

We pause to note that in~\cite{wagner2022landau} some quadratic corrections to the effective theory where considered. The effective theory was given to linear order in the hopping functions, and then these were related to strain working to quadratic order in the strain tensor as above. However we emphasize that it is inconsistent to do this -- one must also include the quadratic corrections in the hopping functions as we do here if one wishes to work to quadratic order in the deformation or else one is clearly missing important contributions.
\footnote{
In `equations' we might say, if the physics $F$ we are interested in is a function of a variable $\delta t$, with an expansion $F(\delta t) = a_1 \delta t + a_2 \delta t^2 + \ldots$, and $\delta t$ is expressed in terms of another variable $\sigma$ perturbatively as $\delta t(\sigma) =  b_1 \sigma + b_2 \sigma^2 + \ldots$ then to express $F$ in terms of $\sigma$ correctly to quadratic order, $F(\sigma) = a_1 b_1 \sigma + ( a_1 b_2 + a_2 (b_1)^2 ) \sigma^2  + \ldots$ we must include the $a_2$ quadratic term in the expression for $F(\delta t)$ above. If we only work with the linear truncation $F^{lin}(\delta t) = a_1 \delta t$, then $F^{lin}(\sigma) = a_1 \delta t(\sigma) = a_0 + a_1 b_1 \sigma + a_1 b_2 \sigma^2 + \ldots$ and we clearly get the quadratic term we are interested in wrong unless $| a_2 (b_1)^2 | \ll | a_1 b_2 |$. In our case here, we see explicitly from equations~\eqref{eq:summarymetric} and~\eqref{eq:summarygauge} that the coefficients $a_{1,2}$ are simply $O(1)$, as are the coefficients $b_{1,2}$ from~\eqref{eq:tfnstrain}. Thus neglecting the quadratic behaviour of the lattice model in the hopping functions relative to that induced in the relation of strain cannot be justified -- for example the coefficients for the quadratic terms in the strain in the gauge field in~\eqref{eq:quadexpressions} go from $ \frac{ (\beta - \tau)}{2}  \to \frac{ (4 \beta - \tau)}{2}$ and $- \frac{  (3 \beta + \tau)}{8} \to - \frac{ \tau}{8}$ if we ignore the quadratic terms in~\eqref{eq:summarygauge}.
}

Given that $v^i(\bx)$ and $h(\bx)$ are independent of $\epsilon$ and $a$, a perturbative expansion yields the coefficients defined in~\eqref{eq:texp}.
We may now give the expression for the (purely magnetic) gauge field $A = A_i(\bx) dx^i$,  and electrometric  $ds^2_{electro} = g_{ij}(\bx) dx^i dx^j$, as determined from equations~\eqref{eq:summarymetric} and~\eqref{eq:summarygauge} in terms of the strain tensor as,
\eqn{
\label{eq:quadexpressions}
A_i(\vec{x}) &=& - \frac{\beta \epsilon_{ij}  }{2 a} \Big( K^{jkl} \left(  \sigma_{kl}(\bx) + \frac{ (\beta - \tau)}{2} \sigma_{km}(\bx) \sigma_{ml}(\bx) - \frac{  (3 \beta + \tau)}{8} \sigma_k(\bx) \sigma_l(\bx)   \right) \nl
&& \qquad \qquad \qquad + \frac{a^2}{12} \left( 9 \partial_j \partial_k \sigma_k(\bx) - 3 \partial_k \partial_k \sigma_j(\bx) - 7 K^{klm} \partial_k \partial_l \sigma_{jm}(\bx) \right) + O(  \epsilon a^3,\epsilon^2  a^2, \epsilon^3)  \Big) \nl
g_{ij}(\vec{x}) &=& \delta_{ij} + 2 \beta  \sigma_{ij}(\bx) +    4 \beta^2 \sigma_{ik}(\bx) \sigma_{kj}(\bx) + \frac{ \beta ( \beta + \tau )}{4} \left( \delta_{ij} \left( \sigma_{kk}(\bx)  \right)^2 - 4 \sigma_{ij}(\bx)  \sigma_{kk}(\bx)  -  \sigma_i(\bx) \sigma_j(\bx)  \right)  + O(\epsilon a^2, \epsilon^2 a, \epsilon^3) \nl
}
where we have defined the covector $\sigma_i$ given in lattice coordinates as $\sigma_i = K^{ijk} \sigma_{jk}$.
Quadratic corrections in $\epsilon$ for homogeneous strain were studied in~\cite{oliva2017low}, and restricting to such deformations, our gauge field and metric above are precisely consistent with their results (when expressed in lattice coordinates).
A potentially confusing issue is that the lattice geometry induced by the embedding, $\Sigma_{ind}$, is generally \emph{not the same} as the spatial electronic geometry $\Sigma_{electro}$. One might naively have expected these would coincide, but this is not the case. As we shall see, interestingly even when $\Sigma_{ind}$ is flat, with vanishing height function and only in-plane displacement, the electronic geometry generally is curved at $O(\epsilon^2)$.

Suppose we are interested in the tight-binding model with hopping functions induced from an embedding. The procedure to use the effective theory is:
\begin{itemize} 
\item Work in lattice coordinates to supply the embedding, via the displacement vector field $v^i(\vec{x})$ and bending function $h(\vec{x})$. 
\item Compute the magnetic gauge field, electrometric, and $C$ and $D$ tensors of the effective theory in lattice coordinates using equations~\eqref{eq:quadexpressions} and~\eqref{eq:invariants}. With this data the theory is defined, up to choosing a convenient gauge and frame and constructing the appropriate torsion free Levi-Civita and spin connections. 
\item
One may then choose to perform computations in any coordinate system (and indeed with any frame choice and gauge).
\end{itemize}
Thus whilst we are required by our expressions above to compute the data for the effective theory using lattice coordinates, in the end we are free to use any coordinates we wish. If one is interested in comparison directly to the tight-binding model then it is natural to remain in lattice coordinates. As emphasized in the work of Oliva-Leyva and Naumis in~\cite{oliva2013understanding,oliva2015generalizing}, if the lattice is to be thought of as arising from an embedding, then in order to perform comparison with lab measurements it is most natural to work with lab frame coordinates.
In the case that the embedding is purely in-plane, so $Z = 0$, we may think of it as the diffeomorphism map $\mathrm{R}^2_{lat} \to \mathrm{R}^2_{lab}$, then it is natural to transform to the lab coordinates $X^I = (X, Y)$ to analyse the effective theory. We note that in the case there is bending too, there is no geometric canonical choice of  two lab coordinates -- the simplest choice again is $X^I$, but one could also choose to mix these with $Z$.

Recall the relation between the lattice coordinates  $\bx = (x,y)$ and lab coordinates $\vec{X} = (X, Y)$, is given as,
\eqn{
\label{eq:veqn}
X^I(\vec{x}) = \delta^I_i \left( x^i + \epsilon v^i(\bx) \right)
}
where $v^i(\bx)$ define the components of a vector field $v = v^i(\bx) (\partial / \partial x^i)$ in the coordinate basis given by the lattice coordinates.
We emphasize that the above expression is exact in $\epsilon$ -- recall we are taking $v^i(\bx)$ to be independent of $\epsilon$.
In order to express the lattice coordinates in terms of the lab coordinates we may invert this relation as a power series in $\epsilon$,
\eqn{
x^i(\vec{X}) = \delta^i_I X^I - \epsilon v^i(\vec{X}) + \epsilon^2  \delta^J_j v^j(\vec{X}) \frac{\partial v^i(\vec{X})}{\partial X^J}  + O(\epsilon^3)  
}
where we note that $v^i(\vec{X})$ are the coefficient functions $v^i(\bx)$ defined in~\eqref{eq:veqn} but now evaluated with arguments given by the lab coordinates -- they are \emph{not} the components of the vector field in the lab frame.
From this we define the Jacobian matrix,
\eqn{
\Lambda^i_{~I}(\vec{X}) = \frac{\partial x^i(\vec{X})}{\partial X^I}  \; .
}
Then tensors transform in the usual manner -- for example, for the gauge field we have,
\eqn{
\left. A_i(\bx) dx^i \right|_{x^i = x^i(\vec{X})} =  A_i(\bx(\vec{X})) \Lambda^i_{~I}(\vec{X})  dX^I
= A^{lab}_I(\vec{X}) dX^I  \; .
}
We now illustrate deriving the hopping functions from an embedding in two cases. Firstly we consider the case of an embedding including out-of-plane bending, and working to linear order in $\epsilon$ for the electrometric -- for consistency this requires including the two covariant derivative term in the effective theory. Secondly, restricting to in-plane strain only, so no bending, we work to quadratic order in $\epsilon$ for the electrometric -- and for consistency this requires our full approximation discussed above, including also the three covariant derivative term, together with the non-trivial subleading correction in $a$ to the gauge field at order $O(\epsilon)$. 

\subsection{Leading order including bending}

We preface this section by emphasizing that the status of the tight-binding lattice model as an approximation to graphene is unclear when out-of-plane bending is included, due to the mixing of $sp^2$ and $p_z$ orbitals which are thought to play an important physical role, potentially gapping the theory~\cite{naumov2011gap,lin2015feature}. However, with this in mind, it is still interesting to discuss this case, in part to link to previous results in the literature.

We will work to the order $\epsilon$ in the electrometric, and thus include the two covariant derivative term.
From above we see the induced hopping functions in lattice coordinates, to the order we require them for this approximation, are,
\eqn{
\label{eq:tfromgeometry}
t_n(\bx) = 1 - \beta \sigma_{ij}(\bx) \ell^i_n \ell^j_n + O(\epsilon^2,\epsilon a^2)  = 1  - \epsilon  \beta \, \ell^i_n \ell^j_n   \left(  \frac{\partial v^i}{\partial x^j} + \frac{1}{2} \frac{\partial h}{\partial x^i}  \frac{\partial h}{\partial x^j} \right) + O(\epsilon^2,\epsilon a^2)  
}
which determines,
\eqn{
\delta_{1,0} t_n(\bx) =  - \beta \,  \ell^i_n \ell^j_n   \left(  \frac{\partial v^i}{\partial x^j} + \frac{1}{2} \frac{\partial h}{\partial x^i}  \frac{\partial h}{\partial x^j} \right) \; , \quad \delta_{1,1} t_n(\bx) = 0 \; .
}
Now $\sigma_{ij}$ encodes the perturbative in-plane deformations $v^i(\bx)$ and out of plane bending $h(\bx)$. From the perspective of the induced metric $g^{(ind)}_{ij}$, the $v^i$ generate infinitessimal diffeomorphisms of the undeformed metric $\delta_{ij}$, and therefore do not change the geometry from being flat, but just the coordinates it is presented in. On the other hand, $h(x)$ induces a real change of the geometry and generates curvature.

At least at this linearized level, this relation is invertible.  Given a perturbation of the hopping functions, this uniquely prescribes the induced geometry of the lattice embedding that would generate such a deformation. Explicitly to the same orders in $\epsilon$ and $a$ we have,
\eqn{
\label{eq:inverset}
 \sigma_{ij} (\bx)= - \frac{1}{3 \beta} \sum_n \left( 4 \ell^i_n \ell^j_n - \delta^{ij} \right) \delta_{1,0} t_{n}(\bx) + O(\epsilon^2, \epsilon a)
}
and then the leading order perturbative $v^i$ and $h$ that generate this geometry by straining and bending the pristine embedding may be solved for.
From above the gauge field and electrometric take the simple form,
\eqn{
\label{eq:linearmetric}
A_i(\vec{x}) = - \frac{\beta}{2 a} \epsilon_{ij} \left( K^{jkl} \sigma_{kl}(\bx) + O(\epsilon^2, \epsilon a^2) \right) \; , \quad g_{ij}(\vec{x}) = \delta_{ij} + 2 \beta  \sigma_{ij}(\bx) + O(\epsilon^2, \epsilon a)
}
to this order of approximation, in terms of the lattice coordinates, with $K^{ijk}$ the lattice invariant defined earlier in~\eqref{eq:Ktensor}.
These expressions may be compared to those of de Juan et al~\cite{dejuan2012spacedepfermi}. In that work they choose to work with spinor densities, rather than spinors as we do here, which effectively Weyl rescales their electrometric so that $g_{tt} \ne $constant. As discussed in~\cite{stegmann2016current}, one can Weyl transform back to the ultrastatic frame we use here, and to canonically normalized spinors, and in doing so their metric (given in~\cite{stegmann2016current}) precisely agrees with the above form. Note that to the leading order given above, the gauge field is not affected by this Weyl scaling.
However we stress again that at this order where we first include the non-trivial metric, the contribution of the two derivative terms must also be included for consistency, and this was missed in these previous analyses ~\cite{vozmediano2008gauge,dejuan2012spacedepfermi,manes2013generalized,oliva2015generalizing,wagner2022landau} as we have emphasized in~\cite{Roberts:2021vmt}. 

We clearly see that the electrometric is not equal to the induced metric. Comparing~\eqref{eq:linearmetric} to $g^{ind}_{ij} = \delta_{ij} + 2 \sigma_{ij}$ (from~\eqref{eq:inducedmetric}), we see they differ by a factor of $\beta$ in the perturbation, as observed in~\cite{stegmann2016current}. However, at this order the in-plane displacement field $v_i$ still acts simply as a diffeomorphism for the electrometric (as well as for the induced metric) --  it changes the coordinates, but doesn't induce actual curvature. If we explicitly compute the Ricci scalar of the electrometric we see,
\eqn{
R(\bx)  = 2 \beta \left( \partial_x^2 h(\vec{x})  \, \partial_y^2 h(\vec{x})  - ( \partial_x \partial_y h(\vec{x})  )^2 \right) + O(\epsilon^2, \epsilon a)
}
confirming that curvature is only generated by the out-of-plane deformation due to $h(\bx)$.  The  gauge field does however see this in-plane displacement field $v^i$, and the (gauge-invariant) magnetic field  is,
\eqn{
B(\vec{x})  = F_{xy} = \frac{\epsilon \beta}{2a} \left( -2 \partial_{x} \partial_y v^x(\vec{x})  + ( \partial_y^2 - \partial_x^2 ) v^y(\vec{x})   -2 \partial_x h(\vec{x}) \, \partial_{x} \partial_y h(\vec{x})  + \partial_y h(\vec{x})  ( \partial_y^2 - \partial_x^2 ) h(\vec{x})   \right) + O(\epsilon^2, \epsilon a) \; .
}

Now suppose we wish to analyse the physics of the effective theory using the lab reference frame. Having the theory in lattice coordinates we may then simply coordinate transform to the lab coordinates as detailed above. Since the gauge field is already $O(\epsilon)$, this coordinate transformation is trivial, so that,
\eqn{
A^{lab}_I(\vec{X}) = - \frac{\beta \epsilon}{2 a} \delta^i_I \epsilon_{ij} \left( K^{jkl} \sigma_{kl}(\vec{X}) + O(\epsilon^2, \epsilon a^2) \right) \; .
}
However there is an effect for the electrometric, which becomes,
\eqn{
g^{lab}_{IJ}(\vec{X})  &=& \delta_{IJ} + 2  (\beta - 1) \epsilon \delta^i_I \delta^j_J \sigma_{ij}(\vec{X}) + \epsilon \frac{\partial h(\vec{X})}{\partial X^I}  \frac{\partial h(\vec{X})}{\partial X^J}  + O(\epsilon^2, \epsilon a) \; .
}
We note that the components of the strain tensor do not transform at the leading order $\sigma_{ij} \sim O(\epsilon)$ we require here.
At this order, the Ricci scalar and magnetic field in lab frame take the same form as in lattice coordinates.

Recalling that the tight-binding model likely does not give a good approximation to physics in the case that the graphene is bent out-of-plane, we see that geometrically this leading geometry from purely in-plane strain is rather boring. While the electrometric is non-trivial, it is simply flat space in distorted coordinates. Hence not-withstanding the fact that we must also include two derivative terms which were missed in the treatments of~\cite{dejuan2012spacedepfermi,Zubkov:2013sja,oliva2015generalizing,yang2015dirac,VOLOVIK2015255,si2016strain,khaidukov2016landau,oliva2017low,wagner2019quantum,de2007charge,guinea2008gauge,vozmediano2008gauge,de2013gauge,arias2015gauge,stegmann2016current,castro2017pseudomagnetic,Golkar:2014paa,Golkar:2014wwa}, it doesn't provide an interesting `analog gravity' model, as there is no sense in which the geometry is curved.
We will now discuss how this becomes much more interesting at quadratic order in the strain.

\subsection{Quadratic order for purely in-plane strain }

Now we continue our analysis of purely in-plane strain, where we are hopeful the tight-binding lattice model may provide a good approximation, to order $O(\epsilon^2)$ in the electrometric perturbation. Again we note we are using our simple length model to turn the lattice deformation to hopping function deformations, and this could potentially be made more realistic. To quadratic order in $\epsilon$ the gauge field and metric take the form above in~\eqref{eq:quadexpressions}, where for consistency we must also include both the two and three covariant derivative terms in the effective theory, and also include the subleading corrections in $a$ present in~\eqref{eq:quadexpressions}.
We may then compute the magnetic field and electrometric curvature in the case of in-plane strain, so vanishing height function. These can be written nicely by defining,
\eqn{
V(\vec{x}) = \left(  \partial_x v^x(\vec{x}), \partial_y v^y(\vec{x}) \right) \; ,\quad U(\vec{x}) = \left( \partial_y v^x(\vec{x}), \partial_x v^y(\vec{x}) \right)
}
and
\eqn{
\bar{V}(\vec{x})  = \left(  \partial^2_x v^y(\vec{x}),  \partial^2_y v^y(\vec{x}) , \partial_x \partial_y v^x(\vec{x}) \right) \; ,\quad \bar{U}(\vec{x})  =  \left(  \partial^2_x v^x(\vec{x}), \partial^2_y v^x(\vec{x}) , \partial_x \partial_y v^y(\vec{x}) \right) \; .
}
Then for the magnetic field we find,
\eqn{
B(\vec{x}) & = &  F_{xy} 
=
\frac{\epsilon \beta}{2a} \left( \left( - \partial^2_x v^y +  \partial^2_y v^y - \partial_x \partial_y v^x  \right) 
+ \frac{a^2}{24} \left( 2 \partial^4_y v^y + 17 \partial_x \partial^3_y v^x -21 \partial^2_x \partial^2_y v^y - 11 \partial^3_x \partial_y v^x
+ 5 \partial^4_x v^y
\right) + O(a^3) \right) \nl
&& \qquad + \frac{\epsilon^2 \beta}{2a} \left(  \bar{V}^T(\vec{x}) \cdot \left(\begin{array}{cc}
 \frac{3 \tau}{4}  + \frac{\beta}{4} & -1 + \frac{\tau}{4}   - \frac{5 \beta}{4} \\
\frac{ \tau }{4}  + \frac{3 \beta}{4}& 1 - \frac{5 \tau}{4}   + \frac{\beta}{4} \\
- 2 + \frac{3 \tau}{2}   - \frac{3 \beta}{2}& \frac{\tau}{2} - \frac{\beta}{2}
\end{array} \right) \cdot {V}(\vec{x}) 
+  \bar{U}^T(\vec{x}) \cdot \left(\begin{array}{cc}
- 1 + \frac{3 \tau}{4}  + \frac{\beta}{4} & \frac{3 \tau}{4}  + \frac{\beta}{4}\\
1 + \frac{\tau}{4}  + \frac{3 \beta}{4} &  \frac{\tau}{4}  + \frac{3 \beta}{4} \\
\frac{\tau}{2} - \frac{\beta}{2} & - 2 + \frac{\tau}{2}  - \frac{\beta}{2}
\end{array} \right) \cdot {U}(\vec{x})
+ O(a^2)
\right) \nl
&& \qquad \qquad + O(\epsilon^3)
}
and the Ricci scalar of the electrometric is given by,
\eqn{
R(\vec{x}) & = & \epsilon^2 \beta \Big(
\frac{1}{2} \left( \partial_y v^x + \partial_x v^y\right) \left( ( \tau - 3 \beta ) \partial^3_y v^x  + ( \beta + 3 \tau) \partial^2_x \partial_y v^x \right) 
+ 4 \beta \left( \partial_y v^y - \partial_x v^x \right) \partial^2_y \partial_x v^x
\nl
&& \quad 
- \frac{1}{2} \left( \partial_y v^y - \partial_x v^x \right) \left( ( \beta + \tau ) \partial^3_y v^y + ( 5 \beta - 3 \tau ) \partial^2_x \partial_y v^y \right) 
- 2 \beta \left( \partial_y v^x + \partial_x v^y \right) \left( \partial^3_x v^y - \partial^2_y \partial_x v^y \right) \nl
&&  
+ \bar{V}^T(\vec{x}) \cdot 
\left(
\begin{array}{ccc}
- 2 \beta & 1 - \frac{\tau}{4} + \frac{3 \beta}{4} & - \frac{3 \tau}{4} - \frac{3 \beta}{4} \\
1 - \frac{\tau}{4} + \frac{3 \beta}{4}   &  - \frac{\tau}{2} - \frac{\beta}{2}  &   \frac{\tau}{4} + \frac{9 \beta}{4} \\
 - \frac{3 \tau}{4} - \frac{3 \beta}{4} &  \frac{\tau}{4} + \frac{9 \beta}{4}  & -2 - 4 \beta
\end{array}
\right)
\cdot \bar{V}(\vec{x}) 
+ \bar{U}^T(\vec{x}) \cdot 
\left(
\begin{array}{ccc}
0 & 1 - \frac{3 \tau}{4} + \frac{ \beta}{4}  & - \frac{3 \tau}{4} + \frac{5 \beta}{4} \\
1 - \frac{3 \tau}{4} + \frac{ \beta}{4}  & \frac{\tau}{2} - \frac{3 \beta}{2} &\frac{\tau}{4} + \frac{\beta}{4} \\
 - \frac{3 \tau}{4} + \frac{5 \beta}{4} & \frac{\tau}{4} + \frac{\beta}{4} & -2 + 2 \tau - 2 \beta
\end{array}
\right)
\cdot \bar{U}(\vec{x})
\Big) \nl
&& \qquad + O(\epsilon a^2,\epsilon^2 a,\epsilon^3 )
}
We now see a very interesting phenomena. The geometric deformation of the lattice is purely in-plane so the induced metric $g^{(ind)}$ is flat, simply a coordinate transformation (i.e. a diffeomorphism) of flat space. However since the induced metric and electrometric explicitly differ in form, we find that at quadratic order in $\epsilon$ the electrometric is indeed curved. While it isn't obvious whether the explicit form above for $g_{ij}(\vec{x})$ in~\eqref{eq:quadexpressions} is flat or not, this explicit calculation of the curvature shows it is not. It is worth emphasizing that this curvature is dependent on the bond model we have chosen -- we see the explicit dependence on the parameter $\tau$ from the bond model. Nonetheless, irrespective of the value of $\tau$, we explicitly see it will generally  be curved. 

If we had worked to one higher order in $a$, then already at linear order in $\epsilon$ we would expect to compute a non-trivial metric contribution at order $\epsilon a^2$, and presumably this would also lead to a non-zero curvature at ${O}(\epsilon a^2)$. However to consistently work to this higher order requires including the next higher covariant derivative term, that is with four derivatives, and this is outside of the scope of this work. Depending on the values of $\epsilon$ and $a$, one could expect that either the $O(\epsilon^2)$ contribution to curvature we have computed here dominates, or this subleading term at order  ${O}(\epsilon a^2)$ does. We note that if we think of keeping $\epsilon/a$ fixed, and scaling $a \to 0$, then comparing the two, $\epsilon^2 = \left( \frac{\epsilon}{a} \right)^2 a^2$ and $\epsilon a^2 =  \left( \frac{\epsilon}{a} \right) a^3$, and so the $O(\epsilon^2)$ term we have computed dominates in the continuum limit $a \to 0$.

In summary, an embedding of the tight-binding lattice theory using our simple bond model does indeed have a low energy `analog gravity' description for inhomogeneous in-plane strains, when it is thought to approximate monolayer graphene. This then raises the very interesting possibility that effects known from curved spacetime QFT may play a role in the physics of this model, and indeed graphene, for such inhomogeneous strains.
However, we emphasize that while the low energy physics of the lattice model is described by an `analog gravity' theory, by which we mean a relativistic effective field with a curved spacetime geometry, it must include the Lorentz violating higher derivative terms for consistency. Thus it is an `analog gravity' model with Lorentz violation which contributes at the same order as the effects of curvature. 
\\

Let us consider now writing the theory in lab frame. Firstly we may transform the strain tensor, precisely since it is a tensor, to lab coordinates $\sigma^{lab}_{ij}(\vec{X})$. Working to this order it is important to remember that the invariant $K^{ijk}$ which takes simple $\pm 1, 0$ values in lattice coordinates no longer does so after a spatial coordinate transform. To write our expressions in a convenient form we may define $\sigma^{lab}_{i}(\vec{X})$ to be the transform of the covector field $\sigma_i(\bx) = K^{ijk} \sigma_{jk}(\bx)$. After doing so, we may give expressions for the gauge field and electrometric in lab coordinates for our full approximation, as,
\eqn{
\label{eq:lab_frame_a_g}
A^{lab}_I(\vec{X}) &=& - \frac{\beta \epsilon_{IJ}  }{2 a} \Big(  \sigma^{lab}_{J}(\vec{X}) + K^{JKL} \left(  \frac{ (\beta - \tau)}{2} \sigma^{lab}_{KM}(\vec{X}) \sigma^{lab}_{ML}(\vec{X}) - \frac{  (3 \beta + \tau \textcolor{red}{-8})}{8} \sigma^{lab}_K(\vec{X}) \sigma^{lab}_L(\vec{X})   \right) \nl
&& \qquad \qquad \qquad + \frac{a^2}{12} \left( 9 \partial_J \partial_K \sigma^{lab}_K(\vec{X}) - 3 \partial_K \partial_K \sigma^{lab}_J(\vec{X}) - 7 K^{KLM} \partial_K \partial_L \sigma^{lab}_{JM}(\vec{X}) \right) + O(  \epsilon a^3,\epsilon^2  a^2, \epsilon^3)  \Big) \nl
g^{lab}_{IJ}(\vec{X}) &=& \delta_{IJ} + 2 ( \beta \textcolor{red}{- 1} )  \sigma^{lab}_{IJ}(\vec{X}) +    4 \beta^2 \sigma^{lab}_{IK}(\vec{X}) \sigma^{lab}_{KJ}(\vec{X}) \nl
&& \qquad \qquad + \frac{ \beta ( \beta + \tau )}{4} \left( \delta_{IJ} \left( \sigma^{lab}_{KK}(\vec{X})  \right)^2 - 4 \sigma^{lab}_{IJ}(\vec{X})  \sigma^{lab}_{KK}(\vec{X})  -  \sigma^{lab}_I(\vec{X}) \sigma^{lab}_J(\vec{X})  \right)  + O(\epsilon a^2, \epsilon^2 a, \epsilon^3) \; .
}
Here $\delta_{IJ}$ and $\epsilon_{IJ}$ are the usual Kronecker delta, and antisymmetric Levi-Civita symbol, and we only require the components of $K^{IJK}$ at $O(\epsilon^0)$ in the expression above, and these don't change with the transformation.
Interestingly, written in this form, the only differences are the change in the coefficient of the linear term in strain for the electrometric (which derives from the coordinate transformation of the leading Euclidean metric $\delta_{ij}$, and we saw at linear order) together with a similar change in one quadratic coefficient for the gauge field -- both of these are shown in the above equations in red. However, an important point to emphasize is that to use these quantities in the effective theory we have to remember that the coefficients of the higher covariant derivative terms, built from the metric and lattice invariants, also must be consistently transformed.

\section{Deriving the effective theory of the lattice tight-binding model}
\label{sec:deriveeffective}

For most of the remainder of this paper we will focus on the Hamiltonian~\eqref{eq:Hamiltonian} with perturbatively deformed hopping functions that are slowly varying, and give the derivation of the effective theory summarized above, thinking in terms of the intrinsic description in lattice coordinates. Rather than work with an embedding picture, and bond model, we will simply give results purely in terms of the hopping functions themselves, but note that using the discussion in the previous section, we may always translate to a lab picture if we have a specific embedding and bond model. 
In order to make the somewhat involved computations involved here more accessible, we have made available a {\tt Mathematica} notebook which performs the explicit matching of the effective theory to the lattice model that we describe in what follows.\footnote{This may be downloaded from {\tt https://sites.google.com/view/graphene-effective-theory}.}

\subsection{Continuum limit of undeformed lattice model}
\label{sec:continuum}

We are interested in the band structure, given by the one particle states of the above Hamiltonian. A general one-particle state is given by,
\eqn{
\ket{\Psi(t)} = \left(  \sum_{\bx_A} A_{\bx_A}(t) a^\dagger_{\bx_A}+ \sum_{\bx_B} B_{\bx_B}(t) b^\dagger_{\bx_B} \right) \ket 0
}
and then its time evolution is given by the Schr\"odinger equation,  $ i \hbar \partial_t \ket \Psi = H\ket\Psi $, which can be resolved as,
\eqn{
\label{eq:lattice}
 i \hbar \partial_t  A_{\bx_A} = T \sum_n t_n( \bx_A +\frac{a \bl_n}{2}) B_{\bx_A +a \bl_n} ,\nn \\
 i \hbar \partial_t  B_{\bx_B} =  T \sum_n t_n( \bx_B -\frac{a \bl_n}{2}) A_{\bx_B -a \bl_n} .
}
For the undeformed lattice, so $t_n( \bx) = 1$ then there are two Dirac points, whose wavevectors, defined by the condition $\sum_n e^{ i a \vec{K} \cdot \bl_n} = 0$, can be taken as (see figure \ref{fig:honeycomb_sym_bz}), 
\eqn{\label{eq:defn_Kpoints}
\vec{K} =  \frac{1}{a} \left( - \frac{4 \pi}{3 \sqrt{3}} , 0 \right) , \qquad \vec{K'} = -\vec{K}.
}
Let us first consider the $K$ point. Taking smooth functions $\psi_1(t, \bx)$, $\psi_2(t, \bx)$ of time, and of lattice coordinates so that they spatially vary slowly, so $\partial_i \psi_{1,2} \sim O(1)$, then we may write,
\eqn{
\label{eq:fastslowansatz}
K: \qquad A_{\bx_A}(t) = \psi_1(t,\bx_A) e^{- \frac{i \pi}{4}} e^{+i  \vec{K} \cdot \bx_A} \; , \quad B_{\bx_B}(t) = \psi_2(t,\bx_B)  e^{+ \frac{i \pi}{4}} e^{+i  \vec{K} \cdot \bx_B} \; .
}
Alternatively, near the $K'$ point, we take,
\eqn{
\label{eq:fastslowansatzKprime}
K': \qquad A_{\bx_A}(t) = \psi_2(t,\bx_A) e^{- \frac{i \pi}{4}} e^{-i  \vec{K} \cdot \bx_A} \; , \quad B_{\bx_B}(t) = \psi_1(t,\bx_B)  e^{+ \frac{i \pi}{4}} e^{-i  \vec{K} \cdot \bx_B} \; .
}
Then for both Dirac points we can recast the continuum limit of the above Schr\"odinger system as,
\eqn{\label{eq:undeformed_dirac}
0 & = &  i \hbar  \partial_t  \left( \begin{array}{c}  \psi_1 \\ -\psi_2 \end{array} \right)  - i T \sum_n \left( \begin{array}{cc} 0 & +e^{+ i a \vec{K} \cdot \bl_n}   \\
 - e^{- i a \vec{K} \cdot \bl_n}   & 0 \end{array} \right)  a \bl_n \cdot \vec{\partial} \left( \begin{array}{c} \psi_1 \\ \psi_2 \end{array} \right) + O(a^2) \; .
}
Now we introduce spacetime coordinates $x^\mu$ with index $\mu = 0, 1,2$, which coincide with our lab time and lattice coordinates, so $x^\mu = ( t, \bx)$. Further we introduce a frame $e^\mu_{~A}$ (with frame index $A = 0, 1, 2$) as,
\eqn{
\label{eq:flatframe}
e^\mu_{~A} = \left( \begin{array}{ccc} 
\frac{1}{c_{eff}} & 0 & 0 \\
0 & 1 & 0 \\
0 & 0 & 1
\end{array} \right) 
}
which corresponds to the spacetime metric, $g_{\mu\nu}$, being Minkowski spacetime in usual coordinates,
\eqn{
\label{eq:undefmetric}
g_{\mu\nu} = 
 \left(
 \begin{array}{cc}
-c_{eff}^2 & 0 \\
0 & \delta_{ij}  
\end{array} 
\right)  
}
with $c_{eff} = \frac{3 a T}{2 \hbar}$ giving the effective speed of light. Then for both $K$ and $K'$ we may write the Schr\"odinger system simply in massless Dirac equation form in this flat Minkowski spacetime as,
\eqn{
\label{eq:undeformedDirac}
0 = e^\mu_{~A} \gamma^A \partial_\mu \Psi + O(a)  \; , \quad \Psi =  \left( \begin{array}{c} \psi_1 \\ \psi_2  \end{array} \right)
}
where the Dirac Gamma matrices are,
\eqn{
\gamma^A =  \left( \gamma^0 , \gamma^I \right) = \left( - i \sigma^3 , \sigma^1 , \sigma^2 \right)
}
with $\sigma^I$ the Pauli matrices, where we split the frame index into time and spatial components $A = (0, I)$. 
Note that since the $K$ and $K'$ points are inequivalent, the full low energy effective theory has two flavors of massless Dirac spinors living on the same spacetime and we have picked conventions so the local Lorentz frame is the same for both of them. Since this tight-binding model has no electron-electron interactions, these two flavors are free fields and do not interaction with each other.

Why do we call this the continuum limit? We have assumed that $\psi_{1,2}$ are slowly varying, so that in our units $\partial_{i_1} \ldots \partial_{i_k} \psi_{1,2} \sim O(1)$. This implies that the time dependence in the wavefunctions $A_{\bx_A}(t)$, $B_{\bx_A}(t)$ goes as $\sim O( T a / \hbar )$. Considering smaller wavelength variations would requires the higher order terms in $a$ to be accounted for, and correspond to higher frequencies, and thus higher energies. 
An important point is that this continuum limit describes only low energies/frequencies for the wavefunctions $A_{\bx_A}(t)$, $B_{\bx_A}(t)$, and while $\psi_{1,2}$ are slowly spatially varying, the wavefunctions themselves certainly are not. This proves to be a crucial point in what follows, and we will return to it later.

\subsection{Preliminaries}
\label{sec:preliminaries}

Before we continue to consider perturbed and spatially varying hopping functions, it is convenient to firstly consider the continuum limit of the \emph{undistorted} tight binding model to higher order in the low energy expansion, so given our units, the expansion in $a$. We will also detail the local symmetries that arise in identifying low energy continuum fields with the discrete wavefunctions.

\subsubsection{Expansion to third order}
\label{sec:secondorder}

Taking the same ansatz~\eqref{eq:fastslowansatz} for the wavefunctions as before, and expanding the Schr\"odinger system to the next two orders in $a$ then yields,
 \eqn{
\label{eq:undeformedDiracThirdOrder}
0 =  e^\mu_{~A} \gamma^A \partial_\mu \Psi \pm i a \, \eta_{AB} \gamma^A e^B_{~\sigma} C^{\sigma\mu\nu}  \partial_\mu \partial_\nu \Psi  + a^2 \,\eta_{AB} \gamma^A e^B_{~\sigma} D^{\sigma\mu\nu\rho}  \partial_\mu \partial_\nu \partial_\rho \Psi + O(a^3)  
}
where the constants $C^{\sigma\mu\nu}$,  $D^{\sigma\mu\nu\alpha}$ have only non-vanishing spatial components and are given in the earlier equation~\eqref{eq:invariants} except here we are taking the trivial flat spatial frame~\eqref{eq:flatframe} and hence the metric determinant factor in those expressions is simply $| g_{ij} | = 1$.
Then one finds $C^{1ij} = \left( \begin{array}{cc} -1/4 & 0 \\ 0 & 1/4 \end{array} \right)^{ij}$ and $C^{2ij} =  \left( \begin{array}{cc} 0 & 1/4 \\ 1/4 & 0 \end{array} \right)^{ij}$, and all other components vanish. 
The first term, with two derivatives, was explored in momentum space in~\cite{Iorio:2017vtw,Iorio:2023cmb}.
The sign in front of the two covariant derivative term is determined by the choice of Dirac point -- the upper sign (`$+$') is for the $K$ point, the lower sign ($`-'$) is for the $K'$ point.
We see that the corrections to the continuum limit take the form of higher derivative terms, and retain a memory of the lattice structure through the invariant tensor $K^{ijk}$. From the perspective of effective field theory we may think of these as irrelevant higher dimension operators that break the Lorentz invariance of the leading Dirac term. However we emphasize here that it is precisely such subleading effects in the effective theory, such as curvature of the electrometric, that derive from the microscopic structure of the lattice model that we are interested in here.

\subsubsection{Curved space Dirac equation}
\label{sec:reviewcurved}

Let us now give a quick review of the curved spacetime Dirac equation, in part to outline the conventions we will use. Given a frame $e^\mu_{~A}(x)$ and its inverse coframe $e^A_{~\mu}(x)$, so that $e^\mu_{~A} e^A_{~\nu} = \delta^\mu_\nu$  and  $ e^A_{~\mu} e^\mu_{~B} = \delta^A_B$ at all spacetime points, then the spacetime metric is given by the coframe and the Minkowski metric $\eta_{AB}$ as,
\eqn{
g_{\mu\nu}(x) = \eta_{AB} e^A_{~\mu} e^B_{~\nu} \quad , \qquad \eta_{AB} = \left(  \begin{array}{cc} -1 & 0 \\ 0 & \delta_{ij} \end{array} \right) \; .
}
Having written the metric in terms of a frame introduces a local Lorentz symmetry that acts as,
\eqn{
e^A_{~\mu}(x) \to \Lambda^A_{~B}(x) e^B_{~\mu}(x)
}
with $\Lambda^A_{~B}(x)$ a Lorentz matrix valued function of spacetime. Since Lorentz matrices obey the defining matrix condition $\eta = \Lambda^T \eta \Lambda$, we see this transformation leaves the spacetime metric invariant.

From the metric we have the unique torsion free metric compatible connection, the Levi-Civita symbol, defining the covariant derivative $\nabla_\mu$ on a covector field $v_\mu$ as,
\eqn{
\nabla_\mu v_\nu = \partial_\mu v_\nu - \Gamma^\rho_{~\mu\nu} v_\rho \quad , \qquad \Gamma^\rho_{~\mu\nu} = \frac{1}{2} g^{\rho\sigma} \left( \partial_\mu g_{\nu\sigma} + \partial_\nu g_{\mu\sigma} - \partial_\sigma g_{\mu\nu} \right) \; .
}
Given the set of covector fields $e^A_{~\mu}$, for $A = 0,1,2$, we define the frame connection,
\eqn{
\label{eq:frameconnection}
\Gamma^A_{~\mu B} = -  e^\nu_{~B} \nabla_\mu e^A_{~\nu} 
}
and this allows us to write a covariant derivative for a frame valued field, $v^A(x)$, as,
\eqn{
D_\mu v^A = \partial_\mu v^A + \Gamma^A_{~\mu B} v^B
}
so that under a local Lorentz frame transformation,
\eqn{
D_\mu v^A \to \Lambda^A_{~B} D_\mu v^B \; .
}
We write this with a $D$ rather than $\nabla$ to emphasize that this should be thought of as a gauge covariant derivative associated to the local Lorentz frame symmetry, and it should not be confused with the covariant derivative $\nabla$ for tensor fields. Here the object $v^A$ is simply a function from the perspective of spacetime, carrying no spacetime tensor indices. However these two derivatives are intimately connected; writing $v^A = v^\mu e^A_{~\mu}$ then,
\eqn{
D_\mu v^A = e^A_{~\nu} \nabla_\mu v^\nu \; .
}
Now we may write this covariant derivative in the manner we do for gauge theory, by introducing the spin connection,
\eqn{
 \Omega_{\mu AB} =  \eta_{AC} \Gamma^C_{~\mu B} 
}
which, following from~\eqref{eq:frameconnection}, is antisymmetric in its frame indices, $ \Omega_{\mu AB} =  \Omega_{\mu [AB]}$. This allows us to write the covariant derivative in terms of a basis for the generators of the Lorentz group, $M^{AB}$, as,
\eqn{
D_\mu v^C = \partial_\mu v^C - \frac{i}{2}  \Omega_{\mu AB} ( M^{AB} )^{C}_{~D} v^D \quad \, \qquad ( M^{AB} )^{C}_{~D} =  i \left( \eta^{AC} \delta^B_D - \eta^{BC} \delta^A_D \right)
}
where the Lorentz group valued function, $\Lambda^A_{~B}(x)$, implementing local Lorentz transformations can be written as,
\eqn{
\Lambda(x) = e^{- \frac{i}{2} \lambda_{AB}(x) M^{AB}}
}
suppressing the frame indices, so that $\lambda_{AB}(x)$ is a function valued in the Lorentz algebra.
From the perspective of local Lorentz transformations forming a principle gauge bundle, then a spinor valued function, $\Psi$, is an associated bundle, transforming as,
\eqn{
\Psi \to  \Lambda_{1/2}(x) \Psi \; , \quad  \Lambda_{1/2}(x) =  e^{- \frac{i}{2} \lambda_{AB}(x) S^{AB}} \; , \quad S^{AB} =  \frac{i}{4} \left[ \gamma^A, \gamma^B \right]
}
for the same local Lorentz transformation corresponding to $\lambda_{AB}(x)$ as above. Here $S^{AB}$ are the Lorentz generators for the spinor representation, and $\Lambda_{1/2}(x)$ is a spinor Lorentz transformation valued function.
Then for a spinor function the corresponding covariant derivative is simply given as,
\eqn{
D_\mu \Psi = \partial_\mu \Psi  - \frac{i}{2}  \Omega_{\mu AB} S^{AB} \Psi 
}
so that again $D_\mu \Psi \to \Lambda_{1/2}(x) D_\mu \Psi$. It will be useful to give the following explicit expression for the spin connection. Let us use the notation that $\partial_A = e^\mu_{~A} \partial_\mu$. Then if we define $J_{CAB} =  e^\nu_{~A} \partial_C e_{\nu B}$ a direct calculation shows that,
\eqn{
\label{eq:spinconnectionfull}
\Omega_{\mu AB} = 2 g_{\mu\nu} \partial_{[A} e^\nu_{~B]}   + \frac{1}{2} e_\mu^{C} \left( \sum_{P \in \mathrm{Perm}(A,B,C)} (-1)^P J_{P_1 P_2 P_3} \right)
}
where, noting the sum over signed permutations, we explicitly see the antisymmetry in $A \leftrightarrow B$.

In all that follows we shall be interested in the case that the spacetime metric is both static and also only has non-trivial spatial geometry. Let us first consider the restriction to having only a non-trivial spatial geometry so that the frame and metric can be written,
\eqn{
\label{eq:ultrastatic}
g_{\mu\nu} = \left( 
\begin{array}{cc}
- c_{eff}^2 & 0 \\
0 & g_{ij}(x) 
\end{array}
\right)
\quad , \qquad 
e^{\mu}_{~A} = \left( 
\begin{array}{cc}
\frac{1}{c_{eff}} & 0 \\
0 & e^i_{I}(x) 
\end{array}
\right)
} 
denoting the decomposition of the frame index into time and spatial parts as $A = (0, I)$.  To preserve this form we will restrict our interest to local Lorentz symmetry transformations which are spatial rotations. Taking such a rotation valued function, $R^I_{~J}(x)$, it generates the local symmetry,
\eqn{
e^I_{~i}(x) \to R^I_{~J}(x) e^J_{~i}(x)
}
We note that in two spatial dimensions, this is a one dimensional subgroup of the Lorentz group. Hence this local symmetry is Abelian.
In terms of our algebra valued function $\lambda_{AB}(x)$ above that specifies the local Lorentz transformation, these local rotations are generated by taking,
\eqn{
\lambda_{12}(x) = - \lambda_{21}(x) = \theta(x) 
}
with all other components vanishing. Then $R^I_{~J}(x)$ is given in terms of the function $\theta(x)$ as,
\eqn{
R(x) = e^{- i \theta(x) J} = \left( \begin{array}{cc} \cos(\theta) & \sin(\theta) \\ -\sin(\theta) & \cos(\theta)  \end{array} \right)  \; , \quad J^I_{~J} = i \epsilon_{IJ} 
}
where, as above, $\epsilon_{IJ}$ has non-vanishing components $1 = \epsilon_{12} = -  \epsilon_{12}$.
The spinor generator associated to rotations is,
\eqn{
S^{12} =  \frac{i}{4} \left[ \gamma^1, \gamma^2 \right] = - \frac{1}{2} \sigma^3
}
so that its action on a spinor is,
\eqn{
\Lambda_{1/2}(x) = e^{- \frac{i}{2} \theta(x) \sigma^3} = \left( \begin{array}{cc} e^{ \frac{i \theta}{2}} &0 \\ 0 & e^{- \frac{i \theta}{2}}  \end{array} \right)  \; .
}
Let us now further restrict our attention to the case that the metric is also static, so time independent. Then we may take the frame to be static, $e^i_{~I} = e^i_{~I}(\bx)$, and the metric then has the above form with static spatial geometry $g_{ij} = g_{ij}(\bx)$, and is referred to as an `ultrastatic' geometry (i.e. one that is static with constant $g_{tt}$ and $g_{ti} = 0$).
To preserve this form further restricts our local rotation symmetry to only depend on space, and not time.  

Now looking at the above explicit expression for the spin connection in equation~\eqref{eq:spinconnectionfull}, we notice that the object $J_{CAB} = e^\nu_{~A} \partial_C e_{\nu B}$ only has spatial components for such a frame. Since we are in two spatial dimensions the term that sums over signed permutations of $J_{CAB}$ then must vanish simply leaving $\Omega_{\mu AB} = 2 g_{\mu\nu} \partial_{[A} e^\nu_{~B]}$. The only non-vanishing components of the spin connection can be written as,
\eqn{
\label{eq:spinconnection}
\Omega_{\mu 12} = -\Omega_{\mu 21} = g_{\mu\nu} \omega^\nu \; , \quad \omega^t = 0 \; , \quad \omega^i = \epsilon^{AB}  \partial_A e^i_{~B} 
}
and so we may write the spinor connection in the simple form,
\eqn{
\label{eq:covpsi}
D_\mu \Psi = \partial_\mu \Psi + \frac{i}{2}  \omega_{\mu} \sigma^3 \Psi \; .
}

\subsubsection{Gauge and frame symmetry}
\label{sec:symmetries}

In writing~\eqref{eq:fastslowansatz} and~\eqref{eq:fastslowansatzKprime} we have separated the spatial dependence of the wavefunctions into `slow' variations encoded by $( \psi_1, \psi_2)$ and `fast' ones, governed by the phase factors $e^{\pm i  \vec{K} \cdot \bx_A}$. Since the Dirac point wavevector $\vec{K} \sim O(1/a)$ these are rapidly varying phases, whose scale of variation is the lattice scale itself by design. As a consequence of this, there is a local freedom in making the separation into fast and slow spatial variations for both wavefunctions, that should not affect the physics of the system.
More concretely taking for the $K$ and $K'$ points 
\eqn{
 A_{\bx_A}(t) &=& \psi_1(t,\bx_A) e^{ \frac{i}{2} \left( - \frac{\pi}{2} \pm \phi(\bx_A) \right)} e^{ \pm i  ( \vec{K} \cdot \bx_A -  \lambda(\bx_A) ) } \; , \quad B_{\bx_B}(t) = \psi_2(t,\bx_B)  e^{-  \frac{i}{2} \left( - \frac{\pi}{2} \pm \phi(\bx_B) \right) } e^{\pm i  ( \vec{K} \cdot \bx_B - \lambda(\bx_B) ) } \
}
the upper signs (`$+$') as well as the first subscript $1$, then $2$ are for the $K$ point, and the lower signs (`$-$') and second subscript $2,~1$ are for the $K'$ point
and where the time independent functions we have introduced $\phi(\bx), \lambda(\bx) \sim O(1)$, and are both slowly varying, so that $\partial_{i_1} \ldots \partial_{i_k} \phi(\bx)  \sim O(1)$ and likewise for $\lambda$, then parameterizes the freedom in making this split into fast and slow spatial variation. Different choices for these slowly spatially varying functions $\phi(\bx)$ and $\lambda(\bx)$ are then different parameterizations, and physics should be independent of this. 

This local invariance manifests elegantly in the continuum Dirac limit as local gauge symmetry and frame rotation symmetry. 
We introduce a $U(1)$ gauge field $A_\mu$ which the spinor field $\Psi$ is charged under, and define the gauged covariant derivative on a spinor field as, 
\eqn{
D_\mu \Psi &=& \partial_\mu \Psi \mp i A_\mu \Psi - \frac{i}{2} \Omega_{\mu AB} S^{AB} \Psi 
}
where again the upper (`$-$') sign for the $K$ point, and the lower (`$+$') sign is for the $K'$ point. Thus we see that we assign the Dirac field opposite charges at the two Dirac points.
We also require the action of more covariant derivatives acting on $\Psi$. Letting $\Psi_{;\nu_1 \ldots \nu_n} = D_{\nu_1} \ldots D_{\nu_n} \Psi$, then these are defined recursively by,
\eqn{
D_\mu (\Psi_{;\mu_1 \ldots \mu_{n-1}}) = \nabla_\mu (\Psi_{;\mu_1 \ldots \mu_{n-1}})  \mp i A_\mu (\Psi_{;\mu_1 \ldots \mu_{n-1}})   - \frac{i}{2} \Omega_{\mu BC} S^{BC} (\Psi_{;\mu_1 \ldots \mu_{n-1}}) 
}
where $\nabla_\mu$ is the usual spacetime covariant derivative acting on a tensor field, and as above we are suppressing the spinor indices.

Now using these gauged covariant derivatives, writing the wavefunctions as above, then the previous continuum limit of \emph{undeformed} graphene \eqref{eq:undeformedDiracThirdOrder} can be written as,
 \eqn{
\label{eq:FlatFrameDiracThirdOrder}
0 =  e^\mu_{~A} \gamma^A D_\mu \Psi \pm i a \, \eta_{AB} \gamma^A e^B_{~\sigma} C^{\sigma\mu\nu} D_\mu D_\nu \Psi  + a^2 \,\eta_{AB} \gamma^A e^B_{~\sigma} D^{\sigma\mu\nu\rho}  D_\mu D_\nu D_\rho \Psi + O(a^3)  
}
where the upper (`$+$') sign is for the $K$ point, and the lower (`$-$') sign is for the $K'$ point, and the gauge field, coframe and torsion-free spin connection are given by, 
\eqn{
e^{A}_{~~\mu} = \left( \begin{array}{ccc} 
{c_{eff}} & 0 & 0 \\
0 &  \cos{\phi} & -\sin{\phi} \\
0 & + \sin{\phi} &  \cos{\phi}
\end{array} \right) 
\; ,\qquad 
A_i = \partial_i \lambda  
\; \qquad
\quad \omega_i = \partial_i \phi 
}
and given the ultrastatic form of the frame, and that $\phi$ is only a function of the spatial coordinates, the covariant derivative takes the form,
\eqn{
D_t \Psi  = \partial_t \Psi \quad , \qquad D_i \Psi = \partial_i \Psi \mp i A_i \Psi + \frac{i}{2}  \omega_{i} \sigma^3 \Psi \; , 
}
and we have analogous results for $D_\mu D_\nu \Psi$ and $D_\mu D_\nu D_\rho \Psi$. We see the simple structure of the spin connection term arising from the Abelian local frame rotations, but note  the important point that the gauge connection and spin connection have a distinct spinor structure from each other.

We see the gauge field is pure gauge, so is unphysical and can be removed by simply setting $\lambda =0$ as we had previously. Further the freedom in $\phi$ simply results in a spatial rotation of the frame field, and correspondingly a non-trivial spin connection. However we emphasize here that this rotation is purely a local freedom in rotating the frame bundle, and doesn't affect the spacetime metric at all, which remains as in the previous equation~\eqref{eq:undefmetric}, so in fact the connection $\Gamma^\rho_{~\mu\nu} = 0$ so that $\nabla_\mu = \partial_\mu$ here.

\subsection{Continuum of the spatially deformed lattice model}
\label{sec:deformed}

Now we turn to the continuum limit when the hopping functions are deformed to be slowly spatially varying as in the Hamiltonian in equation~\eqref{eq:Hamiltonian}. 
We may write the resulting Schr\"odinger system~\eqref{eq:lattice} for one particles states as an expansion in the  parameters $\epsilon$ and $a$,
\eqn{
\Upsilon &=&  \frac{i \hbar}{T} \partial_t  A_{\bx_A} - \sum_n t_n( \bx_A +\frac{a \bl_n}{2}) B_{\bx_A +a \bl_n} = 0 \\
\Upsilon' &=&   \frac{i \hbar}{T} \partial_t  B_{\bx_B} -  \sum_n t_n( \bx_B -\frac{a \bl_n}{2}) A_{\bx_B -a \bl_n} = 0 
}
noting the expansion of the hopping functions introduced earlier in equation~\eqref{eq:texp}.
We expand the $\Upsilon$ equation as,
\eqn{
\Upsilon =  \sum_{n=0}^\infty \epsilon^n \Upsilon_n \; , \quad  \Upsilon_n = \sum_{m=0}^\infty a^m \Upsilon_{n,m}
}
and then the solution $\Upsilon = 0$ implies that order by order $\Upsilon_{n,m} = 0$, and we do similarly for $\Upsilon'$.
As above we introduce slowly spatially varying wavefunctions, $\psi_{1,2}(t,\bx)$, and now also a slowly varying phase modulation function $\Phi(\bx)$ and wavefunction rescaling $f(\bx)$, so $\partial_{i_1} \ldots \partial_{i_k} \Phi \sim O(1)$ and likewise for $f(\bx)$, and use these to write,
\eqn{\label{eq:general_redef_kpoint}
A_{\bx_A}(t) = \psi_{1,2}(t,\bx_A) f(\bx) e^{ \frac{i}{2} ( - \frac{\pi}{2} \pm \phi(\bx_A) ) } e^{\pm \frac{i \Phi(\bx_A)}{a} } \; , \quad B_{\bx_B}(t) = \psi_{2,1}(t,\bx_B)  f(\bx)  e^{- \frac{i}{2}  ( - \frac{\pi}{2} \pm \phi(\bx_B) ) } e^{\pm \frac{i \Phi(\bx_B)}{a}  } 
}
and as above, the upper signs (`$+$') as well as the first subscript $1$, then $2$ are for the $K$ point, and the lower signs (`$-$') and second subscript $2,~1$ are for the $K'$ point.

While $\Phi$ slowly varies, the inverse factor of $a$ multiplying it in the exponential means that the phase rapidly varies, changing on lattice scales. Note that a natural choice of $f$ is $f=(\det g_{ij})^{1/4}$, which will ensure that as we mentioned previously, the number density of the continuum Dirac field is the same as the microscopic electron density,  $\sqrt{|g_{ij}|} \bar\Psi \gamma^t \Psi = |A|^2 + |B|^2$.  In fact we will find that such a choice is also necessary to recover the torsion-free spin connection. We note that the earlier work \cite{dejuan2012spacedepfermi,oliva2015generalizing} specifically worked with the Weyl rescaled field $\hat\Psi = (\det g_{ij})^{-1/4} \Psi$, which as discussed before can be thought of as working in a different Weyl frame with only nontrivial $g_{tt}$ \cite{stegmann2016current}. However the higher derivative terms are \emph{not} Weyl invariant, and so we can only think of this as working with a spinor density instead of a canonically normalized spinor.

We perturbatively expand about $\epsilon = 0$, the undeformed model as, 

\eqn{\label{eq:phi_and_Phi}
\Phi(\bx) =  - \frac{4 \pi}{3 \sqrt{3}} x +  \sum^\infty_{n=1}  \epsilon^n  \delta_n\Phi(\bx) \; , \quad \phi(\bx) =  \sum^\infty_{n=1}  \epsilon^n  \delta_n\phi(\bx) \; , \quad f(\bx) = 1 + \sum^\infty_{n=1}  \epsilon^n  \delta_nf(\bx) 
}
so that for $\epsilon = 0$ then $\frac{1}{a} \Phi(\bx) = \vec{K} \cdot \bx$, and further expand the perturbative functions $\delta_n\Phi$ in $a$ as,
\eqn{
\delta_n\Phi = \sum_{m=0}^\infty a^m \delta_{n,m}\Phi  
}
and likewise for $\delta_n\phi$ and $\delta_nf$. Since we are first expanding in the deformation parameter $\epsilon$, and only afterwards we expand in $a$, we may expand the exponential factor above as,
\eqn{\label{eq:expanding_phase_Phi}
e^{\frac{i \Phi(\bx)}{a} } \simeq e^{i \vec{K} \cdot \vec{x} } \left( 1 +  \frac{i \epsilon}{a} \delta_1\Phi + \frac{i \epsilon^2}{a} \delta_2\Phi  -  \frac{\epsilon^2}{2 a^2} (\delta_1\Phi)^2  + O(\epsilon^3) \right)  
}
and then for each term in this expansion, we expand the $\delta_n\Phi$ in powers of $a$. 
Having performed this expansion also in $a$, so we have a double expansion in both $\epsilon$ and $a$, it is convenient to introduce a new expansion parameter,
\eqn{
\lambda = \frac{\epsilon}{a}
}
so that we may write,
\eqn{
e^{\frac{i \Phi(\bx)}{a} } \simeq e^{i \vec{K} \cdot \vec{x} } \left( 1 +  i \lambda \left( \delta_{1,0} \Phi  + a \delta_{1,1} \Phi  + a^2 \delta_{1,2} \Phi   + O(a^3) \right) - \lambda^2 \left( \frac{1}{2} (\delta_{1,0} \Phi)^2 + a \delta_{1,0} \Phi \delta_{1,1} \Phi - 2 i a \delta_{2,0} \Phi + O(a^2)  \right)  + O(\lambda^3) \right)  \; . \nl
}
While written in $\epsilon$ and $a$ the two limits $\epsilon \to 0$ and $a \to 0$ do not commute -- taking $\epsilon \to 0$ with $a$ finite allows the expansion of the exponential above, but the reverse, $a \to 0$ with finite $\epsilon$, gives a diverging phase and the exponential cannot be expanded. 
Thus we justify our earlier statements, that the expansion in $\epsilon$ should be performed first, and then afterwards the one in $a$, so that this exponential can be expanded. Alternatively we may view the condition that we may expand the exponential in $\epsilon$ and $a$ as being that both $\lambda$ and $a$ are small. Thus $\epsilon/a$ must be held small as we take the continuum limit $a \to 0$ as stated earlier.

It is interesting to consider the magnitude of $\epsilon$ in rippling suspended graphene, even though this involves out of plane displacement, which, as discussed above, may not be well captured by the simple tight-binding model. For such ripples the height is approximately $\sim 0.5$nm and the wavelength is $\sim 5$nm and these configurations are frozen in time, as deduced from STM microscopy~\cite{zan2012scanning}. Thus in our units $L = 1$ corresponds to $5$nm, and so the graphene lattice spacing, which is $\sim 0.25$nm gives approximately $a \sim 0.05$. 
On the other hand, the height function $h$ can be written as $h \sim \sqrt{\epsilon} \cos\left( \frac{x}{2\pi} \right)$, where $\sqrt{\epsilon} \sim 0.1$ to give a ripple height of $0.5$nm. Hence $\epsilon \sim 0.01$, leading to a ratio $\lambda = \epsilon/a \sim 0.2$, which is  small, but not very small. Thus even for these seemingly low amplitude ripples, corrections in $\lambda$ will likely be important.

Finally the Schr\"odinger system can then be written in the form,
\eqn{
0 = \sum_{p=0}^\infty   \sum_{q=0}^\infty a^p \lambda^q  \mathcal{O}_{p,q}(\bx)   \left(
\begin{array}{c}
\psi_1(\bx) \\
\psi_2(\bx)
\end{array}
\right)
}
for spatial differential operators $\mathcal{O}_{p,q}$ which depend only on $\bx$ (with the single exception of $\mathcal{O}_{1,0}$, which contains the one time derivative), and on the various functions $\delta_m t_n$, $\delta_{n,m}\Phi$, $\delta_{n,m}\Phi$, $\delta_{n,m}f$ and their derivatives, but not on $\epsilon$ or $a$. The terms $\mathcal{O}_{p,0}$ are those of the undeformed model, giving, in lattice coordinates,
\eqn{
\mathcal{O}_{1,0} = \frac{1}{c_{eff}} \gamma^0 \partial_t + \delta^i_I \gamma^I \partial_i 
\; , \quad 
\mathcal{O}_{2,0} = \pm i  \, \eta_{AB} \gamma^A  \delta^B_\sigma C^{\sigma\mu\nu} \partial_\mu \partial_\nu
\; , \quad 
\mathcal{O}_{3,0} =  \eta_{AB} \gamma^A \delta^B_\sigma  D^{\sigma\mu\nu\rho} \partial_\mu \partial_\nu \partial_\rho
}
as we saw above. 
Now we must match this to a continuum description.

\subsubsection{Structure of the effective theory}

A key requirement of the continuum description is that it should have local frame rotation and gauge symmetry.  The effective theory therefore contains a gauge field and frame, with its associated spin connection, and all derivatives must be covariant with respect to these. Anticipating the correct form, we perform the double expansion of the frame and gauge field firstly in $\epsilon$ and then $a$ as,
\eqn{
A_\mu =  \left( 0 , A^i \right)  \; , \quad A_i = \sum_{n=1}^\infty \epsilon^n \delta_nA_i \; , \quad \delta_nA_i = \frac{1}{a} \sum_{m=0}^\infty a^m \delta_{n,m}A_i 
}
where we emphasize that the expansion in $a$ above starts with the power $a^{-1}$,
and for the frame,
\eqn{
e^\mu_{~A} = \left( \begin{array}{cc} 
\frac{1}{c_{eff}} & 0  \\
0 & e^i_{~I}
\end{array} \right) \; , \quad e^i_{~I} = \delta^i_I + \sum_{n=1}^\infty \epsilon^n  \delta_{n}e^i_{~I}  \; , \quad \delta_{n}e^i_{~I} =   \sum_{m=0}^\infty a^m \delta_{n,m}e^i_{~I} \; .
}
This frame determines the metric which then has $g_{tt} = - c_{eff}^2$, $g_{ti} = 0$ and an analogous expansion for its spatial components,
\eqn{
g_{ij} = e^I_{~i} e^I_{~j} = \delta_{ij} + \sum_{n=1}^\infty \epsilon^n  \delta_{n}g_{ij}  \; , \quad \delta_{n}g_{ij}=   \sum_{m=0}^\infty a^m \delta_{n,m}g_{ij}
}
and likewise for the Christoffel symbols and torsion free spin connection, $\omega_\mu$, as in equation~\eqref{eq:spinconnection}, which have only non-vanishing spatial components, again with expansions as above,
\eqn{
\Gamma^k_{~ij} =  \sum_{n=1}^\infty \epsilon^n  \left( \sum_{m=0}^\infty a^m \delta_{n,m}\Gamma^k_{~ij} \right) \; , \quad \omega_i =  \sum_{n=1}^\infty \epsilon^n  \left( \sum_{m=0}^\infty a^m \delta_{n,m}\omega_i\right) \; .
}
The key feature of this expansion is that the leading behaviour of the magnetic gauge field goes inversely with $a$, so $A_i \sim \epsilon/a$. 
As we have emphasized earlier, first expanding in infinitessimal $\epsilon$, and then in $a$, this leading behaviour $A_i \simeq \epsilon \delta_1 A_i$ is perturbatively small. However, formally the function $\delta_1 A_i \sim 1/a$ itself diverges in the continuum limit $a \to 0$ and hence we term this the \emph{large magnetic field}, understanding that we should ensure $\epsilon/a$ is finite and small as we take the continuum limit $a \to 0$ to ensure we can perform a perturbative expansion.

Terms in the effective continuum description will involve quantities constructed from lattice invariants and their derivatives, and covariant derivatives of the spinor field $\Psi = (\psi_1, \psi_2)$. Let's consider a term which we schematically write as,
\eqn{
a^M Q^{\mu_1 \ldots \mu_M} D_{\mu_1} \cdots D_{\mu_M}  \Psi
}
where we have suppressed all spinor indices and Gamma matrices. Here the tensor $Q$ is constructed from the lattice data, so from the coupling functions $t_n$ and the lattice vectors $\vec{\ell}_n$. Since the $t_n$ have an expansion in $\epsilon$ and $a$, then we may write,
\eqn{
Q^{\mu_1 \ldots \mu_M}(\bx) = Q_{0,0}^{\mu_1 \ldots \mu_M} + \sum_{n=1}^\infty \sum_{m=0}^\infty \epsilon^n a^m Q_{n,m}^{\mu_1 \ldots \mu_M}(\bx)
}
where the leading term, $Q_{0,0}^{\mu_1 \ldots \mu_M}$ will comprise constants that are independent of $\bx$, but the subleading terms $Q_{n,m}^{\mu_1 \ldots \mu_M}(\bx)$ for $n \ge 1$ will be slowly varying functions of position via the couplings that slowly spatially vary. 
The covariant derivative for the $K$ point may be expanded as,
\eqn{
D  &=& \nabla  -  i  A  - \frac{i}{2} \omega \sigma_3 \\
& = &  \partial - i \epsilon \left(  \left( \frac{1}{a}  \delta_{1,0}A + \left( i \delta_{1,0}\Gamma + \delta_{1,1}A \right) + O(a) \right) + \frac{1}{2}  \left(  \delta_{1,0} \omega + O(a) \right) \sigma_3 \right)  - i  \epsilon^2 \left(   \frac{1}{a}  \delta_{2,0}A +O(a^0) \right) + O(\epsilon^3) 
}
where we have suppressed all indices, and we have a similar expansion for the derivative at the $K'$ point (which differs by signs).
The presence of the gauge field contributing terms going as $1/a$ changes the structure of the effective field theory from the more usual case in relativistic QFT, and implies that higher covariant derivative terms may contribute at lower orders in the derivative expansion due to these gauge terms. Thus the terms above can be expanded as,
\eqn{
a^M Q^{\mu_1 \ldots \mu_M} D_{\mu_1} \cdots D_{\mu_M}  \Psi = a^M  T_{M,0} + \sum_{p=0}^\infty \sum_{q=1}^\infty a^p \lambda^q  T_{p,q} 
}
for an $M$-th derivative term, where we have,
\eqn{
T_{p,q} = 0 \quad \forall \quad p  < M
}
but are otherwise non-trivial.
If $A$ did not have large leading behaviour, so $\delta_{n,0}A = 0$, then one would have $T_{p,q} = 0$ for $p-q < M$. However due to the presence of the large magnetic field, gauge field terms such as,
\eqn{
a^M Q^{\mu_1 \ldots \mu_M} A_{\mu_1} \partial_{\mu_2} \cdots \partial_{\mu_M}  \Psi \; , \quad a^M Q^{\mu_1 \ldots \mu_M} \partial_{\mu_1} A_{\mu_2} \partial_{\mu_3} \cdots \partial_{\mu_M}  \Psi   \; , \quad a^M Q^{\mu_1 \ldots \mu_M} A_{\mu_1} A_{\mu_2} \partial_{\mu_3} \cdots \partial_{\mu_M}  \Psi 
}
contribute to lower orders in the $a$ expansion. While each gauge field contributes an inverse $a$, it also comes with an additional factor $\epsilon$. Thus this term has zero contributions for $p \le M$ but will generally have non-trivial contributions $T_{p,q}$ otherwise. Pictorially we have,\\

\centerline{
\begin{tabular}{c||c|c|c|c|c|c|c|cc}
& $a^0$             &   $a^1$             & \hspace{0.3cm}  $\ldots$   \hspace{0.3cm} & $a^{M-1}$               & $a^M$             & $a^{M+1}$   & $a^{M+2}$          & $\ldots$                \\
\hline\hline
$\epsilon^0$&       0               &           0               &        \ldots          &            0                    & $T_{M,0}$       &           0                 &          0                & \hspace{0.3cm}  \ldots  \\
$\epsilon^1$&      0                &          0                &         \ldots         &  $T_{M,1}$          & $T_{M+1,1}$       & $T_{M+2,1}$     & $T_{M+3,1}$     &   \hspace{0.3cm}  \ldots \\
		&                     &                          &                  &                                &                        &                            &                          &   \\
\vdots         &       \vdots  &                \vdots          &  \reflectbox{$\ddots$}    &    $\vdots$          & \vdots              &      $\vdots$         &     $\vdots$        &  \\
                     &                          &                  &                                &                        &                            &                          &   \\                      
$\epsilon^{M-1}$&        0              &  $T_{M,M-1}$    &  $\ldots$   &  $T_{2 M-2,M-1}$      & $T_{2 M-1,M-1}$   & $T_{2 M,M-1}$ & $T_{2 M+1,M-1}$  &   \hspace{0.3cm} \ldots \\
$\epsilon^M$& $T_{M,M}$    &  $T_{M+1,M}$        & $\ldots$   &  $T_{2 M-1,M}$         & $T_{2 M,M}$       & $T_{2 M+1,M}$    & $T_{2 M+2,M}$  &  \hspace{0.3cm}  \ldots \\
$\epsilon^{M+1}$&$T_{M+1,M+1}$ &  $T_{M+2,M+1}$    & $\ldots$   &  $T_{2 M,M+1}$     & $T_{2 M+1,M+1}$   & $T_{2M+2,M+1}$ & $T_{2 M+3,M+1}$  &   \hspace{0.3cm} \ldots \\
                     &                          &                  &                                &                        &                            &                          &   \\
$\vdots$         &    $\vdots$        &                  &  $\vdots$    & $\vdots$          &    $\vdots$     &    $\vdots$       & &\\
\end{tabular}
}

\noindent
A corollary of this discussion is that if we consider a term with $M$ derivatives of the above form, then,
\eqn{
a^M Q^{\mu_1 \ldots \mu_M} D_{\mu_1} \cdots D_{\mu_M}  \Psi \quad \mathrm{contributes\;to} \quad \mathcal{O}_{p,q} \quad \mathrm{for} \quad p \ge M
}
unlike in usual effective field theory where these terms would only contribute to $\mathcal{O}_{p,q}$ with $p-q \ge M$. 
Alternatively we may say that the equations $\mathcal{O}_{p,q}$ with $p \le M$ only involve contributions from terms in the equation of motion with $M$ derivatives or less.
Let us now see how this works in practice.

\subsubsection{Leading order - $\mathcal{O}_{p,q}$ for $p \le 1$ - flat Dirac with large magnetic field}

Considering all the lattice equations $\mathcal{O}_{p,q}$ for $p \le 1$, yields the non-trivial equations $\mathcal{O}_{1,0}$ and $\mathcal{O}_{1,1}$, and these involve only the one covariant derivative term, i.e. the leading Dirac equation term. We find that to this leading order the continuum theory matching the lattice model is,
 \eqn{
\label{eq:CurvedDiracLeading}
0 =  a e^\mu_{~A} \gamma^A D_\mu \Psi + O(\epsilon^2, \epsilon a, a^2)  
}
where we have a flat undeformed frame and non-trivial magnetic gauge field, found from solving $\mathcal{O}_{1,0}$ and $\mathcal{O}_{1,1}$,
\eqn{
e^\mu_{~A} = \left( \begin{array}{ccc} 
\frac{1}{c_{eff}} & 0 & 0 \\
0 &  1 &0 \\
0 & 0 & 1
\end{array} \right) + O( \epsilon ) \; , \quad A_i = \frac{\epsilon}{a} \left( \delta_{1,0} A_i + O(a) \right) + O( \frac{\epsilon^2}{a}  )  \; , \quad f = 1 + O( {\epsilon}   )
}
where,
\eqn{
\delta_{1,0} A_i & = & \frac{1}{3}  \left( \begin{array}{c} 
 \delta_{1,0} t_1 +  \delta_{1,0} t_2 - 2  \delta_{1,0} t_3  \\
- \sqrt{3} \left(  \delta_{1,0} t_1  -  \delta_{1,0} t_2 \right)
\end{array} \right) - \partial_i (\delta_{1,0} \Phi) 
}
and we see the gauge transformation enter, with gauge parameter $\delta_{1,0} \Phi$.
Subleading corrections in powers of $\epsilon$ or $a$ to the frame and gauge field then only affect the higher order equations $\mathcal{O}_{p,q}$ for $p > 1$, and these also contain contributions from higher covariant derivative terms -- thus these subleading corrections cannot be considered consistently without also including these higher covariant derivative terms too.

This was a key conclusion of our paper~\cite{Roberts:2021vmt}, namely that for the leading order effective theory -- the Dirac equation coupled to the strain gauge field -- whilst the gauge field is non-trivial in general, the frame, and thus the metric, is undeformed.

\subsubsection{Second order - $\mathcal{O}_{p,q}$ for $p \le 2$ - curved space Dirac and higher covariant derivative term with large magnetic field}

Now we take the lattice equations $\mathcal{O}_{p,q}$ for $p \le 2$, which give the previous equations $\mathcal{O}_{1,0}$ and $\mathcal{O}_{1,1}$, and at next order also $\mathcal{O}_{2,0}$, $\mathcal{O}_{2,1}$ and $\mathcal{O}_{2,2}$. The equation $\mathcal{O}_{2,0}$ is solved by adding the two derivative term for the undeformed case given above in equation~\eqref{eq:undeformedDiracThirdOrder}, and so the full effective description up to two derivatives is,
 \eqn{
0 =  a e^\mu_{~A} \gamma^A D_\mu \Psi \pm i a^2 \, \eta_{AB} \gamma^A e^B_{~\sigma} C^{\sigma\mu\nu} D_\mu D_\nu \Psi + O(\epsilon^3, \epsilon^2 a, \epsilon a^2, a^3)  
}
with the upper sign for the $K$ point, and the lower for the $K'$ point. We note that the coefficient $C^{\sigma\mu\nu}$, given in equation~\eqref{eq:invariants} has a factor involving $\det(g_{ij})$, but this doesn't contribute here -- at this order we could consistently simply take the non-zero components $C^{ijk} = -\epsilon_{kl} K^{ijl}/4$, so they are just given by the lattice invariants.

In order to solve the remaining equations, $\mathcal{O}_{2,1}$ and $\mathcal{O}_{2,2}$, we must introduce a linear deformation to the spatial frame, corrections to the gauge field, and also to the rescaling function $f$,
\eqn{
e^i_{~I} &=& \delta^i_I + \epsilon \left( \delta_{1,0}e^i_{~I} + O(a) \right) + O( \epsilon^2 ) \; , \quad f = 1 + \epsilon \left( \delta_{1,0} f + O(a) \right) +  O( {\epsilon^2}  )
\nl
 A_i &=& \frac{\epsilon}{a} \left( \delta_{1,0} A_i + a \delta_{1,1} A_i + O(a^2) \right) + \frac{\epsilon^2}{a} \left( \delta_{2,0} A_i + O(a) \right) + O( \frac{\epsilon^3}{a}  )   \; .
}
The terms in these expansions determined from the leading order equations, here just $\delta_{1,0} A_i$, are as above. The new subleading corrections are then determined by examining the new equations at this order i.e. $\mathcal{O}_{p,q}$ with $p = 2$.
The equation $\mathcal{O}_{2,0}$ is the one from the undeformed theory and is satisfied by the choice for the two derivative term.
Next we consider $\mathcal{O}_{2,1}$ where the one derivative terms on $\Psi$ fix the frame correction as,
\eqn{
\delta_{1,0} e^i_{~I} & = & \left(
\begin{array}{cc}
\frac{2}{3} \delta_{1,0} t_1 + \frac{2}{3} \delta_{1,0} t_2 - \frac{1}{3} \delta_{1,0} t_3 & \frac{1}{\sqrt{3}} \delta_{1,0} t_1  - \frac{1}{\sqrt{3}} \delta_{1,0} t_2 \\
 \frac{1}{\sqrt{3}} \delta_{1,0} t_1  - \frac{1}{\sqrt{3}} \delta_{1,0} t_2 & \delta_{1,0} t_3
\end{array}
\right) + \left(
\begin{array}{cc}
0 &  \delta_{1,0} \phi \\
 - \delta_{1,0} \phi & 0
\end{array}
\right) 
}
and we recognize  the term involving $  \delta_{1,0} \phi$ as a perturbative frame rotation.
The terms in $\mathcal{O}_{2,1}$ with no derivatives on $\Psi$ determine the rescaling function to be,
\eqn{
\delta_{1,0}f = - \frac{1}{3} \left( \delta_{1,0} t_1 + \delta_{1,0} t_2 + \delta_{1,0} t_3 \right)
}
which is the leading correction in the expansion of $f=(\det g_{ij})^{1/4}$. Continuing, we determine the correction $\delta_{1,1} A_i$ to the gauge field from $\mathcal{O}_{1,1}$, obtaining,
\eqn{
\delta_{1,1} A_i & = & \frac{1}{3}  \left( \begin{array}{c} 
 \delta_{1,1} t_1 +  \delta_{1,1} t_2 - 2  \delta_{1,1} t_3  \\
- \sqrt{3} \left(  \delta_{1,1} t_1  -  \delta_{1,1} t_2 \right)
\end{array} \right) - \partial_i (\delta_{1,1} \Phi) \; .
}
Finally, the remaining equations $\mathcal{O}_{2,2}$ determine the $\epsilon^2$ correction to the gauge field,
\eqn{
\delta_{2,0} A_i & = & \frac{1}{3}  \left( \begin{array}{c} 
 \delta_{2,0} t_1 +  \delta_{2,0} t_2 - 2  \delta_{2,0} t_3  \\
- \sqrt{3} \left(  \delta_{2,0} t_1  -  \delta_{2,0} t_2 \right)
\end{array} \right) 
+
 \frac{1}{18}  \left( \begin{array}{c} 
 \delta_{1,0} t_1^2 + \delta_{1,0} t_2^2  - 2 \delta_{1,0} t_3^2   + 8  \delta_{1,0} t_1  \delta_{1,0} t_3+ 8  \delta_{1,0} t_2  \delta_{1,0} t_3 - 16 \delta_{1,0} t_1 \delta_1 t_2  \\
- \sqrt{3} \left(   \delta_{1,0} t_1^2  - \delta_{1,0} t_2^2 + 8 \delta_{1,0} t_2 \delta_{1,0} t_3 - 8 \delta_{1,0} t_1 \delta_{1,0} t_3  \right)
\end{array} \right)  \nn \\
&&
- \partial_i (\delta_{2,0} \Phi) 
}
where again we see the gauge freedom associated to the choice $\delta_{2,0} \Phi$, and now we see this quadratic correction to the gauge field has contributions from the quadratic deformation of the couplings, $\delta_{2,0} t_n$, but also non-linear terms in the leading deformations $\delta_{1,0} t_n$.

We explicitly see another key conclusion of our paper~\cite{Roberts:2021vmt}, namely that while the frame becomes perturbed from being trivial, so that we may consider the Dirac term to live in a curved space,
at the same time one must also introduce a second covariant derivative term for consistency. One cannot truncate to a curved space Dirac equation (plus strain gauge field), since the higher derivative term must be included to match to the microscopic lattice theory at the subleading order where the frame becomes non-trivial.

\subsubsection{Third order - $\mathcal{O}_{p,q}$ for $p \le 3$ - $O(\epsilon^2)$ corrections to the metric}

We now give the theory to third order. It takes the form above, with additional corrections to the metric, gauge field, rescaling function $f$, and to the higher covariant derivative terms. To this third order the equations $\mathcal{O}_{p,q}$ for $p \le 3$ are solved by the continuum theory,
 \eqn{
\label{eq:third order}
0 =  a e^\mu_{~A} \gamma^A D_\mu \Psi \pm i a^2 \, \eta_{AB} \gamma^A e^B_{~\sigma} D_\mu \left( C^{\sigma\mu\nu} D_\nu \Psi \right) 
+ a^3 \,\eta_{AB} \gamma^A e^B_{~\sigma} D^{\sigma\mu\nu\rho}  D_\mu D_\nu D_\rho \Psi 
  + O(\epsilon^4, \epsilon^3 a, \epsilon^2 a^2, \epsilon a^3, a^4)  \nn \\
}
where now we have added a three covariant derivative term with coefficient given by the invariant $D^{\sigma\mu\nu\rho}$ as defined in~\eqref{eq:invariants}, and again the upper sign is for the $K$ point and the lower one is for the $K'$ point.
The spatial frame, gauge field and rescaling function now must have expansions as,
\eqn{
e^i_{~I} &=& \delta^i_I + \epsilon \left( \delta_{1,0}e^i_{~I} + a \delta_{1,1}e^i_{~I} + O(a^2) \right)  + \epsilon^2 \left( \delta_{2,0}e^i_{~I} + O(a) \right) + O( \epsilon^3 ) \nl
 A_i &=& \frac{\epsilon}{a} \left( \delta_{1,0} A_i + a \delta_{1,1} A_i + a^2 \delta_{1,2} A_i + O(a^3) \right) + \frac{\epsilon^2}{a} \left( \delta_{2,0} A_i + a \delta_{2,1} A_i + O(a^2)  \right) + \frac{\epsilon^3}{a}  \left( \delta_{3,0} A_i  + O(a) \right) + O( \frac{\epsilon^4}{a}  ) \nl
  f &=& 1 + \epsilon \left( \delta_{1,0} f + a \delta_{1,1} f + O(a^2)  \right) + \epsilon^2 \left( \delta_{2,0} f + O(a) \right) +  O( {\epsilon^3} ) 
}
where the terms $ \delta_{1,0}e^i_{~I}$, $\delta_{1,0} A_i$,  $\delta_{1,1} A_i$, $\delta_{2,0} A_i$ and $\delta_{1,0} f$ are as for the previous order. The new terms are then determined by the equations $\mathcal{O}_{p,q}$ where $p=3$, namely $\mathcal{O}_{3,0}$, $\mathcal{O}_{3,1}$, $\mathcal{O}_{3,2}$  and $\mathcal{O}_{3,3}$.

Now the $\sqrt{\det(g_{ij})}$ factor in the expression for $C^{ijk}$ in equation~\eqref{eq:invariants} plays an important role at this order -- the derivatives acting on this factor give contributions to the equations $\mathcal{O}_{3,1}$ and $\mathcal{O}_{3,2}$ that are crucial to  allow the equations to be satisfied consistently. We note that at this order, the same factor in the coefficient tensor $D^{ijkl}$ does not play a role.

At order $\mathcal{O}_{3,1}$ we find the frame and rescaling function $f$ are,
\eqn{
\delta_{1,1} e^i_{~I} & = & \left(
\begin{array}{cc}
\frac{2}{3} \delta_{1,1} t_1 + \frac{2}{3} \delta_{1,1} t_2 - \frac{1}{3} \delta_{1,1} t_3 & \frac{1}{\sqrt{3}} \delta_{1,1} t_1  - \frac{1}{\sqrt{3}} \delta_{1,1} t_2 \\
 \frac{1}{\sqrt{3}} \delta_{1,1} t_1  - \frac{1}{\sqrt{3}} \delta_{1,1} t_2 & \delta_{1,1} t_3
\end{array}
\right) + \left(
\begin{array}{cc}
0 & + \delta_{1,1} \phi \\
 - \delta_{1,1} \phi & 0
\end{array}
\right) \\
\delta_{1,1}f &=& - \frac{1}{3} \left( \delta_{1,1} t_1 + \delta_{1,1} t_2 + \delta_{1,1} t_3 \right) 
}
and again we see $\delta_{1,1}\phi$ parameterizes the infinitessimal frame rotation freedom at this order.
The gauge field receives a correction from $\mathcal{O}_{3,1}$ giving,
\eqn{
\delta_{1,2} A_i & = &  \frac{1}{288}  \left( \begin{array}{c} 
\partial_x^2 ( - 27  \delta_1 t_1 -27  \delta_1 t_2 - 36  \delta_1 t_3 ) 
+  \partial_y^2 ( 47 \delta_1 t_1 + 47 \delta_1 t_2 - 3 \delta_1 t_3 ) - 14 \sqrt{3} \partial_x\partial_y ( \delta_1 t_1 -  \delta_1 t_2 ) \\
17 \sqrt{3} \partial_x^2 ( -   \delta_1 t_1 +  \delta_1 t_2  ) 
+ 3 \sqrt{3}  \partial_y^2 ( - \delta_1 t_1 +  \delta_1 t_2  ) + \partial_x\partial_y ( 46 \delta_1 t_1 + 46  \delta_1 t_2 + 88  \delta_1 t_3 ) 
\end{array} \right) \nl
&& \qquad + \frac{1}{3}  \left( \begin{array}{c} 
 \delta_{1,2} t_1 +  \delta_{1,2} t_2 - 2  \delta_{1,2} t_3  \\
- \sqrt{3} \left(  \delta_{1,2} t_1  -  \delta_{1,2} t_2 \right)
\end{array} \right) - \partial_i (\delta_{1,1} \Phi) 
- \partial_i (\delta_{1,2} \Phi) \; . 
} 
Now the equations $\mathcal{O}_{3,2}$ determine the interesting $\epsilon^2$ correction to the spatial frame components, $\delta_{2,0} e^i_{~I}$ and rescaling function $\delta_{2, 0} f$. 
These are;
\eqn{
\delta_{2,0} e^i_{~I} & = & \left(
\begin{array}{cc}
\frac{2}{3} \delta_{2,0} t_1 + \frac{2}{3} \delta_{2,0} t_2 - \frac{1}{3} \delta_{2,0} t_3 & \frac{1}{\sqrt{3}} \delta_{2,0} t_1  - \frac{1}{\sqrt{3}} \delta_{2,0} t_2 \\
 \frac{1}{\sqrt{3}} \delta_{2,0} t_1  - \frac{1}{\sqrt{3}} \delta_{2,0} t_2 & \delta_{2,0} t_3
\end{array}
\right)  \nl
&& \qquad + \left(
\begin{array}{cc}
- \frac{1}{18} \left( \delta_{1,0} t_1 + \delta_{1,0} t_2 - 2 \delta_{1,0} t_3 \right)^2 & \frac{1}{6 \sqrt{3}} \left( \delta_{1,0} t_1 - \delta_{1,0} t_2 \right)  \left( \delta_{1,0} t_1 + \delta_{1,0} t_2 - 2 \delta_{1,0} t_3 \right) \\
\frac{1}{6 \sqrt{3}} \left( \delta_{1,0} t_1 - \delta_{1,0} t_2 \right)  \left( \delta_{1,0} t_1 + \delta_{1,0} t_2 - 2 \delta_{1,0} t_3 \right) & - \frac{1}{6} \left( \delta_{1,0} t_1 - \delta_{1,0} t_2 \right)^2 
\end{array}
\right) \nl
&& \qquad +
 \left(
\begin{array}{cc}
\frac{2}{3} \delta_{1,0} t_1 + \frac{2}{3} \delta_{1,0} t_2 - \frac{1}{3} \delta_{1,0} t_3 & \frac{1}{\sqrt{3}} \delta_{1,0} t_1  - \frac{1}{\sqrt{3}} \delta_{1,0} t_2 \\
 \frac{1}{\sqrt{3}} \delta_{1,0} t_1  - \frac{1}{\sqrt{3}} \delta_{1,0} t_2 & \delta_{1,0} t_3
\end{array}
\right)  \cdot\left(
\begin{array}{cc}
0 & \delta_{1,0} \phi \\
-  \delta_{1,0} \phi & 0
\end{array}
\right) 
+ \left(
\begin{array}{cc}
- \frac{1}{2} (\delta_{1,0} \phi)^2 & \delta_{2,0} \phi \\
 - \delta_{2,0} \phi & - \frac{1}{2} (\delta_{1,0} \phi)^2
\end{array}
\right)  \nl
}
and
\eqn{
\delta_{2,0}f &=& - \frac{1}{3} \left( \delta_{2,0} t_1 + \delta_{2,0} t_2 + \delta_{2,0} t_3 \right) + \frac{4}{9} \left( (\delta_{1,0} t_1)^2 + (\delta_{1,0} t_2)^2 + (\delta_{1,0} t_3)^2 \right) \nl
&& \qquad - \frac{1}{9} \left( (\delta_{1,0} t_1) (\delta_{1,0} t_2) + (\delta_{1,0} t_1) (\delta_{1,0} t_3)  + (\delta_{1,0} t_2) (\delta_{1,0} t_3)  \right)  \; .
}
They also determine the gauge field as,
\eqn{
\delta_{2,1} A_i & = & \frac{1}{3}  \left( \begin{array}{c} 
 \delta_{2,1} t_1 +  \delta_{2,1} t_2 - 2  \delta_{2,1} t_3  \\
- \sqrt{3} \left(  \delta_{2,1} t_1  -  \delta_{2,1} t_2 \right)
\end{array} \right) 
+
 \frac{1}{9}  \left( \begin{array}{c} 
( \delta_{1,0} t_1)( \delta_{1,1} t_1) + ( \delta_{1,0} t_2)( \delta_{1,1} t_2) - 2  ( \delta_{1,0} t_3)( \delta_{1,1} t_3)
 \\
 - \sqrt{3} \left( ( \delta_{1,0} t_1)( \delta_{1,1} t_1) - ( \delta_{1,0} t_2)( \delta_{1,1} t_2)      \right)
\end{array} \right) - \partial_i (\delta_{2,1} \Phi) 
 \nl
&&  + 
 \frac{1}{9}  \left( \begin{array}{c} 
 4  \left( ( \delta_{1,0} t_1)( \delta_{1,1} t_3)  + ( \delta_{1,0} t_3)( \delta_{1,1} t_1)  \right) +  4  \left( ( \delta_{1,0} t_2)( \delta_{1,1} t_3)  + ( \delta_{1,0} t_3)( \delta_{1,1} t_2)  \right) - 8  \left( ( \delta_{1,0} t_1)( \delta_{1,1} t_2)  + ( \delta_{1,0} t_2)( \delta_{1,1} t_1)  \right) 
 \\
 - \sqrt{3} \left(  4  \left( ( \delta_{1,0} t_2)( \delta_{1,1} t_3)  + ( \delta_{1,0} t_3)( \delta_{1,1} t_2)  \right)    -  4  \left( ( \delta_{1,0} t_1)( \delta_{1,1} t_3)  + ( \delta_{1,0} t_3)( \delta_{1,1} t_2)  \right)   \right)
\end{array} \right) \; . \nl
}
Finally at the last order $\mathcal{O}_{3,3}$ we determine the $\epsilon^3$ correction to the gauge field,
\eqn{
\delta_{3,0} A_i & = & \frac{1}{3}  \left( \begin{array}{c} 
 \delta_{3,0} t_1 +  \delta_{3,0} t_2 - 2  \delta_{3,0} t_3  \\
- \sqrt{3} \left(  \delta_{3,0} t_1  -  \delta_{3,0} t_2 \right)
\end{array} \right) 
+
 \frac{1}{9}  \left( \begin{array}{c} 
( \delta_{1,0} t_1)( \delta_{2,0} t_1) + ( \delta_{1,0} t_2)( \delta_{2,0} t_2) - 2  ( \delta_{1,0} t_3)( \delta_{2,0} t_3)
 \\
 - \sqrt{3} \left( ( \delta_{1,0} t_1)( \delta_{2,0} t_1) - ( \delta_{1,0} t_2)( \delta_{2,0} t_2)      \right)
\end{array} \right) - \partial_i (\delta_{3,0} \Phi) 
 \nl
&&  + 
 \frac{1}{9}  \left( \begin{array}{c} 
 4  \left( ( \delta_{1,0} t_1)( \delta_{2,0} t_3)  + ( \delta_{1,0} t_3)( \delta_{2,0} t_1)  \right) +  4  \left( ( \delta_{1,0} t_2)( \delta_{2,0} t_3)  + ( \delta_{1,0} t_3)( \delta_{2,0} t_2)  \right) - 8  \left( ( \delta_{1,0} t_1)( \delta_{2,0} t_2)  + ( \delta_{1,0} t_2)( \delta_{2,0} t_1)  \right) 
 \\
 - \sqrt{3} \left(  4  \left( ( \delta_{1,0} t_2)( \delta_{2,0} t_3)  + ( \delta_{1,0} t_3)( \delta_{2,0} t_2)  \right)    -  4  \left( ( \delta_{1,0} t_1)( \delta_{2,0} t_3)  + ( \delta_{1,0} t_3)( \delta_{2,0} t_2)  \right)   \right) 
 \end{array} \right) \nl
&&  +
 \frac{1}{27}  \left( \begin{array}{c} 
( \delta_{1,0} t_1)^3 + ( \delta_{1,0} t_2)^3 - 2  (\delta_{1,0} t_3 )^3 \\
- \sqrt{3} \left(  (\delta_{1,0} t_1)^3  -  ( \delta_{3,0} t_2)^3  \right)
\end{array} \right)  
\nl
&&  +
 \frac{1}{27}  \left( \begin{array}{c} 
12   \left( ( \delta_{1,0} t_1)^2 ( \delta_{1,0} t_2)  + ( \delta_{1,0} t_1)( \delta_{1,0} t_2)^2  \right) 
-18   \left( ( \delta_{1,0} t_1)^2 +( \delta_{1,0} t_2)^2  \right)   \delta_{1,0} t_3
+ 6 \left( \delta_{1,0} t_1 + \delta_{1,0} t_2  \right)  ( \delta_{1,0} t_3)^2
 \\
- \sqrt{3} \left(  
-8   \left( ( \delta_{1,0} t_1)^2 ( \delta_{1,0} t_2)  - ( \delta_{1,0} t_1)( \delta_{1,0} t_2)^2  \right) 
+2   \left( ( \delta_{1,0} t_1)^2 -( \delta_{1,0} t_2)^2  \right)   \delta_{1,0} t_3
+ 10 \left( \delta_{1,0} t_1 - \delta_{1,0} t_2  \right)  ( \delta_{1,0} t_3)^2
  \right)
\end{array} \right)  \; .
\nl
}
As described in the summary Section~\ref{sec:summary}, we may give compact and elegant expressions for the electrometric and gauge field that encompass all the subleading corrections detailed above. We begin by defining the quantity, 
\eqn{
\Delta^2 = ( \sum_n  t_n^2 )^2 - 2 ( \sum_m t_m^4 )
}
and then to this order we find that the determinant of the metric is given by the expansion of the expression,
\eqn{
\det( g_{ij} ) = \frac{3}{\Delta^2} + O(\epsilon^3,  \epsilon^2 a,  \epsilon a^2 )
}
As mentioned previously, the rescaling function is given by the quarter power of this determinant of the metric,
\eqn{
f = ( \det{g_{ij}} )^{1/4} + O(\epsilon^3,  \epsilon^2 a,  \epsilon a^2 )
}
and the electrometric is given by the remarkably simple expression,
\eqn{
\label{eq:electrometric}
g_{ij} =  \frac{3}{\Delta^2} \sum_n ( \delta_{ij} - \frac{4}{3} \ell_n^i \ell_n^j ) t_n^2 + O(\epsilon^3,  \epsilon^2 a,  \epsilon a^2 )
}
to this order, by which we mean that if we write the metric as, $g_{ij} = \delta_{ij} + \epsilon \left( \delta_{1,0}g_{ij} + a \delta_{1,1}g_{ij}  + O(a^2)  \right) + \epsilon^2 \left( \delta_{2,0}g_{ij} + O(a) \right) + O(\epsilon^3)$, then this expression correctly encodes the subleading behaviours from $\delta_{1,0}g_{ij}$, $\delta_{1,1}g_{ij}$ and $\delta_{2,0}g_{ij}$.
Remarkably this is precisely the same expression for the electrometric as in our paper~\cite{Roberts:2021vmt} which we derived when the hopping functions were fine tuned to ensure the gauge field vanished. In that case one can compute the metric fully non-perturbatively in $\epsilon$, and we obtained the above expression with no corrections. It is then a very interesting question here whether, in the presence of a large gauge field,  the form~\eqref{eq:electrometric} still holds to all orders in $\epsilon$ and $a$.
We similarly find a compact expression for the strain gauge field. Firstly we define,
\eqn{
\Delta t_n = t_n - 1
}
and then we find that, 
\eqn{
A_i &=& \frac{1}{a \Delta^2} \epsilon_{ij} \sum_m \Big[ \ell^j_m \Delta t_m \left( 2  + \sum_n \left( 3 \delta_{mn}  \Delta t_n \right) + \sum_{n,p} \left( \left( \frac{1}{3} + 2 \delta_{mn} - 3 \delta_{np} \right)  \Delta t_n  \Delta t_p  \right) \right) \nl
&& \qquad + a^2 \left( \frac{1}{4} \ell^j_m \ell^k_m \ell^l_m - \frac{3}{8} K^{jkl} + \frac{1}{6}  \delta^{jk}  \ell^l_m \right) \partial_k \partial_l \Delta t_m \Big] + O(\frac{\epsilon^4}{a},\epsilon^3 , \epsilon^2 a, \epsilon a^2)
}
elegantly encodes all the gauge field contributions detailed above, namely $\delta_{1,0} A_i$, $\delta_{1,1} A_i$, $\delta_{1,2} A_i$, $\delta_{2,0} A_i$, $\delta_{2,1} A_i$ and $\delta_{3,0} A_i$.

\section{Example: Armchair deformation and comparison to exact diagonalization}
\label{sec:armchair}

As a check of our continuum effective theory we compare its solution to the low energy spectrum of the distorted tight-binding model found by numerical diagonalization. For the moment we will work in terms of the lattice model hopping functions in the lattice coordinates $\bx = (x,y)$.
We  consider an ``Armchair'' deformation, so one that varies only in the $y$ direction and preserves the reflection symmetry in $x \to -x$. Then the hopping functions have the functional form
\eqn{
t_1(\vec{x}) = t_2(\vec{x}) = t(y) \; , \quad t_3(\vec{x}) = \tau(y) \; .
}
Already one may consider a homogeneous deformation, taking $t(y)$ and $\tau(y)$ constant and not equal to one. However here we are not so interested in the physics of homogeneous anisotropic deformations -- which cannot lead to non-trivial strain magnetic fields or electrometric curvature -- but rather that of inhomogeneity, and thus we choose the functions $t(y)$ and $\tau(y)$ to be non-constant and periodic, and slowly varying so that $t^{(n)}(y), \tau^{(n)}(y) \sim O(1)$.
We are then interested in the low lying energy spectrum as we approach the limit $a \to 0$, corresponding to periodic distortions that slowly vary over many unit cells.
In lattice coordinates, the $y$ periodicity of the $A$ or $B$ sublattices is $3a/2$, and so we require the periodicities $t(y) = t(y+3aM/2)$, $\tau(y) = \tau(y+3aM/2)$ for some integer $M$ which must be taken so that $M \gg 1$ to ensure a slow variation.

 For this case there is a different lattice basis that is more natural,
\eqn{
\bw_1= a \left( \bl_1-\bl_2 \right) = \bv_1-\bv_2  \; , \qquad\qquad \bw_2 = a \left( \bl_1-\bl_3 \right) = \bv_1\;, \nonumber \\
 \bc_1 = -\bb_2 \; , \qquad\qquad \bc_2 = \bb_1+\bb_2 \; ,  \qquad\qquad \bw_i \cdot \bc_j = 2\pi \delta_{ij} \; .
}
It is then natural to take a supercell generated by $\bw_1$ and $M \bw_2$, with $M$ $A$ and $B$ sites within the supercell. We proceed with the standard Bloch wave decomposition. We define our lattice positions as,
\eq{
{\bx_{A,B} = n_1 \bw_1 + n_2 M \bw_2 + m \bw_2 \mp \frac{a}{2} \bl_3 \; , \qquad m = 0 , \ldots M-1 \;.}
 }
 \begin{figure}[h!]
\begin{center}
\includegraphics[scale=.6]{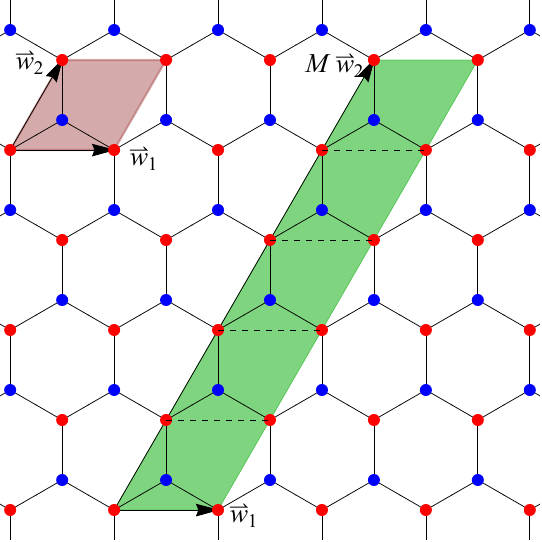} \qquad \qquad \qquad
\includegraphics[scale=.6]{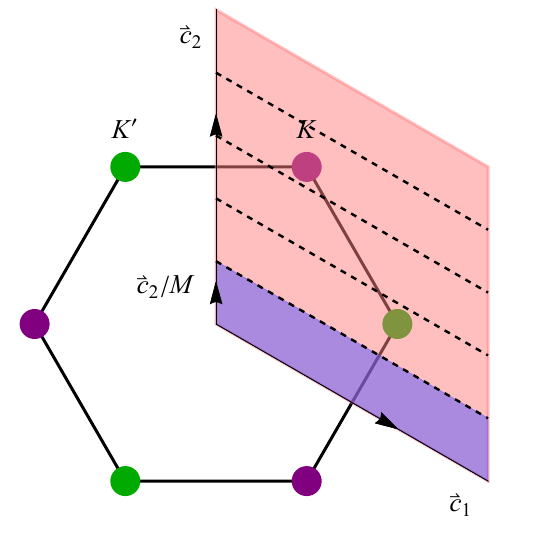}
\caption{Left: the honeycomb lattice, with red A and blue B sites, related by translations $a \vec{\ell}_{1,2,3}$. We show the standard unit cell generated by $\bw_i$ and the supercell generated by $ \bw_1, M\bw_2$ for $M=5$. Right: the Brillouin zone for a single unit cell generated by the dual basis $\bc_i$ and its coverings by the dual lattice and folded Brillouin zone of the supercell.}
\label{fig:armchair_bz}
\end{center}
\end{figure}
This leads to the usual folding of the Brillouin zone, with dual generators $(\bc_1, \tfrac{1}{M} \bc_2)$. We then have crystal momentum,
\eq{
\bk = \left( \lambda_1 \bc_1 + \tfrac{\lambda_2}{M} \bc_2\right) \; , 
}
where the periodicity of the Brillouin zone is given by $\lambda_i \sim \lambda_i+1$.
We will focus on the $K$ point, taking,
\eq{
A(\bx)= e^{i( \bK + \bk) \cdot \bx} u(\bx), \qquad
B(\bx)= e^{i( \bK + \bk)  \cdot \bx} v(\bx),
}
where $u,v$ are periodic on the supercell, so that $\bk$ gives the crystal momentum relative to the $K$  Dirac point. One could equivalently choose to look at the momentum relative to the $K'$ point.  We then find the discrete equations,
\eqn{\label{eq:latticesystem}
\hbar \omega u_m =2e^{+\frac{iak_y}{2}}\cos[\tfrac{\sqrt{3}a}{2} (K_x + k_x)] \, t\left(\tfrac{3a}{4}(2m+1)\right)v_{m+1}+ e^{-iak_y}\tau\left(\tfrac{3am}{2} \right)v_m , \nonumber \\
\hbar \omega v_m =2e^{-\frac{iak_y}{2}}\cos[\tfrac{\sqrt{3}a }{2} (K_x + k_x)] \, t\left(\tfrac{3a}{4}(2m-1)\right)u_{m-1}+ e^{+iak_y}\tau\left(\tfrac{3am}{2} \right)u_m,
}
where the $u_m,~v_m$ are periodic in $M$, ie $u_{M+1}=u_1$,\footnote{One can of course then do a discrete Fourier transform on the $u_m$ and $v_m$, however this position space formulation leads to very sparse arrays and so is preferable for numerical diagonalization.}
 and $K_x = -  \frac{1}{a} \frac{4 \pi}{3 \sqrt{3}}$.%

\subsection{Hopping functions from an armchair distortion}

We could now simply choose some functional form for the hopping functions $t(y)$ and $T(y)$, numerically diagonalize and compare to our effective theory. However in order to illustrate our earlier discussion relating to  deriving hopping functions from an embedding, we now consider a specific in-plane displacement field which, following our previous discussion in Section~\ref{sec:embedding}, gives a particular instance of the armchair hopping functions. 
In terms of the earlier equation~\eqref{eq:straindef} we consider the following in-plane displacement ,\footnote{One can consider shifting this distortion by an arbitrary phase, e.g.~$\sin(\cK y + \varphi a)$, but this will only give corrections subleading in $a$ and not affect the leading behaviour.}
\eqn{\label{eq:displacement_armchair_KMrel}
v^x (\vec{x})= 0 \; , \quad v^y (\vec{x}) = - \frac{1}{\cK} \sin(\cK y) \; , \qquad \cK = \frac{4\pi}{3aM} \;.
}
Then the lattice and lab frame coordinates $x$ and $X$ are trivially related as $x = X$, and the $y$ and $Y$ coordinates are related (exactly) as $Y = y - \frac{\epsilon}{\cK} \sin{\cK y}$, or inverting, $y = Y + \frac{\epsilon}{\cK} \sin( \cK Y) + \frac{\epsilon^2}{2 \cK} \sin(2 \cK Y) + O( \epsilon^3 ) $.
We assume an exactly exponential bond model~\eqref{eq:expbondmodel} depending only on the bond length~\eqref{eq:bondlen} and this in-plane displacement field leads to hopping functions given in lattice coordinates as,
\eqn{
\label{eq:armchairEmbed}
t(y + \frac{a}{4}) = e^{- \beta \left( \frac{1}{2} \int_0^1 d\lambda \sqrt{ 3 + (1 - \epsilon \cos( \cK ( y + \frac{a \lambda}{2} ) ))^2}   - 1 \right) }  \; , \quad
\tau(y - \frac{a}{2}) =  e^{- \beta \left(  \int_0^1 d\lambda (1 - \epsilon \cos( \cK (y - a \lambda))) - 1 \right)}  \; .
}
The integral in the second expression is simple to compute, and in fact the one in the first can be found in terms of elliptic functions.
Now to make contact with our continuum description we expand these in $\epsilon$ and $a$ which yields,
\eqn{
\label{eq:armchairPert}
t(y) &=& 1 +  \frac{ \epsilon \beta}{4} \left(  \cos(\cK y)  - \frac{a^2 \cK^2}{96}  \cos(\cK y) + O(a^3) \right) + \frac{ \epsilon^2 \beta ( \beta -3)}{64}  \left(  1 + \cos(2 \cK y) + O(a^2) \right)  + O(\epsilon^3) 
\nl
\tau(y) &=& 1 + \epsilon \beta \left(  \cos(\cK y) - \frac{a^2 \cK^2}{24}  \cos(\cK y) + O(a^3) \right) + \epsilon^2  \frac{\beta^2}{4} \left( 1 +  \cos(2 \cK y)  + O(a^2) \right)  + O(\epsilon^3) \; .
}
In the comparison we detail shortly, we will take the phenomenological value $\beta = 3$.
While we have included the corrections in $a$ to these hopping functions, we emphasize that these corrections are dependent on the bond model, as we have discussed earlier.
This embedding yields the following gauge field and electrometric for the effective theory in lattice coordinates, 
\eqn{
A^i &=& \left( - \frac{\epsilon\beta}{2 a} \left( \cos(\cK y) + \frac{a^2 \cK^2}{12} \cos(\cK y) + O(a^3) \right) - \frac{\epsilon^2 \beta (1 + 4 \beta)}{16 a} \left( \cos^2(\cK y) + O(a^2) \right) + O(\frac{\epsilon^3}{a}) , 0 \right)  \nl
g_{ij} & = & \left( 
\begin{array}{cc}
1 + \frac{\epsilon^2 \beta ( 1 + 2 \beta)}{4} \cos^2(\cK y) & 0  \\
0 & 1 - 2 \epsilon \beta \cos(\cK y) + 2 \epsilon^2 \beta^2  \cos^2(\cK y)
\end{array}
\right)   + O(\epsilon^3 , \epsilon^2 a, \epsilon a^2)
}
and corresponds to the following magnetic field and electrometric Ricci scalar curvature, again in lattice coordinates,
\eqn{
B(y) & = & \frac{\epsilon \cK \beta}{2a} \left( \sin(\cK y) + \frac{a^2 \cK^2}{12} \sin(\cK y) + O(a^3) \right) + \frac{\epsilon^2 \cK \beta (1 + 4 \beta)}{16 a} \left( \sin(2 \cK y)+ O(a^2) \right)  + O(\frac{\epsilon^3}{a} ) \nl
R(y) & = & \frac{\epsilon^2 \cK^2 \beta ( 1 + 2\beta )}{2} \cos(2 \cK y) + O(\epsilon^3 , \epsilon^2 a , \epsilon a^2) \; .
}
Transforming the gauge field and electrometric to the lab frame gives,
\eqn{
A^I_{lab} &=& \left( - \frac{\epsilon\beta}{2 a}  \left( \cos(\cK y) + \frac{a^2 \cK^2}{12} \cos(\cK y) \right) + \frac{\epsilon^2 \beta \left( (7 -  4 \beta) -  (9 + 4 \beta) \cos(2 \cK Y) \right) }{32 a}  , 0 \right) + O(\frac{\epsilon^3}{a} , \epsilon^2 a , \epsilon a^2) \nl
g_{IJ}^{lab} & = & \left( 
\begin{array}{cc}
1 + \frac{\epsilon^2 \beta ( 1 + 2 \beta )}{4} \cos^2(\cK Y) & 0  \\
0 & 1 + 2 \epsilon (\beta - 1) \cos(\cK Y) + \frac{\epsilon^2}{2} \left(  (1 - 2 + 2 \beta^2 ) + (5 - 6 \beta + 2 \beta^2 ) \cos(2 \cK Y) \right)
\end{array}
\right)   + O(\epsilon^3 , \epsilon^2 a, \epsilon a^2) \nl
}
which we note can be derived as a special case of the expressions given in \eqref{eq:lab_frame_a_g}. Then the strain magnetic field and Ricci curvature in lab coordinates are,
\eqn{
B^{lab}(Y) & = & \frac{\epsilon \cK \beta}{2a} \left( \sin(\cK Y) + \frac{a^2 \cK^2}{12} \sin(\cK Y) +O(a^3) \right) + \frac{\epsilon^2 \cK \beta (9 + 4 \beta)}{16a} \left( \sin(2 \cK Y) +O(a^2) \right) + O(\frac{\epsilon^3}{a} ) \nl
R^{lab}(Y) & = & \frac{\epsilon^2 \cK^2 \beta ( 1 + 2\beta )}{2} \cos(2 \cK Y)  + O(\epsilon^3 , \epsilon^2 a, \epsilon a^2) \; .
}
We again emphasize that while the deformation of the lattice is purely in-plane, and so its embedding remains flat, the electrometric that governs the effective  theory nonetheless  becomes curved at quadratic order $O(\epsilon^2)$ in the deformation. Furthermore, this curvature is sensitive to the bond model taken.

Before we turn to the numerical comparison we briefly take an aside to discuss another simple deformation -- the zig-zag, where again we have $t_1 = t_2 = t$ and $t_3 = \tau$ but these are now functions of $x$ rather than $y$. The zig-zag deformation can also be generated by an in-plane displacement by taking, $v^x = v^x(x)$ and $v^y = 0$. However the crucial difference with the armchair case is that for this deformation the gauge field is pure gauge, and the electrometric is simply a coordinate transform of flat space, although interestingly not the diffeomorphism generated by the vector $v^i = (v^x, v^y)$, but nonetheless the geometry that the effective description sees is simply flat. Thus there is not the interesting physics that arises in the armchair case, and hence we do not discuss the zig-zag case further.

\subsection{Comparison to effective theory}
 
 We wish to compare numerical diagonalization with our continuum effective theory. In order to do so we focus on certain low energy states in the lattice theory and follow them as the above  armchair strain perturbation is applied, and compare to an analytic solution in the effective theory. For simplicity we restrict to energy eigenstates with definite momentum $k_x$ in the $x$ direction and zero momentum in the $y$ direction. Clearly this is simpler than including also $k_y$ momentum. However also since the lattice is undistorted in the $x$ direction, and the distortion only depends on $y$, the lattice and lab frame $X$ coordinate are the same $x = X$, and hence the lattice momentum $k_x$ is the same as that measured in the $X$ direction in lab frame. Thus $k_x$ is easy to physically interpret. 
  
For the undeformed continuum Dirac theory to leading order in $a$ the low energy eigenstates forming the Dirac cone at the $K$ point, with $k_y = 0$, are,  
\eqn{
\label{eq:leadingwavefn}
\Psi(t,\vec{x}) = e^{-i \omega t} e^{+ i k_x x} \left( \begin{array}{c}
1 \\
\pm i
\end{array} \right) \; , \quad \frac{\hbar\omega}{T} = \pm \frac{3  }{2} a k_x
}
which solves the continuum leading order Dirac equation. In this undeformed case there are then subleading corrections in $a$ due to the higher derivative terms, which up to and including the three derivative term, go as,
\eqn{
\frac{\hbar\omega}{T}  = \pm \left(  \frac{3 }{2} a  k_x  + \frac{3}{8} (a k_x)^2 - \frac{3}{16} (a  k_x )^3 + O(a^3) \right) \; .
}
Now consider turning on the specific armchair strain deformation~\eqref{eq:armchairPert} above.
We write an ansatz for the deformed eigenstate wavefunction as,
\eqn{
\label{eq:wavefnansatz}
\Psi(t,\vec{x}) = e^{-i \omega t}  e^{+ i k_x x} \left( \begin{array}{c}
\psi_1(y) \\
\psi_2(y)
\end{array} \right) 
}
where $\psi_{1,2}$ are periodic functions,
and consider the flow of the solution~\eqref{eq:leadingwavefn} above, taking, 
\eqn{
\omega &=& \sum_{m = 0}  \sum_{n=0} \epsilon^m a^n \delta_{m,n}\omega \\
\psi_{1,2} & = &  \sum_{m = 0}  \sum_{n=-m} \epsilon^m a^n \delta_{m,n}\psi_{1,2} \; .
}
We may then straightforwardly solve the continuum theory for all $\mathcal{O}_{p,q}$ for $p \le 3$, finding that the energy up to order $O(\epsilon^2)$ goes as, 
\eqn{
&&\pm \frac{\hbar\omega}{T}  =  \frac{3 }{2}  k_x + \frac{3}{8} a k_x^2 - \frac{3}{16} a^2  k_x ^3 + O(a^3) \nl
&& \quad + \frac{ \epsilon^2}{\cK^2} \left( 
-  \frac{ \frac{3 \beta^2}{8} k_x}{ a }  
 +  \left(  \frac{3 \beta}{64}  \cK^2  -  \frac{ 33 \beta^2 }{32} k_x^2 \right)
+  \left( - \tfrac{ \beta (9 + 16 \beta) }{128} k_x \cK^2  - \tfrac{ 123 \beta^2 }{64} (k_x)^3 \right) a + O(a^2)
\right) + O(\epsilon^3) \; .
}
We note that this correction to the energy due to this armchair deformation is quadratic in deformation parameter $\epsilon$, since at leading $\epsilon$ order the hopping functions are purely harmonic, with no constant component -- a constant shift in the hopping functions  would have shifted the energy at $O(\epsilon)$.
This is an explicit manifestation of the fact that, due to translation symmetry, the dispersion relation only responds to inhomogeneity at $O(\epsilon^2)$.
We find that the leading correction on the righthand side going as $\sim \frac{ \epsilon^2 }{a}$ is solely due to the leading gauge field, $\delta_{1,0} A_i$ and is proportional to $k_x$. The subleading correction, $\sim   \epsilon^2 a^0$  is composed from three contributions, including from the frame being non-trivial,
\eqn{
\begin{array}{ccc}
 \underbrace{    \frac{3  \beta ( 1 + 4 \beta )}{64 }  \epsilon^2
  }
   \; , \quad& 
\underbrace{  - \frac{ 3 \beta^2 }{8}   \epsilon^2
}
   \; , \quad& 
\underbrace{   \left(   \frac{3}{16} - \frac{33 k_x^2}{32 \cK^2}    \right) \beta^2 \epsilon^2
} \nl
 &&\nl
 \delta_{1,0} A_i &  \mathrm{frame} &  \mathrm{two\;derivative}
 \end{array} \; .
 }
We may separate all the contributions from the leading gauge field, two derivative term, three derivative term and frame corrections by defining, 
\eqn{
\tilde{\lambda} = \frac{\beta \lambda}{ \cK} = \frac{\epsilon \beta }{a \cK} \; , \quad \tilde{a} = a \cK
} 
and then writing $\omega$ as, 
\eqn{
\label{eq:analyticapprox}
&& \pm \frac{\hbar\omega}{T}  = \tilde{\lambda}^2  c_0(\tilde{a}) + \left( \frac{3 }{2} +   \tilde{\lambda}^2  c_1(\tilde{a}) \right)  (a k_x) +  \left(  \frac{3 }{8} +   \tilde{\lambda}^2  c_2(\tilde{a}) \right) (a k_x)^2 +  \left( - \frac{3 }{16} +  \tilde{\lambda}^2  c_3(\tilde{a}) \right) (a k_x)^3 + O(\epsilon^3)
}
and then we find the various terms give contributions as,
\eqn{
\label{eq:contributions}
\begin{array}{ccccccc}
c_0(\tilde{a}) = &  \frac{3 (1 + 4 \beta) }{64 \beta} \tilde{a}^2 & - \frac{3 }{8 } \tilde{a}^2 & + \frac{3 }{16 } \tilde{a}^2 &  & &+ O(\tilde{a}^3) \\
c_1(\tilde{a}) = & - \frac{3}{8} -  \frac{1}{16}  \tilde{a}^2  & - \frac{3 (1 + 2 \beta) }{32 \beta} \tilde{a}^2 & + \frac{3 (1 + 32 \beta)}{128\beta} \tilde{a}^2 & - \frac{1}{16} \tilde{a}^2 & - \frac{9}{16} \tilde{a}^2 & + O(\tilde{a}^3) \\
c_2(\tilde{a}) = & & & - \frac{33}{32 } & & &+ O(\tilde{a}) \\
c_3(\tilde{a}) = & & & - \frac{33}{16} & + \frac{9}{64 }& &+ O(\tilde{a})\\
& \underbrace{\hspace{2cm}} &  \underbrace{\hspace{2cm}} & \underbrace{\hspace{2cm}} &  \underbrace{\hspace{2cm}} &  \underbrace{\hspace{2cm}} \\
&  \delta_{1,0} A_i &  \mathrm{frame} &  \mathrm{two\;derivative} &  \mathrm{three\;derivative} &  \mathrm{mixed}
 \end{array}
 }
 where we have separated the contributions from different terms in the expansion, and in particular distinguish the leading part of the strain gauge field from the leading correction to the frame, as well as contributions from the higher derivative terms.
 More precisely we distinguish these contributions by writing the effective theory as,
 \eqn{
0 =  a e^\mu_{~A} \gamma^A D_\mu \Psi \pm i q_2 a^2 \, \eta_{AB} \gamma^A e^B_{~\sigma} D_\mu \left( C^{\sigma\mu\nu} D_\nu \Psi \right) 
+ q_3 a^3 \,\eta_{AB} \gamma^A e^B_{~\sigma} D^{\sigma\mu\nu\rho}  D_\mu D_\nu D_\rho \Psi 
  + O(\epsilon^4, \epsilon^3 a, \epsilon^2 a^2, \epsilon a^3, a^4)  \nn \\
 }
 where we have artificially added coefficients $q_{2,3}$ to the two and three covariant derivative terms, and then we modify the frame and gauge field expansions as,
 \eqn{
 A_i = \frac{\epsilon q_{gauge}}{a} \delta_{1,0}A_i + q_{subleading} \left(  \mathrm{remaining}\;  \delta_{n,m} \;\mathrm{A_i\;contributions} \right) \; , \quad e^i_{~I} = \delta^i_I + q_{frame} \sum_{n=1}^\infty \epsilon^n \delta_n e^i_{~I}
 }
 so that for $q_{2,3} = q_{gauge} = q_{subleading} = q_{frame} = 1$ then we recover our full effective theory. We solve this modified effective theory with the hopping function deformation above, computing the dispersion relation $\omega(a k_x)$, and computing the $c_{0,1,2,3}(\tilde{a})$ which now depend on these constants. Setting $q_{subleading} = q_{frame} = q_{2,3} = 0$ gives the leading consistent effective theory approximation, that from truncating only to one covariant derivative, i.e. the flat Dirac equation with $O(\epsilon/a)$ magnetic strain field, denoted `$\delta_{1,0} A_i$' in the table above. Setting $q_{subleading} = q_{gauge} = q_{2,3} = 0$ yields the contribution from curved frame corrections, denoted `frame' above. Keeping only $q_{2}$ or $q_3$ non-zero isolates the two or three derivative terms shown above. Finally the `mixed' corrections are ones that involves more than one source. For example, for $c_1$ the mixed contribution comes from frame corrections originating from the two derivative term, and so is proportional to both $q_2$ and $q_{frame}$.

Quite interestingly $c_2(\tilde{a})$ receives no correction from the frame -- in fact this is due to the specific choice of leading hopping deformation; more generally for $t = 1 + \epsilon \alpha \cos(\cK y) + \ldots$ and $\tau = 1 + \epsilon \beta \cos(\cK y) + \ldots$ one finds,
\eqn{
c_2(\tilde{a}) &=& \underbrace{ - \frac{11(\beta - \alpha)^2 }{6 \beta^2} } + \underbrace{  \frac{2 ( \beta -\alpha) ( \beta - 4 \alpha  ) }{3 \beta^2} } 
\nl
 &&\nl
&& { \mathrm{two\;derivative}} \qquad { \mathrm{frame}}
} 
but for our in-plane strain, we have $\alpha = \beta/4$ so the frame contribution vanishes. Precisely the same is true also for $c_3$ where again more generally we would obtain a frame correction and a mixed correction along with the two and three derivative contributions, but precisely for the in-plain embedding case where $\alpha = \beta/4$ the frame correction vanishes.

The $k_x^0$ terms shift the energy and result in the Dirac point $K$ moving from $k_x = 0$ to,
\eqn{
 k_x = \epsilon^2 K_D = - \frac{2 \tilde{\lambda}^2}{ 3 a}   c_0(\tilde{a})  =  - \frac{ \epsilon^2 \beta}{32  a} + O(\frac{\epsilon^3}{a}, \epsilon^2 a) 
}
and receive contributions from the frame and the two derivative term, as well as from the gauge field $\delta_{1,0} A_i$.
Shifting to the deformed $K$ Dirac point by taking $k_x =  \epsilon^2 K_D + \tilde{k}_x$, and then considering the linear terms in $\tilde{k}_x$  yields an  effective speed of light in the $x$-direction at order $O(\epsilon^2)$ going as,
\eqn{
c_{eff, x} = \frac{3 a T}{2 \hbar} \left( 1 - \frac{\epsilon^2 \beta^2}{4 a^2 \cK^2} - \frac{ \epsilon^2  \beta (3 + {4} \beta )}{48}  + O(\frac{\epsilon^3}{a}, \epsilon^2 a) \right) \; .
}
where the first subleading term is due only to the gauge field correction, $ \delta_{1,0} A_i$, and the second subsubleading term has contributions from the frame and higher derivative terms.
Thus we see very explicitly that if we are interested only in the Dirac cone itself, then the leading correction is simply due to the flat space Dirac theory coupling to the large gauge field. The frame correction is subleading to this and comes along at the same order as the higher derivative terms. One the other hand, if we are interested in the band structure more generally, not simply the very low energy Dirac cone behaviour going as $\tilde{k}_x$, but also $\tilde{k}_x^2$ and higher powers, then the gauge field is just one contributing correction, and is not the leading one.
In particular as we see above, it contributes to the shift of the Dirac point at the \emph{same} order as everything else.

\subsection{Numerical diagonalization}

We now compare our low energy effective theory to a direct computation of $\omega$. We take the full hopping functions in~\eqref{eq:armchairEmbed} with the phenomenological value $\beta = 3$, and compute $\omega$ by numerically diagonalizing the tight-binding lattice model (we will term this $\omega^{num}$) in equation~\eqref{eq:latticesystem}. We extract the appropriate lowest (absolute value) eigenvalues as a function of $k_x$ (with $k_y = 0$) and the deformation amplitude $\epsilon$ and period given by $K$. From this we compute the derivatives, 
\eqn{
c^{num}_{n}(\tilde{a}) = \frac{1}{n!} \frac{\hbar}{ T \tilde{\lambda}^2  } \left. \frac{\partial^n \left( \omega^{num}(a k_x) - \omega^0(a k_x) \right)}{a^n \partial k_x^n} \right|_{k_x = 0, \tilde{\lambda}} \; , \quad \frac{\hbar\omega^0(a k_x)}{T}  =  \frac{3 }{2}  ak_x + \frac{3}{8} (a k_x)^2 - \frac{3}{16} (ak_x)^3 + \ldots
}
where $\omega^0(a k_x)$ is the dispersion relation for the undeformed theory (in the $k_x$ direction).
We then obtain the analog expression to the analytic approximation~\eqref{eq:analyticapprox} above,
\eqn{
&& \frac{\hbar\, \omega^{num}( a k_x) }{T}  =  \tilde{\lambda}^2  c^{num}_0(\tilde{a}) + \left( \frac{3 }{2} +    \tilde{\lambda}^2  c^{num}_1(\tilde{a}) \right)  (a k_x) +  \left( \frac{3 }{8} +    \tilde{\lambda}^2  c^{num}_2(\tilde{a}) \right) (a k_x)^2 +  \left( - \frac{3 }{16} +   \tilde{\lambda}^2  c^{num}_3(\tilde{a}) \right) (a k_x)^3 + \ldots \nl
}
as an expansion about $k_x = 0$ whose coefficients must be computed numerically. We may then compare our approximation at order $O(\tilde{\lambda}^2)$ above to the full numerical diagonalization at each order in powers of $k_x$ expanded about $k_x = 0$ by simply comparing the numerical $c^{num}_n(\tilde{a})$ to the analytic approximations $c_n(\tilde{a})$ truncated up to the powers of $\tilde{a}$ indicated above.

For this numerical diagonalization it is convenient to choose units so that the lattice scale $a = 1$, and $T = 1$, and vary the periodicity $M$ (and hence $\tilde{a} = a \cK = \frac{4 \pi}{3 M}$), and the dimensionless deformation $\epsilon$ (and hence coupling $\tilde{\lambda}$). We numerically compute the derivatives above using standard finite differencing stencils. For example, for the first derivative we approximate,
\eqn{
\frac{1}{a} \left.  \frac{\partial \omega^{num}(a k_x)}{\partial k_x}\right|_{k_x = 0}  = \frac{\omega^{num}(\delta) - \omega^{num}(-\delta)}{2 \delta}
}
taking a small $\delta = 10^{-10}$. In order to ensure accurate results we find we must use higher precision arithmetic in determining the matrix eigenvalues, and hence $\omega^{num}(a k_x)$. 

\begin{figure}
 \centerline{
  \includegraphics[width=9cm]{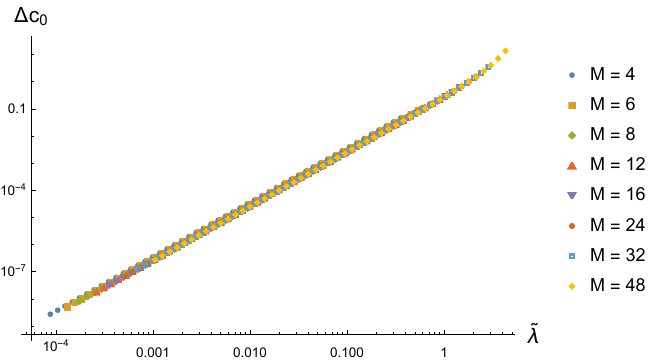}    \hspace{0.25cm}  \includegraphics[width=9cm]{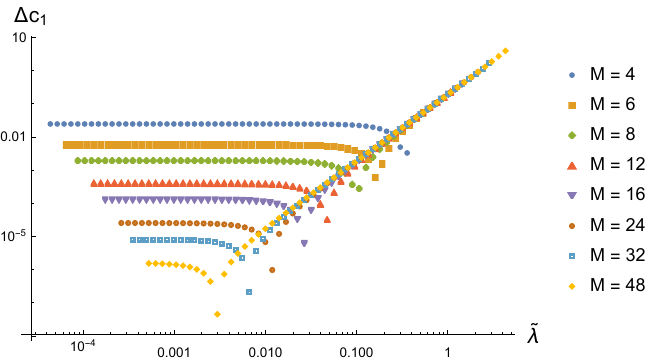}
  }
  \vspace{0.25cm}
  \centerline{
  \includegraphics[width=9cm]{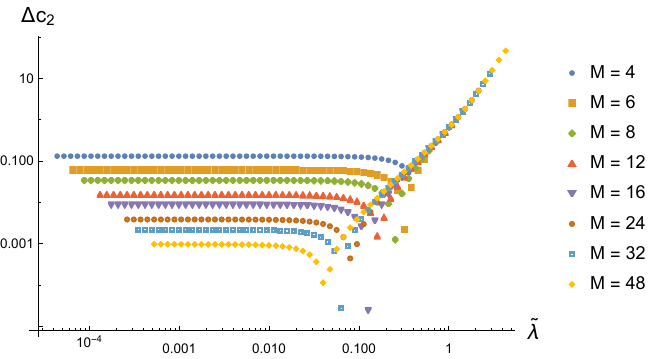}   \hspace{0.25cm}   \includegraphics[width=9cm]{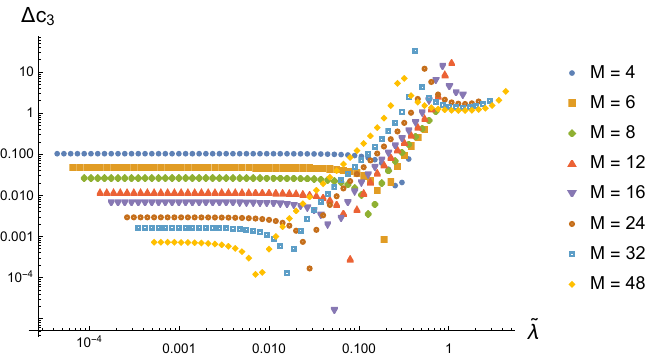}
  }
  \caption{ \label{fig:compare1}
Plot showing the quantities $\Delta c_{0,1,2,3}$ against our deformation parameters, $\tilde{\lambda}$ and $M = 4\pi/(3 \tilde{a})$. This compares the numerical diagonalization of the tight binding model to the approximation from the low energy effective theory. We clearly see that as we decrease the perturbation parameters, $\tilde{\lambda}$ and $\tilde{a}$ (so increase $M$) these quantities $\Delta c_{0,1,2,3}$ decrease indicating our effective theory is better approximating the full tight binding model. The markers show the actual numerically computed data points. 
We note that the sudden `dips' in some of these plots are due to  the derivatives of $\omega^{num}$ and the undeformed $\omega^0$ coinciding for specific values of $\tilde{\lambda}$, leading to an accidentally zero of the fractional differences $\Delta c_{0,1,2,3}$.
}
\end{figure}

In figure~\ref{fig:compare1} we show the results of comparing our approximations for $c_{0,1,2,3}$ to the numerical results $c^{num}_{0,1,2,3}$ for various values of $M$ (and hence $\tilde{a}$), and as a function of $\tilde{\lambda}$. We plot the fractional error between the approximation and the  numerical result,
\eqn{
\Delta c_n = \left| \frac{c_n - c^{num}_n}{c^{num}_n} \right| \; .
}
We expect that this fractional error goes to zero as we take both $\tilde{\lambda}$ and $\tilde{a}$ to zero, and indeed this is what we see. We see in  figure \ref{fig:compare1} that for fixed $M$, and hence fixed $\tilde{a}$, that for $c_{1,2,3}$ the error in the approximation initially decreases as a small $\tilde{\lambda}$ is further decreased. However for sufficiently small $\tilde{\lambda}$ this error saturates due to the error from higher terms in $\tilde{a}$ beyond our approximation. However, we see that the saturation value is consistent with correctly scaling to zero as $M$ is increased (or equivalently $\tilde{a}$ is decreased).  The only difference is $c_0$ which doesn't appear to receive corrections in $\tilde{a}$ (at least at order $O(\tilde{\lambda}^2)$) and so the error simply scales to zero with decreasing $\tilde{\lambda}$, independent of $M$ (or equivalently $\tilde{a}$).
We emphasize again that contributions from the gauge field, frame corrections, the two covariant derivative term, are all required to obtain this approximation for $c_0$, Leaving out any one of them destroys the convergence of the approximation seen above. Likewise the two and three derivative terms are crucial for obtaining the correct behaviour for $c_{2}$ and $c_{3}$. 

\begin{figure}
 \centerline{
  \includegraphics[width=14cm]{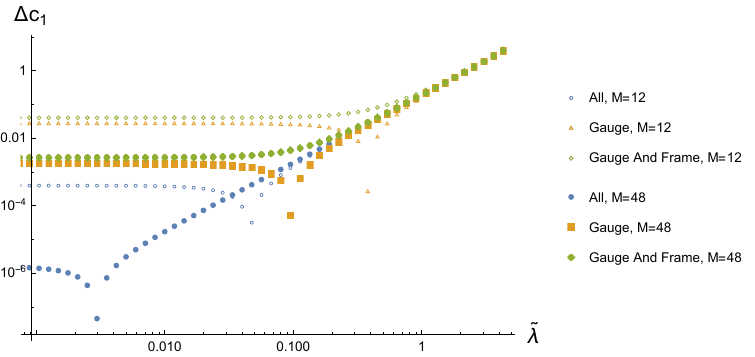}  
  }
  \caption{ \label{fig:compare2}
  This plot illustrates many of the ideas in this paper. It shows the quantity $\Delta c_1$ which compares the numerical computation of the first derivative in $k_x$ of the band structure near the Dirac point to that predicted by our effective theory approximation. The orange data points are computed using only the leading effective description, the Dirac equation in flat space with strain gauge field. The blue data points are for our full approximation. The green ones are an approximation which ignores the two and three covariant derivative terms, so the (inconsistent) truncation to the curved space Dirac equation and strain field. We show data for two deformation periodicities, $M = 12$ (open markers) and $M = 48$ (closed markers). Firstly we see the leading flat space Dirac plus gauge field gives a good approximation that improves as we take $\tilde{\lambda}$ and $\tilde{a}$ to zero (or increase $M$). Secondly, taking the curved space Dirac equation plus gauge field, but ignoring higher derivative terms doesn't improve the approximation -- in fact here it makes it worse. Thirdly, the full approximation is superior to the leading one, since it correctly and consistently includes the subleading corrections -- those from the curved frame, but also those from higher covariant derivative terms, whose contributions we see are equally important.
}
\end{figure}

The case of $c_1$ is interesting as we see from~\eqref{eq:contributions} that the gauge field gives the leading contribution, with $c_1 \simeq -3/8$, and then the frame, two and three derivative terms contribute the subleading $\sim O(\tilde{a}^2)$ behaviour. We may ask how good an approximation this leading gauge field alone is.  In figure~\ref{fig:compare2} we plot $\Delta c_1$ for two values of the periodicity $M$, taking $M = 12$ and $48$ but only taking the leading correction in $c_1$ from the gauge field (the curves labelled `Gauge'). This would correspond to artificially setting to zero the coefficients of the higher covariant derivative terms, and also artificially forcing the frame to be undeformed, before solving the effective theory for $\omega(a k_x)$. We also show $\Delta c_1$ for our full approximation, including the frame and two and three covariant derivative terms (the curves labelled `All'). Finally we plot $\Delta c_1$ for an approximation where the two and three covariant derivative contributions are ignored, so we only take the first two columns of~\eqref{eq:contributions} (the curves labelled `Gauge and Frame').
We again see the fractional error decrease as we scale $\lambda$ and $\tilde{a}$ to zero using only the leading gauge field approximation -- this is precisely the statement that the leading effective description is the flat Dirac equation plus magnetic strain gauge field. However, importantly we see no improvement in the level of approximation if we include the frame corrections, but ignore the higher covariant derivative terms. On the other hand, including these terms too, we indeed see a much better approximation, as we are correctly accounting for the subleading behaviour. This clearly illustrates our main assertions, firstly that the corrections from the curved frame are subleading to those of the gauge field, and further that they come at the same order as corrections from the higher
covariant derivative terms, and thus if you wish to consider these subleading corrections you must include them all. 

\begin{figure}
 \centerline{
  \includegraphics[width=10cm]{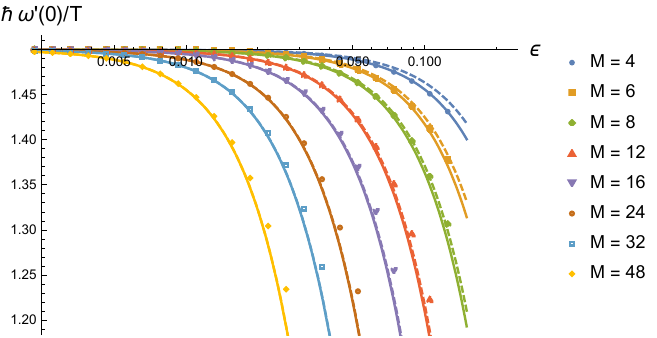} 
  }
  \caption{ \label{fig:compare3}
This plot shows the first derivative of the dispersion relation, $\frac{\hbar}{T} {\omega^{num}}'(0)$,  near the Dirac point (recall the Dirac point is slightly shifted to non-zero $k_x$), against the deformation amplitude $\epsilon$, as computed numerically from the tight-binding model (data points), and for various deformation periodicities $M$. We also plot the effective theory approximations -- both the leading one from the flat space Dirac equation with magnetic gauge field (dashed curves), and our full one consistently including spatial curvature and higher two and three covariant derivative terms. We see for this first derivative the leading flat space Dirac description gives a good approximation, even when the theory is deformed so that this quantity changes considerably -- for example the curves do well even when the value has changed by $\sim 20\%$. The full approximation is better for smaller $M$, as we would expect, but both work well.
}
\end{figure}

Finally it is interesting to consider how good the full approximation including gauge fields, curved frame and two and three covariant derivative terms is, compared to simply truncating to the flat space Dirac theory and leading gauge field contribution. The answer is that it depends on what quantity one computes. We see above that the leading correction to  $c_1$, and hence to the wavespeed, is from the gauge field. In figure~\ref{fig:compare3} we show this by plotting $\frac{\hbar}{T} {\omega^{num}}'(a k_x)$ at $k_x = 0$ as a function of the deformation $\epsilon$ and the periodicity of the deformation, $M$, computed numerically and shown as data points. We compare these values from numerical diagonalization to our full analytic approximation, drawn as solid curves, and also to the  approximation coming only from the leading gauge contribution, drawn as dashed curves. They are very close by eye, although interestingly for small $M$, where subleading $\tilde{a}$ corrections are relatively more important, one can see by eye the improvement in taking the full approximation.
 Perhaps the most interesting point to note is that for the quite large deformations of the theory shown, where we see a $\sim 20\%$ change in this quantity compared to the undeformed theory,  these analytic approximations still give a reasonable result.

\begin{figure}
 \centerline{
  \includegraphics[width=10cm]{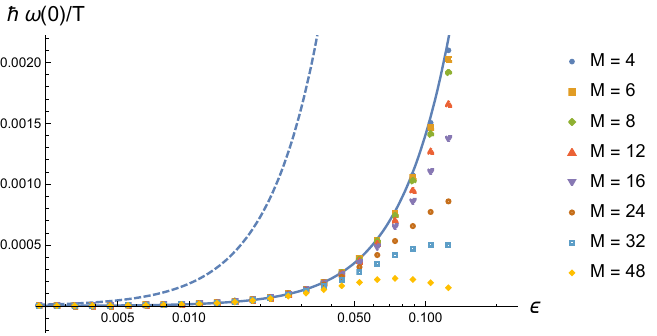} 
  }
  \caption{ \label{fig:compare4}
This is a similar plot to the previous one, but now looking at the value of $\frac{\hbar}{T} \omega^{num}(0)$ which determines the leading shift in the Dirac point. For this quantity there is no leading gauge field contribution, and instead the gauge field, curved frame and higher derivative corrections all contribute to the leading behaviour. The dashed curve shows the contribution ignoring the higher derivative terms (so a truncation to the curved space Dirac equation plus gauge field) and we see there is no agreement with the numerical results. Including these higher covariant derivative terms yields the full approximation, the solid curve, which well approximates the numerical results when $\epsilon$ becomes small (for larger $M$ then $\epsilon$ must be smaller, since the real perturbation parameter is $\tilde{\lambda} \sim \epsilon/\tilde{a}$).
}
\end{figure}

 If we look at the shift in the Dirac point, which to leading order in $\tilde{\lambda}$ is determined simply by the value of $\omega^{num}(0)$, then the gauge field, frame and two derivative terms all contribute at leading order as we have discussed above. Plotting this value from the numerical diagonalization in figure~\ref{fig:compare4} as data points, again we can compare to our full analytic approximation (solid curve), and also to the approximation truncating to the curved space Dirac equation, so including gauge field and frame corrections but ignoring the higher two covariant derivative term (dashed curve).
Note that these are single curves, as it happens there is no $M$ (or $\tilde{a}$) dependence in the approximations for this quantity.  We see that the full approximation gives good agreement for suitably small $\epsilon$ (which must be smaller for larger $M$, as really the perturbation parameter is $\tilde{\lambda}$), while the approximation ignoring the two derivative term does not agree at all.

\begin{figure}
 \centerline{
 \includegraphics[width=10cm]{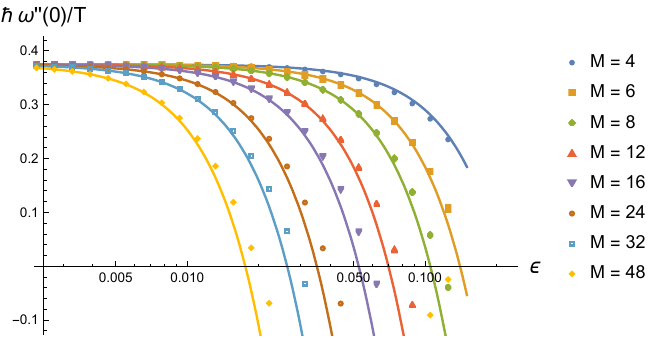}
  }
  \caption{ \label{fig:compare5}
A similar plot to the previous two, now for the second derivative  $\frac{\hbar}{T} {\omega^{num}}''(0)$. The full approximation is the solid curve, with the two and three covariant derivative terms giving this leading behaviour. Again it is interesting that the approximation is rather good even when the quantity is deformed far from its undistorted value.
}
\end{figure}

Lastly suppose we are interested in the subleading behaviour at the Dirac point, going as $k_x^2$. This is irrelevant on large scales compared to the leading Dirac behaviour, but is certainly of key interest if we are considering the full band structure. This is determined from the second derivative of $\omega^{num}$. In figure~\ref{fig:compare5} we plot $\frac{\hbar}{T}  {\omega^{num}}''(0)$ and compare to our analytic approximation. The leading contribution at $O(\tilde{\lambda}^2)$ comes from the two covariant derivative term. Here we see again that, somewhat surprisingly, even for quite small $\epsilon$ there is a very large deviation of this second derivative from its undeformed value -- for example for $\epsilon > 0.1$ and $M < 8$ we see that it actually becomes negative, as compared to its undeformed positive value of $3/8$. While this deviation is very large, certainly being $\sim O(1)$, the leading analytic approximation does a remarkably good job of reproducing the behaviour.

Thus to summarize, the leading effective theory taking the flat space Dirac equation coupled to the gauge field can correctly reproduce some features of the dispersion relation for in-plane inhomogeneous deformations. In particular it gives the leading correction to the wavespeed. However it cannot account for more detailed phenomena, such as the shift in the Dirac point, or the subleading $k_x^2$ deformation to the Dirac cone. Taking our effective theory with non-trivial frame and large strain gauge field, and including up to three covariant derivatives, allows a consistent derivation of the shift in the Dirac point, the leading  $k_x^2$ and $k_x^3$ corrections to the dispersion relation, as well as the subleading (and thus small) correction to the wavespeed. However it is crucial to emphasize again that one cannot pick and choose between these contributions -- the frame, two and three covariant derivative terms, and subleading gauge field corrections all enter with the same parametric dependence. Taking only a subset of these contributions will simply yield the wrong results.

\section{General Effective Theory}
\label{sec:EffectiveTheory}

So far our discussion has been centered on deriving the low energy effective theory from the (nearest neighbour) microscopic tight binding model. However this lattice model cannot capture the full detail of monolayer graphene. It is then natural to wonder what would be the appropriate effective theory of actual monolayer graphene in the presence of an in-plane strain, where, at least for reasonably small deformations we might expect Dirac points to persist. 

The main lessons we have learned from the tight binding lattice model are that its low energy effective theory, given above in~\eqref{eq:summarytheory},~\eqref{eq:summarymetric} and~\eqref{eq:summarygauge}, is gauge and frame covariant, that the power counting is modified from the usual structure of a relativistic effective theory, due to the large gauge field, and that it is entirely determined by the frame and gauge field, with the remaining necessary higher covariant derivative terms being controlled by lattice invariants with the geometric connection being the torsion free one. Then using a bond model where hopping functions are determined purely by bond lengths under a deformed embedding, we have written the gauge field and metric purely in terms of the strain tensor in~\eqref{eq:quadexpressions}.

It is then natural to postulate that the effective theory of a Dirac point for monolayer graphene, deformed by in-plane inhomogeneous strain, has the same features and again can be written in terms of the strain tensor $\sigma_{ij}$. 
We conjecture that this effective description takes the same form as for that of the tight-binding model, since it is determined by the lattice structure, but the numerical coefficients will differ from those in the case of the nearest-neighbour lattice model. Thus we expect the theory describing deformations up to linear order $O(\epsilon)$, and to leading order in the lattice scale $O(a)$, to simply be the flat space Dirac equation with large magnetic gauge field, which in lattice coordinates is,
\eqn{
0 =  a e^\mu_{~A} \gamma^A D_\mu \Psi + O(\epsilon^2 , \epsilon a) \; , \quad A_i(\vec{x}) = - \frac{\beta}{2a} \epsilon_{ij} K^{ijk} \sigma_{jk}(\bx) + O(\frac{\epsilon^2}{a} , \epsilon)
}
so that $e^\mu_{~A}$ is a flat frame, giving the flat electrometric $ds^2_{effective} = - c_{eff}^2 dt^2 + \delta_{ij} dx^i dx^j$, and $D_\mu \Psi$ is the covariant derivative of equation~\eqref{eq:covderiv}. Here the constants $c_{eff}$ and $\beta$ must be measured from the monolayer graphene system; the speed $c_{eff}$ may be measured from the undeformed Dirac cone, and $\beta$ from some particular deformation, for example a homogeneous strain. 
 The scale $a$ represents the lattice scale, but its actual value can be rescaled by appropriate scaling of the purely numerical constant $\beta$, as well as the wavefunction $\Psi$. Naturalness then implies that choosing $a$ again to be the lattice bond length, the numerical constant $\beta$ should be of order $O(1)$.
Once these constants are determined, then the theory should describe the leading order low energy behaviour at leading order in the deformation, $O(\epsilon)$, for any inhomogeneous strain field. As mentioned earlier, we expect in-plane inhomogeneous strain have low energy Dirac points, but for out-of-plane bending this is less clear due to the interaction of $\sigma$ orbitals.

This leading theory is a flat space theory, and as we have discussed, in order to consistently include curvature we must work to higher order in covariant derivatives. 
The structure of these again will be determined by the underlying lattice  symmetry, and so should take the same form as for the tight-binding model, but with different numerical coefficients. An important point is that for realistic graphene, even for a pristene monolayer, when going beyond linear dispersion near the Dirac points the conductance and valence bands are \emph{not} symmetric around the Fermi energy, in contrast to the behaviour of the simple lattice tight binding model \cite{reich2002tight,neto2009electronic}. Thus our effective theory should refer to a specific band, and the coefficients of the higher order corrections will be specific to this band. Here we will consider the effective theory of the valance band. 
Then from our discussions above, we expect the effective theory governing a Dirac point, working to order $O(\epsilon^2)$ and leading order in $a$ in the metric deformation, is,
 \eqn{
\label{eq:general}
0 =  a e^\mu_{~A} \gamma^A D_\mu \Psi \pm c_2 i a^2 \, \eta_{AB} \gamma^A e^B_{~\sigma} D_\mu \left( C^{\sigma\mu\nu} D_\nu \Psi \right) 
+ c_3 a^3 \,\eta_{AB} \gamma^A e^B_{~\sigma} D^{\sigma\mu\nu\rho}  D_\mu D_\nu D_\rho \Psi 
  + O(\epsilon^4, \epsilon^3 a, \epsilon^2 a^2, \epsilon a^3, a^4)  \nn \\
}
with the magnetic gauge field and electrometric in lattice coordinates being,
\eqn{
\label{eq:generalAandMet}
A_i(\vec{x}) &=& - \frac{\beta}{2a} \epsilon_{ij} \Big( K^{jkl} \left( \sigma_{kl}(\bx) + \xi_1 \sigma_{km}(\bx) \sigma_{ml}(\bx) + \xi_2 \sigma_k(\bx)   \sigma_l(\bx)   \right) \nl
&& \qquad \qquad + a^2 \left( \alpha_1 \partial_j \partial_k \sigma_k(\bx) +\alpha_2 \partial_k \partial_k \sigma_j(\bx) + \alpha_3 K^{klm} \partial_k \partial_l \sigma_{jm}(\bx) \right)  + O(\epsilon^3, \epsilon^2 a, \epsilon a^2) \Big) \nl
g_{ij}(\vec{x}) &=& \delta_{ij} + \beta\left( \chi_1  \sigma_{ij}(\bx) +   \chi_2 \sigma_{ik}(\bx) \sigma_{kj}(\bx) + \chi_3  \delta_{ij} \left( \sigma_{kk}(\bx)  \right)^2 + \chi_4 \sigma_{ij}(\bx)  \sigma_{kk}(\bx)  + \chi_5  \sigma_i(\bx) \sigma_j(\bx)  \right)  + O(\epsilon^3, \epsilon^2 a, \epsilon a^2) \nl
}
where again $\sigma_i = K^{ijk} \sigma_{jk}$, and these expressions are consistent to the $O(\epsilon^2)$ in the electrometric.
We emphasize again that in order to work consistently to $O(\epsilon^2)$ in the metric perturbation, we must include the two higher covariant derivatives in the effective theory and also the subleading $O(a^2)$ corrections at $O(\epsilon)$ in the gauge field (noting that $O(a)$ corrections to both the metric and gauge field vanish).
Here $\beta$, $c_{eff}$ are as for the leading theory, and now $c_{2,3}$, $\xi_{1,2}$, $\alpha_{1,2,3}$  and $\chi_{1, \ldots, 5}$ are more numerical constants that should be fixed by matching to the valence band of the monolayer graphene theory. Again it is natural to choose the scale $a$ to be the lattice bond length, but its precise value can be adjusted by rescaling these numerical constants. We reiterate that the effective theory for the conduction band will take the same form, but we should expect the numerical constants will have different values.

For the nearest neighbour tight-binding model with our bond-model~\eqref{eq:bondmodel} we see that they take the values,
\eqn{\label{eq:tb_eft_params}
c_2 & = & 1 \; , \quad c_3 = 1 \nl
\xi_1 & = &  \frac{(\beta - \tau)}{2} \; , \quad \xi_2 = - \frac{(3 \beta + \tau)}{8} \; , \quad \alpha_1 = \frac{3}{4} \; , \quad \alpha_2 = - \frac{1}{4}   \; , \quad \alpha_3 =  - \frac{7}{12}  \nl
 \chi_1 &=& 2 \; , \quad \chi_2 = 4 \beta \; , \quad \chi_3 =  \frac{( \beta + \tau )}{4} \; , \quad \chi_4 = - ( \beta + \tau ) \; , \quad \chi_5 = - \frac{( \beta + \tau )}{4} 
}
and further have $\tau = \beta + 1$ for an exponential bond model.
One may then adjust the constants $c_2$ and $c_3$ by fitting the band structure to the undeformed model. Then $\xi_{1,2}$, $\alpha_{1,2,3}$  and $\chi_{1, \ldots, 5}$ would be fixed by comparing to specific deformations. 

Specifically we may determine $\xi_{1,2}$ and $\chi_{1, \ldots, 5}$ from homogeneous, anisotropic strains as we now demonstrate. Suppose we perform a strain induced by,
\eqn{
X = x + \epsilon ( v_1 x + v_3 y ) \; , \quad Y = y + \epsilon v_2 y 
}
for constants $v_{1,2,3}$, which corresponds to the constant strain tensor (in lattice coordinates),
\eqn{
\sigma_{ij} = \left( 
\begin{array}{cc}
 \epsilon v_1 + \frac{\epsilon^2}{2} v_1^2 &  \epsilon v_3 + \frac{\epsilon^2}{2} v_1 v_3 \\
  \epsilon v_3 + \frac{\epsilon^2}{2} v_1 v_3  &  \epsilon v_2 + \frac{\epsilon^2}{2} ( v_2^2 + v_3^2 )
\end{array}
\right) \; .
}
The terms with coefficients given by the $\alpha_{1,2,3}$ in the gauge field all drop out as they involve derivative of this constant strain tensor. Thus we will be insensitive to these constants. However as we now show, we will be able to fix the other constants.

We may then solve the continuum theory in a similar manner to that in Section~\ref{sec:armchair}. We write a similar ansatz for the wavefunction as in~\eqref{eq:wavefnansatz}, taking,
\eqn{
\label{eq:wavefnansatztwo}
\Psi(t,\vec{x}) = e^{-i \omega t}  e^{+ i ( k_x x + k_y y)} \left( \begin{array}{c}
\psi_1 \\
\psi_2
\end{array} \right)
}
for constant components $\psi_{1,2}$.  For the undeformed theory the Dirac point is simply at $k_i = 0$ -- when uniform strain is turned on we define the location of the Dirac point, where $\omega = 0$, to be at $k_i = k^D_i$. Since the strain is constant, the magnetic gauge field is constant, and hence pure gauge, and directly corresponds to a shifting of the Dirac point, so that $k^D_i = A_i$.
Thus we consider the momentum shift from the Dirac point by writing,
\eqn{
k_i = k^D_i + \Delta k_i 
}
so that $(\Delta k_x, \Delta k_y) = (0,0)$ is the Dirac point. Further we may transform these lattice momenta to the lab frame via,
\eqn{
\Delta k_x =  \Delta k^{lab}_x + \epsilon v_1   \Delta k^{lab}_x \; , \quad \Delta k_y =  \Delta k^{lab}_y + \epsilon ( v_2  \Delta k^{lab}_y + v_3  \Delta k^{lab}_x ) 
}
and likewise for the shift of the Dirac point, given by $k^{D,lab}_i$, the transform of $k^D_i$ (which is simply given by the gauge field). Firstly we see that this shift, given simply by the transform of the (constant) gauge vector to lab frame, is,
\eqn{
k^{D,lab}_x & = & \frac{\beta}{a} \left(  \frac{\epsilon}{2} (v_2 - v_1) + \frac{\epsilon^2}{4} \left( (v_1 - v_2)^2 + v_3^2 - 2 \xi_1 (v_1^2 - v_2^2)  + 2 \xi_2 \left( (v_1 - 2 v_2)^2 - v_3^2 \right)  \right) \right) \\
k^{D,lab}_y & = & \frac{\beta}{a} \left(  \frac{\epsilon}{2} v_3 + \frac{\epsilon^2}{2} v_3 \left( 2(v_1 - v_2) + \xi_1 ( v_1 + v_2 ) + 2 \xi_2 (v_1 - v_2) \right) \right) \; .
}
Thus measuring this shift of the valence band Dirac point (in lab frame) as the homogeneous strain is turned on, at leading order $O(\epsilon)$ we may deduce the constant $\beta$, and at order $O(\epsilon^2)$ we find $\xi_{1,2}$. Note that we can extract these coefficients from some specific strain configuration, say $v_1 = v_2 = 2 v_3$. Assuming the effective theory is a consistent description it will then predict the dependence of $k^{D,lab}_{i}$ for any other values of $v_{1,2,3}$, so it may already be tested in this homogeneous setting.

To determine the remaining constants, $\chi_{1, \ldots , 5}$ we compute the dispersion relation purely as a function of $\Delta k^{lab}_x$, assumed to be positive for notational simplicity, and setting $\Delta k^{lab}_y = 0$.
For this we find, 
\eqn{
\pm\frac{a \omega}{ c_{eff}}  &=& \left( 1+\epsilon   \frac{v_1 ( 2 - \beta \chi_1) }{2} + \epsilon^2  \left(\frac{(2-\beta \chi_1)(v_3^2 ( 2 - \beta \chi_1) - 3 v_1^2 \beta \chi_1)}{8} \right. \right. \nonumber \\
& - &\left. \left. \frac{\beta}{2} \left( \frac{4 v_1^2 + v_3^2}{4} \chi_2 + (v_1 + v_2)^2 \chi_3 + v_1 (v_1 + v_2) \chi_4 + v_3^2 \chi_5  \right)  \right)\right)a \Delta k^{lab}_x \nonumber \\
&+&c_2\left(  \frac{1}{4}  + \frac{( 4 v_1 - v_2 \beta \chi_1 ) }{8} \right) (a \Delta k^{lab}_x)^2 - \frac{1}{8} c_3  (a \Delta k^{lab}_x)^3 + O(a^4, \epsilon a^3, \epsilon^2 a^2, \epsilon^3)  \; .
}
Note that when restricting to the tight binding model \eqref{eq:tb_eft_params} we find agreement with \cite{oliva2013understanding,oliva2015generalizing} at leading order. This result is slightly different than the expressions given in \cite{dejuan2012spacedepfermi}, because we have explicitly translated back to laboratory frame coordinates when defining momentum.

The energy $\omega$ as a function of the momentum $k^{lab}_x$ for the undeformed theory determines the constants $c_{eff}$, $c_2$ and $c_3$ through its second and third derivatives at the Dirac point $\Delta k^{lab}_x = 0$, as discussed above. At order $O(\epsilon)$  measuring the linear $\Delta k^{lab}_x$ behaviour should then determine the constant $\chi_1$. Then the $(\Delta k^{lab}_x)^2$ behaviour should be predicted by the effective theory. Finally at order $O(\epsilon^2)$ we may determine the remaining constants $\chi_{2,3,4,5}$ by measuring the linear $\Delta k^{lab}_x$ behaviour for various deformations. For example, setting $v_1 = v_3 = 0$ and having only $v_2$ will yield $\chi_3$. Taking $v_1 = - v_2$, $v_3 = 0$ will yield $\chi_2$, and so on. Once these are determined, the general dependence on $v_{1,2,3}$ should be predicted if the effective theory works, and thus it can again be tested already for these homogeneous strains. 

Once all these constants have been fixed, and the theory tested for consistency with homogeneous deformations, it then remains to fix the three remaining constants $\alpha_{1,2,3}$ at this order of approximation. This should be done using measurements made on the valence band for a particular inhomogeneous strain field, such as the periodic armchair deformation in the previous section. Once these too are fixed, we expect this effective theory of the valence band will determine its low energy behaviour for  \emph{any} inhomogeneous in-plane strain field up to the order of approximation we are truncating the theory to. As discussed above, one may use the same procedure to find the effective theory for the conduction band, and we expect the constants controlling the higher order corrections to be different to those of the valence band.
It would be very interesting to test these ideas using DFT calculations of monolayer graphene with inhomogeneous in-plane deformations.

\section{Conclusion}
\label{sec:Conclusion}

In this paper we have demonstrated that due to the lattice scale magnetic field resulting from lattice strain, the standard power counting of a derivative expansion does not apply to the continuum effective field theory of the graphene nearest-neighbor tight binding lattice model.
An important result of this is that while the continuum effective theory of this lattice model couples to a nontrivial curved electrometric, this curvature is subleading to the gauge field and appears at the same order in the effective theory as a higher covariant derivative correction, and so there is no consistent truncation to simply the curved space Dirac equation coupled to the strain gauge field. However, when considering perturbative elastic strain, it is possible to follow a double series expansion, in both the distortion and lattice spacing, which allows us to consistently organize an effective description. The leading truncation is the flat space Dirac equation coupled to the gauge field,  where there is no correction to the frame, and thus metric, and this applies only to linear order in the deformation amplitude. If we wish to go beyond linear order, or incorporate the interesting subleading effects of a curved `electrometric', we must include these higher covariant derivative terms, and also be careful to include the appropriate  corrections in the gauge field and electrometric which are subleading in the lattice spacing.

In this work we have carried out this expansion explicitly to quadratic order in the electrometric deformation, which requires including two higher covariant derivative terms. This is the order where we first see  corrections to the low energy dispersion relation due to spatial variation of the hopping functions. We have performed explicit numerical comparison with the lattice model to show that our effective theory correctly reproduces this low energy band structure. We see the effect of the subleading electrometric curvature corrections in the dispersion relation, and confirm that these contribute at the same order as corrections from higher derivative terms.

An interesting feature of the effective theory is that the electrometric, which the Dirac field propagates on, is not simply the induced metric of the graphene membrane. In particular, even for the case of pure in-plane distortions, which give a flat induced metric, 
going to quadratic order in the deformation one obtains an electrometric with nonzero curvature. This curvature is a subleading correction to the large magnetic field, and contributes at the same order as the Lorentz violating higher derivative terms, but still may provide an interesting testbed for `analog gravity'.

One of the most interesting results we find is that, at least to the order we have explored, torsion vanishes for the electrometric geometry -- this geometry is simply derived from the curved metric with the torsion-free Levi-Civita connection.
This is perhaps not surprising, as we are not changing the lattice topology, and so are not adding lattice dislocations, but given that we can generate curvature without adding lattice disclinations, it is not immediately apparent why we must couple to the torsion-free connection only.  In particular, using the lattice invariants and strain tensor one can define an object, 
$\epsilon_{jk} K^{ilm} \sigma_{lm}$
with the same index structure as spatial torsion, $T^i{}_{jk}$.  This tensor has a trace proportional to the leading strain gauge field, but minimal coupling to torsion has a different structure than to the strain gauge field - in particular, both components of $\Psi$ have the same charge under the local strain gauge symmetry, but opposite charge under frame rotations. It would be very nice if there were a symmetry argument for vanishing torsion, which we leave to future work. On a related note, it is interesting to follow what happened to the microscopic time reversal symmetry. Since this symmetry exchanges the $K$ and $K'$ points, it is clear that it appears in the continuum theory as the action of standard time reversal acting on a Dirac fermion combined with an exchange of the two Dirac fields. It is crucially only by considering the total theory of both Dirac cones that we recover a theory that is time reversal invariant, as it must be as elastic strain does not break time reversal.

Here we have explicitly derived the effective theory for the tight binding lattice model. Having seen its elegant structure, and in particular its local gauge and frame symmetry, and how it is constructed from lattice invariants, it is natural to conjecture that the effective theory for the valence and conduction bands of monolayer elastically strained graphene take a similar form, albeit with different numerical coefficients that would have to be determined through comparison to measurements.
It would be very interesting to compare our effective theory to ab initio numerical approaches used to study elastically deformed graphene. In particular it would be interesting to try to numerically extract  these effective field theory parameters. One might also use numerical methods to study the validity of the tight binding lattice model to bent graphene. It would be very interesting to understand how much of the effective theory remains valid, and what the effects of mixing between $p_z$ and $sp^2$ orbitals are on the continuum theory. One can also hope to make contact with experimental constructions of optical honeycomb lattices  \cite{lee2009ultracold}, which should have a similar tight-binding lattice model.

\subsection*{Acknowledgments}
We would like to thank Arash Mostofi for providing valuable insights. We also wish to thank Johan J\"onsson, Pablo Morales and Gerado Naumis for useful discussion.
This work was supported by STFC Consolidated Grant ST/T000791/1, the National Research Foundation of Korea (NRF) grant funded by the Korea government (MSIT) (No. 2023R1A2C1006975) and an appointment to the JRG Program at the APCTP through the Science and Technology Promotion Fund and Lottery Fund of the Korean Government.

\bibliography{Hofstadter.bib}

\begin{thebibliography}{52}%
\makeatletter
\providecommand \@ifxundefined [1]{%
 \@ifx{#1\undefined}
}%
\providecommand \@ifnum [1]{%
 \ifnum #1\expandafter \@firstoftwo
 \else \expandafter \@secondoftwo
 \fi
}%
\providecommand \@ifx [1]{%
 \ifx #1\expandafter \@firstoftwo
 \else \expandafter \@secondoftwo
 \fi
}%
\providecommand \natexlab [1]{#1}%
\providecommand \enquote  [1]{``#1''}%
\providecommand \bibnamefont  [1]{#1}%
\providecommand \bibfnamefont [1]{#1}%
\providecommand \citenamefont [1]{#1}%
\providecommand \href@noop [0]{\@secondoftwo}%
\providecommand \href [0]{\begingroup \@sanitize@url \@href}%
\providecommand \@href[1]{\@@startlink{#1}\@@href}%
\providecommand \@@href[1]{\endgroup#1\@@endlink}%
\providecommand \@sanitize@url [0]{\catcode `\\12\catcode `\$12\catcode
  `\&12\catcode `\#12\catcode `\^12\catcode `\_12\catcode `\%12\relax}%
\providecommand \@@startlink[1]{}%
\providecommand \@@endlink[0]{}%
\providecommand \url  [0]{\begingroup\@sanitize@url \@url }%
\providecommand \@url [1]{\endgroup\@href {#1}{\urlprefix }}%
\providecommand \urlprefix  [0]{URL }%
\providecommand \Eprint [0]{\href }%
\providecommand \doibase [0]{https://doi.org/}%
\providecommand \selectlanguage [0]{\@gobble}%
\providecommand \bibinfo  [0]{\@secondoftwo}%
\providecommand \bibfield  [0]{\@secondoftwo}%
\providecommand \translation [1]{[#1]}%
\providecommand \BibitemOpen [0]{}%
\providecommand \bibitemStop [0]{}%
\providecommand \bibitemNoStop [0]{.\EOS\space}%
\providecommand \EOS [0]{\spacefactor3000\relax}%
\providecommand \BibitemShut  [1]{\csname bibitem#1\endcsname}%
\let\auto@bib@innerbib\@empty
\bibitem [{\citenamefont {Novoselov}\ \emph {et~al.}(2005)\citenamefont
  {Novoselov}, \citenamefont {Geim}, \citenamefont {Morozov}, \citenamefont
  {Jiang}, \citenamefont {Katsnelson}, \citenamefont {Grigorieva},
  \citenamefont {Dubonos},\ and\ \citenamefont {Firsov}}]{novoselov2005two}%
  \BibitemOpen
  \bibfield  {author} {\bibinfo {author} {\bibfnamefont {K.~S.}\ \bibnamefont
  {Novoselov}}, \bibinfo {author} {\bibfnamefont {A.~K.}\ \bibnamefont {Geim}},
  \bibinfo {author} {\bibfnamefont {S.~V.}\ \bibnamefont {Morozov}}, \bibinfo
  {author} {\bibfnamefont {D.}~\bibnamefont {Jiang}}, \bibinfo {author}
  {\bibfnamefont {M.~I.}\ \bibnamefont {Katsnelson}}, \bibinfo {author}
  {\bibfnamefont {I.}~\bibnamefont {Grigorieva}}, \bibinfo {author}
  {\bibfnamefont {S.}~\bibnamefont {Dubonos}},\ and\ \bibinfo {author}
  {\bibfnamefont {A.}~\bibnamefont {Firsov}},\ }\bibfield  {title} {\bibinfo
  {title} {Two-dimensional gas of massless dirac fermions in graphene},\
  }\href@noop {} {\bibfield  {journal} {\bibinfo  {journal} {nature}\ }\textbf
  {\bibinfo {volume} {438}},\ \bibinfo {pages} {197} (\bibinfo {year}
  {2005})}\BibitemShut {NoStop}%
\bibitem [{\citenamefont {Zhang}\ \emph {et~al.}(2005)\citenamefont {Zhang},
  \citenamefont {Tan}, \citenamefont {Stormer},\ and\ \citenamefont
  {Kim}}]{Zhang:2005zz}%
  \BibitemOpen
  \bibfield  {author} {\bibinfo {author} {\bibfnamefont {Y.}~\bibnamefont
  {Zhang}}, \bibinfo {author} {\bibfnamefont {Y.-W.}\ \bibnamefont {Tan}},
  \bibinfo {author} {\bibfnamefont {H.~L.}\ \bibnamefont {Stormer}},\ and\
  \bibinfo {author} {\bibfnamefont {P.}~\bibnamefont {Kim}},\ }\bibfield
  {title} {\bibinfo {title} {{Experimental observation of the quantum Hall
  effect and and Berry's phase in graphene}},\ }\href
  {https://doi.org/10.1038/nature04235} {\bibfield  {journal} {\bibinfo
  {journal} {Nature}\ }\textbf {\bibinfo {volume} {438}},\ \bibinfo {pages}
  {201} (\bibinfo {year} {2005})}\BibitemShut {NoStop}%
\bibitem [{\citenamefont {Lee}\ \emph {et~al.}(2008)\citenamefont {Lee},
  \citenamefont {Wei}, \citenamefont {Kysar},\ and\ \citenamefont
  {Hone}}]{lee2008measurement}%
  \BibitemOpen
  \bibfield  {author} {\bibinfo {author} {\bibfnamefont {C.}~\bibnamefont
  {Lee}}, \bibinfo {author} {\bibfnamefont {X.}~\bibnamefont {Wei}}, \bibinfo
  {author} {\bibfnamefont {J.~W.}\ \bibnamefont {Kysar}},\ and\ \bibinfo
  {author} {\bibfnamefont {J.}~\bibnamefont {Hone}},\ }\bibfield  {title}
  {\bibinfo {title} {Measurement of the elastic properties and intrinsic
  strength of monolayer graphene},\ }\href@noop {} {\bibfield  {journal}
  {\bibinfo  {journal} {science}\ }\textbf {\bibinfo {volume} {321}},\ \bibinfo
  {pages} {385} (\bibinfo {year} {2008})}\BibitemShut {NoStop}%
\bibitem [{\citenamefont {Naumis}\ \emph {et~al.}(2017)\citenamefont {Naumis},
  \citenamefont {Barraza-Lopez}, \citenamefont {Oliva-Leyva},\ and\
  \citenamefont {Terrones}}]{naumis2017electronic}%
  \BibitemOpen
  \bibfield  {author} {\bibinfo {author} {\bibfnamefont {G.~G.}\ \bibnamefont
  {Naumis}}, \bibinfo {author} {\bibfnamefont {S.}~\bibnamefont
  {Barraza-Lopez}}, \bibinfo {author} {\bibfnamefont {M.}~\bibnamefont
  {Oliva-Leyva}},\ and\ \bibinfo {author} {\bibfnamefont {H.}~\bibnamefont
  {Terrones}},\ }\bibfield  {title} {\bibinfo {title} {Electronic and optical
  properties of strained graphene and other strained 2d materials: a review},\
  }\href@noop {} {\bibfield  {journal} {\bibinfo  {journal} {Reports on
  Progress in Physics}\ }\textbf {\bibinfo {volume} {80}},\ \bibinfo {pages}
  {096501} (\bibinfo {year} {2017})}\BibitemShut {NoStop}%
\bibitem [{\citenamefont {Sasaki}\ \emph {et~al.}(2005)\citenamefont {Sasaki},
  \citenamefont {Kawazoe},\ and\ \citenamefont {Saito}}]{sasaki2005local}%
  \BibitemOpen
  \bibfield  {author} {\bibinfo {author} {\bibfnamefont {K.-i.}\ \bibnamefont
  {Sasaki}}, \bibinfo {author} {\bibfnamefont {Y.}~\bibnamefont {Kawazoe}},\
  and\ \bibinfo {author} {\bibfnamefont {R.}~\bibnamefont {Saito}},\ }\bibfield
   {title} {\bibinfo {title} {Local energy gap in deformed carbon nanotubes},\
  }\href@noop {} {\bibfield  {journal} {\bibinfo  {journal} {Progress of
  theoretical physics}\ }\textbf {\bibinfo {volume} {113}},\ \bibinfo {pages}
  {463} (\bibinfo {year} {2005})}\BibitemShut {NoStop}%
\bibitem [{\citenamefont {Suzuura}\ and\ \citenamefont
  {Ando}(2002)}]{PhysRevB.65.235412}%
  \BibitemOpen
  \bibfield  {author} {\bibinfo {author} {\bibfnamefont {H.}~\bibnamefont
  {Suzuura}}\ and\ \bibinfo {author} {\bibfnamefont {T.}~\bibnamefont {Ando}},\
  }\bibfield  {title} {\bibinfo {title} {Phonons and electron-phonon scattering
  in carbon nanotubes},\ }\href {https://doi.org/10.1103/PhysRevB.65.235412}
  {\bibfield  {journal} {\bibinfo  {journal} {Phys. Rev. B}\ }\textbf {\bibinfo
  {volume} {65}},\ \bibinfo {pages} {235412} (\bibinfo {year}
  {2002})}\BibitemShut {NoStop}%
\bibitem [{\citenamefont {Vozmediano}\ \emph {et~al.}(2010)\citenamefont
  {Vozmediano}, \citenamefont {Katsnelson},\ and\ \citenamefont
  {Guinea}}]{vozmediano:physrev}%
  \BibitemOpen
  \bibfield  {author} {\bibinfo {author} {\bibfnamefont {M.}~\bibnamefont
  {Vozmediano}}, \bibinfo {author} {\bibfnamefont {M.}~\bibnamefont
  {Katsnelson}},\ and\ \bibinfo {author} {\bibfnamefont {F.}~\bibnamefont
  {Guinea}},\ }\bibfield  {title} {\bibinfo {title} {Gauge fields in
  graphene},\ }\href
  {https://doi.org/https://doi.org/10.1016/j.physrep.2010.07.003} {\bibfield
  {journal} {\bibinfo  {journal} {Physics Reports}\ }\textbf {\bibinfo {volume}
  {496}},\ \bibinfo {pages} {109} (\bibinfo {year} {2010})}\BibitemShut
  {NoStop}%
\bibitem [{\citenamefont {de~Juan}\ \emph {et~al.}(2012)\citenamefont
  {de~Juan}, \citenamefont {Sturla},\ and\ \citenamefont
  {Vozmediano}}]{dejuan2012spacedepfermi}%
  \BibitemOpen
  \bibfield  {author} {\bibinfo {author} {\bibfnamefont {F.}~\bibnamefont
  {de~Juan}}, \bibinfo {author} {\bibfnamefont {M.}~\bibnamefont {Sturla}},\
  and\ \bibinfo {author} {\bibfnamefont {M.~A.~H.}\ \bibnamefont
  {Vozmediano}},\ }\bibfield  {title} {\bibinfo {title} {Space dependent fermi
  velocity in strained graphene},\ }\href
  {https://doi.org/10.1103/PhysRevLett.108.227205} {\bibfield  {journal}
  {\bibinfo  {journal} {Phys. Rev. Lett.}\ }\textbf {\bibinfo {volume} {108}},\
  \bibinfo {pages} {227205} (\bibinfo {year} {2012})}\BibitemShut {NoStop}%
\bibitem [{\citenamefont {Zubkov}\ and\ \citenamefont
  {Volovik}(2015)}]{Zubkov:2013sja}%
  \BibitemOpen
  \bibfield  {author} {\bibinfo {author} {\bibfnamefont {M.~A.}\ \bibnamefont
  {Zubkov}}\ and\ \bibinfo {author} {\bibfnamefont {G.~E.}\ \bibnamefont
  {Volovik}},\ }\bibfield  {title} {\bibinfo {title} {{Emergent gravity in
  graphene}},\ }\href {https://doi.org/10.1088/1742-6596/607/1/012020}
  {\bibfield  {journal} {\bibinfo  {journal} {J. Phys. Conf. Ser.}\ }\textbf
  {\bibinfo {volume} {607}},\ \bibinfo {pages} {012020} (\bibinfo {year}
  {2015})},\ \Eprint {https://arxiv.org/abs/1308.2249} {arXiv:1308.2249
  [cond-mat.str-el]} \BibitemShut {NoStop}%
\bibitem [{\citenamefont {Oliva-Leyva}\ and\ \citenamefont
  {Naumis}(2015)}]{oliva2015generalizing}%
  \BibitemOpen
  \bibfield  {author} {\bibinfo {author} {\bibfnamefont {M.}~\bibnamefont
  {Oliva-Leyva}}\ and\ \bibinfo {author} {\bibfnamefont {G.~G.}\ \bibnamefont
  {Naumis}},\ }\bibfield  {title} {\bibinfo {title} {Generalizing the fermi
  velocity of strained graphene from uniform to nonuniform strain},\
  }\href@noop {} {\bibfield  {journal} {\bibinfo  {journal} {Physics Letters
  A}\ }\textbf {\bibinfo {volume} {379}},\ \bibinfo {pages} {2645} (\bibinfo
  {year} {2015})}\BibitemShut {NoStop}%
\bibitem [{\citenamefont {Yang}(2015)}]{yang2015dirac}%
  \BibitemOpen
  \bibfield  {author} {\bibinfo {author} {\bibfnamefont {B.}~\bibnamefont
  {Yang}},\ }\bibfield  {title} {\bibinfo {title} {Dirac cone metric and the
  origin of the spin connections in monolayer graphene},\ }\href@noop {}
  {\bibfield  {journal} {\bibinfo  {journal} {Physical Review B}\ }\textbf
  {\bibinfo {volume} {91}},\ \bibinfo {pages} {241403} (\bibinfo {year}
  {2015})}\BibitemShut {NoStop}%
\bibitem [{\citenamefont {Volovik}\ and\ \citenamefont
  {Zubkov}(2015)}]{VOLOVIK2015255}%
  \BibitemOpen
  \bibfield  {author} {\bibinfo {author} {\bibfnamefont {G.}~\bibnamefont
  {Volovik}}\ and\ \bibinfo {author} {\bibfnamefont {M.}~\bibnamefont
  {Zubkov}},\ }\bibfield  {title} {\bibinfo {title} {Emergent geometry
  experienced by fermions in graphene in the presence of dislocations},\ }\href
  {https://doi.org/https://doi.org/10.1016/j.aop.2015.03.005} {\bibfield
  {journal} {\bibinfo  {journal} {Annals of Physics}\ }\textbf {\bibinfo
  {volume} {356}},\ \bibinfo {pages} {255} (\bibinfo {year}
  {2015})}\BibitemShut {NoStop}%
\bibitem [{\citenamefont {Si}\ \emph {et~al.}(2016)\citenamefont {Si},
  \citenamefont {Sun},\ and\ \citenamefont {Liu}}]{si2016strain}%
  \BibitemOpen
  \bibfield  {author} {\bibinfo {author} {\bibfnamefont {C.}~\bibnamefont
  {Si}}, \bibinfo {author} {\bibfnamefont {Z.}~\bibnamefont {Sun}},\ and\
  \bibinfo {author} {\bibfnamefont {F.}~\bibnamefont {Liu}},\ }\bibfield
  {title} {\bibinfo {title} {Strain engineering of graphene: a review},\
  }\href@noop {} {\bibfield  {journal} {\bibinfo  {journal} {Nanoscale}\
  }\textbf {\bibinfo {volume} {8}},\ \bibinfo {pages} {3207} (\bibinfo {year}
  {2016})}\BibitemShut {NoStop}%
\bibitem [{\citenamefont {Khaidukov}\ and\ \citenamefont
  {Zubkov}(2016)}]{khaidukov2016landau}%
  \BibitemOpen
  \bibfield  {author} {\bibinfo {author} {\bibfnamefont {Z.~V.}\ \bibnamefont
  {Khaidukov}}\ and\ \bibinfo {author} {\bibfnamefont {M.~A.}\ \bibnamefont
  {Zubkov}},\ }\bibfield  {title} {\bibinfo {title} {Landau levels in graphene
  in the presence of emergent gravity},\ }\href@noop {} {\bibfield  {journal}
  {\bibinfo  {journal} {The European Physical Journal B}\ }\textbf {\bibinfo
  {volume} {89}},\ \bibinfo {pages} {1} (\bibinfo {year} {2016})}\BibitemShut
  {NoStop}%
\bibitem [{\citenamefont {Oliva-Leyva}\ and\ \citenamefont
  {Wang}(2017)}]{oliva2017low}%
  \BibitemOpen
  \bibfield  {author} {\bibinfo {author} {\bibfnamefont {M.}~\bibnamefont
  {Oliva-Leyva}}\ and\ \bibinfo {author} {\bibfnamefont {C.}~\bibnamefont
  {Wang}},\ }\bibfield  {title} {\bibinfo {title} {Low-energy theory for
  strained graphene: an approach up to second-order in the strain tensor},\
  }\href@noop {} {\bibfield  {journal} {\bibinfo  {journal} {Journal of
  Physics: Condensed Matter}\ }\textbf {\bibinfo {volume} {29}},\ \bibinfo
  {pages} {165301} (\bibinfo {year} {2017})}\BibitemShut {NoStop}%
\bibitem [{\citenamefont {Wagner}\ \emph {et~al.}(2019)\citenamefont {Wagner},
  \citenamefont {de~Juan},\ and\ \citenamefont {Nguyen}}]{wagner2019quantum}%
  \BibitemOpen
  \bibfield  {author} {\bibinfo {author} {\bibfnamefont {G.}~\bibnamefont
  {Wagner}}, \bibinfo {author} {\bibfnamefont {F.}~\bibnamefont {de~Juan}},\
  and\ \bibinfo {author} {\bibfnamefont {D.~X.}\ \bibnamefont {Nguyen}},\
  }\bibfield  {title} {\bibinfo {title} {Quantum hall effect in curved space
  realized in strained graphene},\ }\href@noop {} {\bibfield  {journal}
  {\bibinfo  {journal} {arXiv preprint arXiv:1911.02028}\ } (\bibinfo {year}
  {2019})}\BibitemShut {NoStop}%
\bibitem [{\citenamefont {de~Juan}\ \emph {et~al.}(2007)\citenamefont
  {de~Juan}, \citenamefont {Cortijo},\ and\ \citenamefont
  {Vozmediano}}]{de2007charge}%
  \BibitemOpen
  \bibfield  {author} {\bibinfo {author} {\bibfnamefont {F.}~\bibnamefont
  {de~Juan}}, \bibinfo {author} {\bibfnamefont {A.}~\bibnamefont {Cortijo}},\
  and\ \bibinfo {author} {\bibfnamefont {M.~A.}\ \bibnamefont {Vozmediano}},\
  }\bibfield  {title} {\bibinfo {title} {Charge inhomogeneities due to smooth
  ripples in graphene sheets},\ }\href@noop {} {\bibfield  {journal} {\bibinfo
  {journal} {Physical Review B}\ }\textbf {\bibinfo {volume} {76}},\ \bibinfo
  {pages} {165409} (\bibinfo {year} {2007})}\BibitemShut {NoStop}%
\bibitem [{\citenamefont {Guinea}\ \emph
  {et~al.}(2008{\natexlab{a}})\citenamefont {Guinea}, \citenamefont
  {Horovitz},\ and\ \citenamefont {Le~Doussal}}]{guinea2008gauge}%
  \BibitemOpen
  \bibfield  {author} {\bibinfo {author} {\bibfnamefont {F.}~\bibnamefont
  {Guinea}}, \bibinfo {author} {\bibfnamefont {B.}~\bibnamefont {Horovitz}},\
  and\ \bibinfo {author} {\bibfnamefont {P.}~\bibnamefont {Le~Doussal}},\
  }\bibfield  {title} {\bibinfo {title} {Gauge field induced by ripples in
  graphene},\ }\href@noop {} {\bibfield  {journal} {\bibinfo  {journal}
  {Physical Review B}\ }\textbf {\bibinfo {volume} {77}},\ \bibinfo {pages}
  {205421} (\bibinfo {year} {2008}{\natexlab{a}})}\BibitemShut {NoStop}%
\bibitem [{\citenamefont {Vozmediano}\ \emph {et~al.}(2008)\citenamefont
  {Vozmediano}, \citenamefont {de~Juan},\ and\ \citenamefont
  {Cortijo}}]{vozmediano2008gauge}%
  \BibitemOpen
  \bibfield  {author} {\bibinfo {author} {\bibfnamefont {M.~A.}\ \bibnamefont
  {Vozmediano}}, \bibinfo {author} {\bibfnamefont {F.}~\bibnamefont
  {de~Juan}},\ and\ \bibinfo {author} {\bibfnamefont {A.}~\bibnamefont
  {Cortijo}},\ }\bibfield  {title} {\bibinfo {title} {Gauge fields and
  curvature in graphene},\ }in\ \href@noop {} {\emph {\bibinfo {booktitle}
  {Journal of Physics: Conference Series}}},\ Vol.\ \bibinfo {volume} {129}\
  (\bibinfo {organization} {IOP Publishing},\ \bibinfo {year} {2008})\ p.\
  \bibinfo {pages} {012001}\BibitemShut {NoStop}%
\bibitem [{\citenamefont {de~Juan}\ \emph {et~al.}(2013)\citenamefont
  {de~Juan}, \citenamefont {Manes},\ and\ \citenamefont
  {Vozmediano}}]{de2013gauge}%
  \BibitemOpen
  \bibfield  {author} {\bibinfo {author} {\bibfnamefont {F.}~\bibnamefont
  {de~Juan}}, \bibinfo {author} {\bibfnamefont {J.~L.}\ \bibnamefont {Manes}},\
  and\ \bibinfo {author} {\bibfnamefont {M.~A.}\ \bibnamefont {Vozmediano}},\
  }\bibfield  {title} {\bibinfo {title} {Gauge fields from strain in
  graphene},\ }\href@noop {} {\bibfield  {journal} {\bibinfo  {journal}
  {Physical Review B}\ }\textbf {\bibinfo {volume} {87}},\ \bibinfo {pages}
  {165131} (\bibinfo {year} {2013})}\BibitemShut {NoStop}%
\bibitem [{\citenamefont {Arias}\ \emph {et~al.}(2015)\citenamefont {Arias},
  \citenamefont {Hern{\'a}ndez},\ and\ \citenamefont
  {Lewenkopf}}]{arias2015gauge}%
  \BibitemOpen
  \bibfield  {author} {\bibinfo {author} {\bibfnamefont {E.}~\bibnamefont
  {Arias}}, \bibinfo {author} {\bibfnamefont {A.~R.}\ \bibnamefont
  {Hern{\'a}ndez}},\ and\ \bibinfo {author} {\bibfnamefont {C.}~\bibnamefont
  {Lewenkopf}},\ }\bibfield  {title} {\bibinfo {title} {Gauge fields in
  graphene with nonuniform elastic deformations: A quantum field theory
  approach},\ }\href@noop {} {\bibfield  {journal} {\bibinfo  {journal}
  {Physical Review B}\ }\textbf {\bibinfo {volume} {92}},\ \bibinfo {pages}
  {245110} (\bibinfo {year} {2015})}\BibitemShut {NoStop}%
\bibitem [{\citenamefont {Stegmann}\ and\ \citenamefont
  {Szpak}(2016)}]{stegmann2016current}%
  \BibitemOpen
  \bibfield  {author} {\bibinfo {author} {\bibfnamefont {T.}~\bibnamefont
  {Stegmann}}\ and\ \bibinfo {author} {\bibfnamefont {N.}~\bibnamefont
  {Szpak}},\ }\bibfield  {title} {\bibinfo {title} {Current flow paths in
  deformed graphene: from quantum transport to classical trajectories in curved
  space},\ }\href@noop {} {\bibfield  {journal} {\bibinfo  {journal} {New
  Journal of Physics}\ }\textbf {\bibinfo {volume} {18}},\ \bibinfo {pages}
  {053016} (\bibinfo {year} {2016})}\BibitemShut {NoStop}%
\bibitem [{\citenamefont {Castro-Villarreal}\ and\ \citenamefont
  {Ruiz-S{\'a}nchez}(2017)}]{castro2017pseudomagnetic}%
  \BibitemOpen
  \bibfield  {author} {\bibinfo {author} {\bibfnamefont {P.}~\bibnamefont
  {Castro-Villarreal}}\ and\ \bibinfo {author} {\bibfnamefont {R.}~\bibnamefont
  {Ruiz-S{\'a}nchez}},\ }\bibfield  {title} {\bibinfo {title} {Pseudomagnetic
  field in curved graphene},\ }\href@noop {} {\bibfield  {journal} {\bibinfo
  {journal} {Physical Review B}\ }\textbf {\bibinfo {volume} {95}},\ \bibinfo
  {pages} {125432} (\bibinfo {year} {2017})}\BibitemShut {NoStop}%
\bibitem [{\citenamefont {Golkar}\ \emph {et~al.}(2015)\citenamefont {Golkar},
  \citenamefont {Roberts},\ and\ \citenamefont {Son}}]{Golkar:2014paa}%
  \BibitemOpen
  \bibfield  {author} {\bibinfo {author} {\bibfnamefont {S.}~\bibnamefont
  {Golkar}}, \bibinfo {author} {\bibfnamefont {M.~M.}\ \bibnamefont
  {Roberts}},\ and\ \bibinfo {author} {\bibfnamefont {D.~T.}\ \bibnamefont
  {Son}},\ }\bibfield  {title} {\bibinfo {title} {{The Euler current and
  relativistic parity odd transport}},\ }\href
  {https://doi.org/10.1007/JHEP04(2015)110} {\bibfield  {journal} {\bibinfo
  {journal} {JHEP}\ }\textbf {\bibinfo {volume} {04}},\ \bibinfo {pages}
  {110}},\ \Eprint {https://arxiv.org/abs/1407.7540} {arXiv:1407.7540 [hep-th]}
  \BibitemShut {NoStop}%
\bibitem [{\citenamefont {Golkar}\ \emph {et~al.}(2014)\citenamefont {Golkar},
  \citenamefont {Roberts},\ and\ \citenamefont {Son}}]{Golkar:2014wwa}%
  \BibitemOpen
  \bibfield  {author} {\bibinfo {author} {\bibfnamefont {S.}~\bibnamefont
  {Golkar}}, \bibinfo {author} {\bibfnamefont {M.~M.}\ \bibnamefont
  {Roberts}},\ and\ \bibinfo {author} {\bibfnamefont {D.~T.}\ \bibnamefont
  {Son}},\ }\bibfield  {title} {\bibinfo {title} {{Effective Field Theory of
  Relativistic Quantum Hall Systems}},\ }\href
  {https://doi.org/10.1007/JHEP12(2014)138} {\bibfield  {journal} {\bibinfo
  {journal} {JHEP}\ }\textbf {\bibinfo {volume} {12}},\ \bibinfo {pages}
  {138}},\ \Eprint {https://arxiv.org/abs/1403.4279} {arXiv:1403.4279
  [cond-mat.mes-hall]} \BibitemShut {NoStop}%
\bibitem [{\citenamefont {Roberts}\ and\ \citenamefont
  {Wiseman}(2022{\natexlab{a}})}]{Roberts:2021vmt}%
  \BibitemOpen
  \bibfield  {author} {\bibinfo {author} {\bibfnamefont {M.~M.}\ \bibnamefont
  {Roberts}}\ and\ \bibinfo {author} {\bibfnamefont {T.}~\bibnamefont
  {Wiseman}},\ }\bibfield  {title} {\bibinfo {title} {{Curved-space Dirac
  description of elastically deformed monolayer graphene is generally
  incorrect}},\ }\href {https://doi.org/10.1103/PhysRevB.105.195412} {\bibfield
   {journal} {\bibinfo  {journal} {Phys. Rev. B}\ }\textbf {\bibinfo {volume}
  {105}},\ \bibinfo {pages} {195412} (\bibinfo {year} {2022}{\natexlab{a}})},\
  \Eprint {https://arxiv.org/abs/2112.09144} {arXiv:2112.09144
  [cond-mat.mes-hall]} \BibitemShut {NoStop}%
\bibitem [{\citenamefont {Volovik}\ and\ \citenamefont
  {Zubkov}(2014)}]{VOLOVIK2014352}%
  \BibitemOpen
  \bibfield  {author} {\bibinfo {author} {\bibfnamefont {G.}~\bibnamefont
  {Volovik}}\ and\ \bibinfo {author} {\bibfnamefont {M.}~\bibnamefont
  {Zubkov}},\ }\bibfield  {title} {\bibinfo {title} {Emergent horava gravity in
  graphene},\ }\href
  {https://doi.org/https://doi.org/10.1016/j.aop.2013.11.003} {\bibfield
  {journal} {\bibinfo  {journal} {Annals of Physics}\ }\textbf {\bibinfo
  {volume} {340}},\ \bibinfo {pages} {352} (\bibinfo {year}
  {2014})}\BibitemShut {NoStop}%
\bibitem [{\citenamefont {Iorio}\ and\ \citenamefont
  {Pais}(2015)}]{iorio2015revisiting}%
  \BibitemOpen
  \bibfield  {author} {\bibinfo {author} {\bibfnamefont {A.}~\bibnamefont
  {Iorio}}\ and\ \bibinfo {author} {\bibfnamefont {P.}~\bibnamefont {Pais}},\
  }\bibfield  {title} {\bibinfo {title} {Revisiting the gauge fields of
  strained graphene},\ }\href@noop {} {\bibfield  {journal} {\bibinfo
  {journal} {Physical Review D}\ }\textbf {\bibinfo {volume} {92}},\ \bibinfo
  {pages} {125005} (\bibinfo {year} {2015})}\BibitemShut {NoStop}%
\bibitem [{\citenamefont {Iorio}\ and\ \citenamefont
  {Pais}(2022)}]{iorio2022comment}%
  \BibitemOpen
  \bibfield  {author} {\bibinfo {author} {\bibfnamefont {A.}~\bibnamefont
  {Iorio}}\ and\ \bibinfo {author} {\bibfnamefont {P.}~\bibnamefont {Pais}},\
  }\bibfield  {title} {\bibinfo {title} {Comment on ``curved-space dirac
  description of elastically deformed monolayer graphene is generally
  incorrect''},\ }\href@noop {} {\bibfield  {journal} {\bibinfo  {journal}
  {Physical Review B}\ }\textbf {\bibinfo {volume} {106}},\ \bibinfo {pages}
  {157401} (\bibinfo {year} {2022})}\BibitemShut {NoStop}%
\bibitem [{\citenamefont {Roberts}\ and\ \citenamefont
  {Wiseman}(2022{\natexlab{b}})}]{roberts2022reply}%
  \BibitemOpen
  \bibfield  {author} {\bibinfo {author} {\bibfnamefont {M.~M.}\ \bibnamefont
  {Roberts}}\ and\ \bibinfo {author} {\bibfnamefont {T.}~\bibnamefont
  {Wiseman}},\ }\bibfield  {title} {\bibinfo {title} {Reply to ``comment on
  `curved-space dirac description of elastically deformed monolayer graphene is
  generally incorrect'''},\ }\href@noop {} {\bibfield  {journal} {\bibinfo
  {journal} {Physical Review B}\ }\textbf {\bibinfo {volume} {106}},\ \bibinfo
  {pages} {157402} (\bibinfo {year} {2022}{\natexlab{b}})}\BibitemShut
  {NoStop}%
\bibitem [{\citenamefont {Guinea}\ \emph
  {et~al.}(2008{\natexlab{b}})\citenamefont {Guinea}, \citenamefont
  {Horovitz},\ and\ \citenamefont {Le~Doussal}}]{PhysRevB.77.205421}%
  \BibitemOpen
  \bibfield  {author} {\bibinfo {author} {\bibfnamefont {F.}~\bibnamefont
  {Guinea}}, \bibinfo {author} {\bibfnamefont {B.}~\bibnamefont {Horovitz}},\
  and\ \bibinfo {author} {\bibfnamefont {P.}~\bibnamefont {Le~Doussal}},\
  }\bibfield  {title} {\bibinfo {title} {Gauge field induced by ripples in
  graphene},\ }\href {https://doi.org/10.1103/PhysRevB.77.205421} {\bibfield
  {journal} {\bibinfo  {journal} {Phys. Rev. B}\ }\textbf {\bibinfo {volume}
  {77}},\ \bibinfo {pages} {205421} (\bibinfo {year}
  {2008}{\natexlab{b}})}\BibitemShut {NoStop}%
\bibitem [{\citenamefont {Kim}\ and\ \citenamefont
  {Neto}(2008)}]{kim2008graphene}%
  \BibitemOpen
  \bibfield  {author} {\bibinfo {author} {\bibfnamefont {E.-A.}\ \bibnamefont
  {Kim}}\ and\ \bibinfo {author} {\bibfnamefont {A.~C.}\ \bibnamefont {Neto}},\
  }\bibfield  {title} {\bibinfo {title} {Graphene as an electronic membrane},\
  }\href@noop {} {\bibfield  {journal} {\bibinfo  {journal} {Europhysics
  Letters}\ }\textbf {\bibinfo {volume} {84}},\ \bibinfo {pages} {57007}
  (\bibinfo {year} {2008})}\BibitemShut {NoStop}%
\bibitem [{\citenamefont {Guinea}\ \emph
  {et~al.}(2008{\natexlab{c}})\citenamefont {Guinea}, \citenamefont
  {Katsnelson},\ and\ \citenamefont {Vozmediano}}]{guinea2008midgap}%
  \BibitemOpen
  \bibfield  {author} {\bibinfo {author} {\bibfnamefont {F.}~\bibnamefont
  {Guinea}}, \bibinfo {author} {\bibfnamefont {M.}~\bibnamefont {Katsnelson}},\
  and\ \bibinfo {author} {\bibfnamefont {M.}~\bibnamefont {Vozmediano}},\
  }\bibfield  {title} {\bibinfo {title} {Midgap states and charge
  inhomogeneities in corrugated graphene},\ }\href@noop {} {\bibfield
  {journal} {\bibinfo  {journal} {Physical Review B}\ }\textbf {\bibinfo
  {volume} {77}},\ \bibinfo {pages} {075422} (\bibinfo {year}
  {2008}{\natexlab{c}})}\BibitemShut {NoStop}%
\bibitem [{\citenamefont {Wehling}\ \emph {et~al.}(2008)\citenamefont
  {Wehling}, \citenamefont {Balatsky}, \citenamefont {Tsvelik}, \citenamefont
  {Katsnelson},\ and\ \citenamefont {Lichtenstein}}]{wehling2008midgap}%
  \BibitemOpen
  \bibfield  {author} {\bibinfo {author} {\bibfnamefont {T.}~\bibnamefont
  {Wehling}}, \bibinfo {author} {\bibfnamefont {A.}~\bibnamefont {Balatsky}},
  \bibinfo {author} {\bibfnamefont {A.}~\bibnamefont {Tsvelik}}, \bibinfo
  {author} {\bibfnamefont {M.}~\bibnamefont {Katsnelson}},\ and\ \bibinfo
  {author} {\bibfnamefont {A.}~\bibnamefont {Lichtenstein}},\ }\bibfield
  {title} {\bibinfo {title} {Midgap states in corrugated graphene: Ab initio
  calculations and effective field theory},\ }\href@noop {} {\bibfield
  {journal} {\bibinfo  {journal} {Europhysics Letters}\ }\textbf {\bibinfo
  {volume} {84}},\ \bibinfo {pages} {17003} (\bibinfo {year}
  {2008})}\BibitemShut {NoStop}%
\bibitem [{\citenamefont {Naumov}\ and\ \citenamefont
  {Bratkovsky}(2011{\natexlab{a}})}]{PhysRevB.84.245444}%
  \BibitemOpen
  \bibfield  {author} {\bibinfo {author} {\bibfnamefont {I.~I.}\ \bibnamefont
  {Naumov}}\ and\ \bibinfo {author} {\bibfnamefont {A.~M.}\ \bibnamefont
  {Bratkovsky}},\ }\bibfield  {title} {\bibinfo {title} {Gap opening in
  graphene by simple periodic inhomogeneous strain},\ }\href
  {https://doi.org/10.1103/PhysRevB.84.245444} {\bibfield  {journal} {\bibinfo
  {journal} {Phys. Rev. B}\ }\textbf {\bibinfo {volume} {84}},\ \bibinfo
  {pages} {245444} (\bibinfo {year} {2011}{\natexlab{a}})}\BibitemShut
  {NoStop}%
\bibitem [{\citenamefont {Lin}\ \emph {et~al.}(2015)\citenamefont {Lin},
  \citenamefont {Chang}, \citenamefont {Shyu}, \citenamefont {Lu},\ and\
  \citenamefont {Lin}}]{lin2015feature}%
  \BibitemOpen
  \bibfield  {author} {\bibinfo {author} {\bibfnamefont {S.-Y.}\ \bibnamefont
  {Lin}}, \bibinfo {author} {\bibfnamefont {S.-L.}\ \bibnamefont {Chang}},
  \bibinfo {author} {\bibfnamefont {F.-L.}\ \bibnamefont {Shyu}}, \bibinfo
  {author} {\bibfnamefont {J.-M.}\ \bibnamefont {Lu}},\ and\ \bibinfo {author}
  {\bibfnamefont {M.-F.}\ \bibnamefont {Lin}},\ }\bibfield  {title} {\bibinfo
  {title} {Feature-rich electronic properties in graphene ripples},\
  }\href@noop {} {\bibfield  {journal} {\bibinfo  {journal} {Carbon}\ }\textbf
  {\bibinfo {volume} {86}},\ \bibinfo {pages} {207} (\bibinfo {year}
  {2015})}\BibitemShut {NoStop}%
\bibitem [{\citenamefont {L{\'o}pez-Sancho}\ and\ \citenamefont
  {Brey}(2016)}]{lopez2016magnetic}%
  \BibitemOpen
  \bibfield  {author} {\bibinfo {author} {\bibfnamefont {M.~P.}\ \bibnamefont
  {L{\'o}pez-Sancho}}\ and\ \bibinfo {author} {\bibfnamefont {L.}~\bibnamefont
  {Brey}},\ }\bibfield  {title} {\bibinfo {title} {Magnetic phases in
  periodically rippled graphene},\ }\href@noop {} {\bibfield  {journal}
  {\bibinfo  {journal} {Physical Review B}\ }\textbf {\bibinfo {volume} {94}},\
  \bibinfo {pages} {165430} (\bibinfo {year} {2016})}\BibitemShut {NoStop}%
\bibitem [{\citenamefont {Talla}\ and\ \citenamefont
  {Ahmad}(2022)}]{talla2022structural}%
  \BibitemOpen
  \bibfield  {author} {\bibinfo {author} {\bibfnamefont {J.~A.}\ \bibnamefont
  {Talla}}\ and\ \bibinfo {author} {\bibfnamefont {M.~S.}\ \bibnamefont
  {Ahmad}},\ }\bibfield  {title} {\bibinfo {title} {Structural and electronic
  properties of rippled graphene monolayer: density functional theory},\
  }\href@noop {} {\bibfield  {journal} {\bibinfo  {journal} {Journal of
  Electronic Materials}\ }\textbf {\bibinfo {volume} {51}},\ \bibinfo {pages}
  {2464} (\bibinfo {year} {2022})}\BibitemShut {NoStop}%
\bibitem [{\citenamefont {Oliva-Leyva}\ and\ \citenamefont
  {Naumis}(2013)}]{oliva2013understanding}%
  \BibitemOpen
  \bibfield  {author} {\bibinfo {author} {\bibfnamefont {M.}~\bibnamefont
  {Oliva-Leyva}}\ and\ \bibinfo {author} {\bibfnamefont {G.~G.}\ \bibnamefont
  {Naumis}},\ }\bibfield  {title} {\bibinfo {title} {Understanding electron
  behavior in strained graphene as a reciprocal space distortion},\ }\href@noop
  {} {\bibfield  {journal} {\bibinfo  {journal} {Physical Review B}\ }\textbf
  {\bibinfo {volume} {88}},\ \bibinfo {pages} {085430} (\bibinfo {year}
  {2013})}\BibitemShut {NoStop}%
\bibitem [{\citenamefont {Balents}(2019)}]{balents:twisted_bilayer}%
  \BibitemOpen
  \bibfield  {author} {\bibinfo {author} {\bibfnamefont {L.}~\bibnamefont
  {Balents}},\ }\bibfield  {title} {\bibinfo {title} {{General continuum model
  for twisted bilayer graphene and arbitrary smooth deformations}},\ }\href
  {https://doi.org/10.21468/SciPostPhys.7.4.048} {\bibfield  {journal}
  {\bibinfo  {journal} {SciPost Phys.}\ }\textbf {\bibinfo {volume} {7}},\
  \bibinfo {pages} {048} (\bibinfo {year} {2019})}\BibitemShut {NoStop}%
\bibitem [{\citenamefont {Cortijo}\ and\ \citenamefont
  {Vozmediano}(2007)}]{cortijo2007effects}%
  \BibitemOpen
  \bibfield  {author} {\bibinfo {author} {\bibfnamefont {A.}~\bibnamefont
  {Cortijo}}\ and\ \bibinfo {author} {\bibfnamefont {M.~A.}\ \bibnamefont
  {Vozmediano}},\ }\bibfield  {title} {\bibinfo {title} {Effects of topological
  defects and local curvature on the electronic properties of planar
  graphene},\ }\href@noop {} {\bibfield  {journal} {\bibinfo  {journal}
  {Nuclear Physics B}\ }\textbf {\bibinfo {volume} {763}},\ \bibinfo {pages}
  {293} (\bibinfo {year} {2007})}\BibitemShut {NoStop}%
\bibitem [{\citenamefont {Morales}\ and\ \citenamefont
  {Copinger}(2023)}]{morales_timedep_graphene2023}%
  \BibitemOpen
  \bibfield  {author} {\bibinfo {author} {\bibfnamefont {P.~A.}\ \bibnamefont
  {Morales}}\ and\ \bibinfo {author} {\bibfnamefont {P.}~\bibnamefont
  {Copinger}},\ }\bibfield  {title} {\bibinfo {title} {Curvature-induced
  pseudogauge fields from time-dependent geometries in graphene},\ }\href
  {https://doi.org/10.1103/PhysRevB.107.075432} {\bibfield  {journal} {\bibinfo
   {journal} {Phys. Rev. B}\ }\textbf {\bibinfo {volume} {107}},\ \bibinfo
  {pages} {075432} (\bibinfo {year} {2023})}\BibitemShut {NoStop}%
\bibitem [{\citenamefont {Ribeiro}\ \emph {et~al.}(2009)\citenamefont
  {Ribeiro}, \citenamefont {Pereira}, \citenamefont {Peres}, \citenamefont
  {Briddon},\ and\ \citenamefont {Neto}}]{ribeiro2009strained}%
  \BibitemOpen
  \bibfield  {author} {\bibinfo {author} {\bibfnamefont {R.}~\bibnamefont
  {Ribeiro}}, \bibinfo {author} {\bibfnamefont {V.~M.}\ \bibnamefont
  {Pereira}}, \bibinfo {author} {\bibfnamefont {N.}~\bibnamefont {Peres}},
  \bibinfo {author} {\bibfnamefont {P.}~\bibnamefont {Briddon}},\ and\ \bibinfo
  {author} {\bibfnamefont {A.~C.}\ \bibnamefont {Neto}},\ }\bibfield  {title}
  {\bibinfo {title} {Strained graphene: tight-binding and density functional
  calculations},\ }\href@noop {} {\bibfield  {journal} {\bibinfo  {journal}
  {New Journal of Physics}\ }\textbf {\bibinfo {volume} {11}},\ \bibinfo
  {pages} {115002} (\bibinfo {year} {2009})}\BibitemShut {NoStop}%
\bibitem [{\citenamefont {Wagner}\ \emph {et~al.}(2022)\citenamefont {Wagner},
  \citenamefont {de~Juan},\ and\ \citenamefont {Nguyen}}]{wagner2022landau}%
  \BibitemOpen
  \bibfield  {author} {\bibinfo {author} {\bibfnamefont {G.}~\bibnamefont
  {Wagner}}, \bibinfo {author} {\bibfnamefont {F.}~\bibnamefont {de~Juan}},\
  and\ \bibinfo {author} {\bibfnamefont {D.}~\bibnamefont {Nguyen}},\
  }\bibfield  {title} {\bibinfo {title} {Landau levels in curved space realized
  in strained graphene},\ }\href@noop {} {\bibfield  {journal} {\bibinfo
  {journal} {SciPost Physics Core}\ }\textbf {\bibinfo {volume} {5}},\ \bibinfo
  {pages} {029} (\bibinfo {year} {2022})}\BibitemShut {NoStop}%
\bibitem [{\citenamefont {Naumov}\ and\ \citenamefont
  {Bratkovsky}(2011{\natexlab{b}})}]{naumov2011gap}%
  \BibitemOpen
  \bibfield  {author} {\bibinfo {author} {\bibfnamefont {I.}~\bibnamefont
  {Naumov}}\ and\ \bibinfo {author} {\bibfnamefont {A.}~\bibnamefont
  {Bratkovsky}},\ }\bibfield  {title} {\bibinfo {title} {Gap opening in
  graphene by simple periodic inhomogeneous strain},\ }\href@noop {} {\bibfield
   {journal} {\bibinfo  {journal} {Physical Review B}\ }\textbf {\bibinfo
  {volume} {84}},\ \bibinfo {pages} {245444} (\bibinfo {year}
  {2011}{\natexlab{b}})}\BibitemShut {NoStop}%
\bibitem [{\citenamefont {Manes}\ \emph {et~al.}(2013)\citenamefont {Manes},
  \citenamefont {de~Juan}, \citenamefont {Sturla},\ and\ \citenamefont
  {Vozmediano}}]{manes2013generalized}%
  \BibitemOpen
  \bibfield  {author} {\bibinfo {author} {\bibfnamefont {J.~L.}\ \bibnamefont
  {Manes}}, \bibinfo {author} {\bibfnamefont {F.}~\bibnamefont {de~Juan}},
  \bibinfo {author} {\bibfnamefont {M.}~\bibnamefont {Sturla}},\ and\ \bibinfo
  {author} {\bibfnamefont {M.~A.}\ \bibnamefont {Vozmediano}},\ }\bibfield
  {title} {\bibinfo {title} {Generalized effective hamiltonian for graphene
  under nonuniform strain},\ }\href@noop {} {\bibfield  {journal} {\bibinfo
  {journal} {Physical Review B}\ }\textbf {\bibinfo {volume} {88}},\ \bibinfo
  {pages} {155405} (\bibinfo {year} {2013})}\BibitemShut {NoStop}%
\bibitem [{\citenamefont {Iorio}\ \emph {et~al.}(2018)\citenamefont {Iorio},
  \citenamefont {Pais}, \citenamefont {Elmashad}, \citenamefont {Ali},
  \citenamefont {Faizal},\ and\ \citenamefont {Abou-Salem}}]{Iorio:2017vtw}%
  \BibitemOpen
  \bibfield  {author} {\bibinfo {author} {\bibfnamefont {A.}~\bibnamefont
  {Iorio}}, \bibinfo {author} {\bibfnamefont {P.}~\bibnamefont {Pais}},
  \bibinfo {author} {\bibfnamefont {I.~A.}\ \bibnamefont {Elmashad}}, \bibinfo
  {author} {\bibfnamefont {A.~F.}\ \bibnamefont {Ali}}, \bibinfo {author}
  {\bibfnamefont {M.}~\bibnamefont {Faizal}},\ and\ \bibinfo {author}
  {\bibfnamefont {L.~I.}\ \bibnamefont {Abou-Salem}},\ }\bibfield  {title}
  {\bibinfo {title} {{Generalized Dirac structure beyond the linear regime in
  graphene}},\ }\href {https://doi.org/10.1142/S0218271818500803} {\bibfield
  {journal} {\bibinfo  {journal} {Int. J. Mod. Phys. D}\ }\textbf {\bibinfo
  {volume} {27}},\ \bibinfo {pages} {1850080} (\bibinfo {year} {2018})},\
  \bibinfo {note} {[Erratum: Int.J.Mod.Phys.D 27, 1850080 (2023)]},\ \Eprint
  {https://arxiv.org/abs/1706.01332} {arXiv:1706.01332 [physics.gen-ph]}
  \BibitemShut {NoStop}%
\bibitem [{\citenamefont {Iorio}\ \emph {et~al.}(2023)\citenamefont {Iorio},
  \citenamefont {Iveti\'c},\ and\ \citenamefont {Pais}}]{Iorio:2023cmb}%
  \BibitemOpen
  \bibfield  {author} {\bibinfo {author} {\bibfnamefont {A.}~\bibnamefont
  {Iorio}}, \bibinfo {author} {\bibfnamefont {B.}~\bibnamefont {Iveti\'c}},\
  and\ \bibinfo {author} {\bibfnamefont {P.}~\bibnamefont {Pais}},\ }\bibfield
  {title} {\bibinfo {title} {{Turning graphene into a lab for
  noncommutativity}},\ }\href@noop {} {\  (\bibinfo {year} {2023})},\ \Eprint
  {https://arxiv.org/abs/2306.17196} {arXiv:2306.17196 [physics.gen-ph]}
  \BibitemShut {NoStop}%
\bibitem [{\citenamefont {Zan}\ \emph {et~al.}(2012)\citenamefont {Zan},
  \citenamefont {Muryn}, \citenamefont {Bangert}, \citenamefont {Mattocks},
  \citenamefont {Wincott}, \citenamefont {Vaughan}, \citenamefont {Li},
  \citenamefont {Colombo}, \citenamefont {Ruoff}, \citenamefont {Hamilton}
  \emph {et~al.}}]{zan2012scanning}%
  \BibitemOpen
  \bibfield  {author} {\bibinfo {author} {\bibfnamefont {R.}~\bibnamefont
  {Zan}}, \bibinfo {author} {\bibfnamefont {C.}~\bibnamefont {Muryn}}, \bibinfo
  {author} {\bibfnamefont {U.}~\bibnamefont {Bangert}}, \bibinfo {author}
  {\bibfnamefont {P.}~\bibnamefont {Mattocks}}, \bibinfo {author}
  {\bibfnamefont {P.}~\bibnamefont {Wincott}}, \bibinfo {author} {\bibfnamefont
  {D.}~\bibnamefont {Vaughan}}, \bibinfo {author} {\bibfnamefont
  {X.}~\bibnamefont {Li}}, \bibinfo {author} {\bibfnamefont {L.}~\bibnamefont
  {Colombo}}, \bibinfo {author} {\bibfnamefont {R.~S.}\ \bibnamefont {Ruoff}},
  \bibinfo {author} {\bibfnamefont {B.}~\bibnamefont {Hamilton}}, \emph
  {et~al.},\ }\bibfield  {title} {\bibinfo {title} {Scanning tunnelling
  microscopy of suspended graphene},\ }\href@noop {} {\bibfield  {journal}
  {\bibinfo  {journal} {Nanoscale}\ }\textbf {\bibinfo {volume} {4}},\ \bibinfo
  {pages} {3065} (\bibinfo {year} {2012})}\BibitemShut {NoStop}%
\bibitem [{\citenamefont {Reich}\ \emph {et~al.}(2002)\citenamefont {Reich},
  \citenamefont {Maultzsch}, \citenamefont {Thomsen},\ and\ \citenamefont
  {Ordejon}}]{reich2002tight}%
  \BibitemOpen
  \bibfield  {author} {\bibinfo {author} {\bibfnamefont {S.}~\bibnamefont
  {Reich}}, \bibinfo {author} {\bibfnamefont {J.}~\bibnamefont {Maultzsch}},
  \bibinfo {author} {\bibfnamefont {C.}~\bibnamefont {Thomsen}},\ and\ \bibinfo
  {author} {\bibfnamefont {P.}~\bibnamefont {Ordejon}},\ }\bibfield  {title}
  {\bibinfo {title} {Tight-binding description of graphene},\ }\href@noop {}
  {\bibfield  {journal} {\bibinfo  {journal} {Physical Review B}\ }\textbf
  {\bibinfo {volume} {66}},\ \bibinfo {pages} {035412} (\bibinfo {year}
  {2002})}\BibitemShut {NoStop}%
\bibitem [{\citenamefont {Neto}\ \emph {et~al.}(2009)\citenamefont {Neto},
  \citenamefont {Guinea}, \citenamefont {Peres}, \citenamefont {Novoselov},\
  and\ \citenamefont {Geim}}]{neto2009electronic}%
  \BibitemOpen
  \bibfield  {author} {\bibinfo {author} {\bibfnamefont {A.~C.}\ \bibnamefont
  {Neto}}, \bibinfo {author} {\bibfnamefont {F.}~\bibnamefont {Guinea}},
  \bibinfo {author} {\bibfnamefont {N.~M.}\ \bibnamefont {Peres}}, \bibinfo
  {author} {\bibfnamefont {K.~S.}\ \bibnamefont {Novoselov}},\ and\ \bibinfo
  {author} {\bibfnamefont {A.~K.}\ \bibnamefont {Geim}},\ }\bibfield  {title}
  {\bibinfo {title} {The electronic properties of graphene},\ }\href@noop {}
  {\bibfield  {journal} {\bibinfo  {journal} {Reviews of modern physics}\
  }\textbf {\bibinfo {volume} {81}},\ \bibinfo {pages} {109} (\bibinfo {year}
  {2009})}\BibitemShut {NoStop}%
\bibitem [{\citenamefont {Lee}\ \emph {et~al.}(2009)\citenamefont {Lee},
  \citenamefont {Gremaud}, \citenamefont {Han}, \citenamefont {Englert},
  \citenamefont {Miniatura} \emph {et~al.}}]{lee2009ultracold}%
  \BibitemOpen
  \bibfield  {author} {\bibinfo {author} {\bibfnamefont {K.~L.}\ \bibnamefont
  {Lee}}, \bibinfo {author} {\bibfnamefont {B.}~\bibnamefont {Gremaud}},
  \bibinfo {author} {\bibfnamefont {R.}~\bibnamefont {Han}}, \bibinfo {author}
  {\bibfnamefont {B.-G.}\ \bibnamefont {Englert}}, \bibinfo {author}
  {\bibfnamefont {C.}~\bibnamefont {Miniatura}}, \emph {et~al.},\ }\bibfield
  {title} {\bibinfo {title} {Ultracold fermions in a graphene-type optical
  lattice},\ }\href@noop {} {\bibfield  {journal} {\bibinfo  {journal}
  {Physical Review A}\ }\textbf {\bibinfo {volume} {80}},\ \bibinfo {pages}
  {043411} (\bibinfo {year} {2009})}\BibitemShut {NoStop}%
\end{thebibliography}%

\end{document}